\pdfoutput=1
\documentclass[11pt]{article}

\usepackage[english]{babel}
\usepackage{xcolor}
\usepackage{graphicx}

\usepackage[colorlinks,allcolors=blue,linktocpage=true]{hyperref}

\usepackage{cite}
\usepackage{subfigure}
\usepackage{amsmath}

\usepackage{placeins}

\usepackage[top=15mm,bottom=12mm,left=30mm,right=30mm,foot=12mm,includefoot]{geo
metry}

\allowdisplaybreaks[3]

\numberwithin{equation}{section}

\newcommand{\sect}[1]{section~#1}
\newcommand{\sects}[1]{sections~#1}
\newcommand{\app}[1]{appendix~#1}

\newcommand{\fig}[1]{figure~#1}

\newcommand{\tab}[1]{table~#1}
\newcommand{\tabs}[1]{tables~#1}

\newcommand{\eqn}[1]{equation~#1}
\newcommand{\eqs}[1]{equations~#1}

\newcommand{\half}{{\textstyle\frac{1}{2}}}

\newcommand{\ms}{\mskip 1.5mu}
\newcommand{\bs}{\mskip -1.5mu}

\newcommand{\tvec}[1]{\boldsymbol{#1}}

\newcommand{\msbar}{$\overline{\text{MS}}$ }

\newcommand{\msbars}{\scalebox{0.6}{$\overline{\text{MS}}$}}

\newcommand{\conv}[1]{\underset{#1}{\otimes}}

\newcommand{\SA}{S\bs A}
\newcommand{\AS}{A\bs S}
\newcommand{\tenbar}{\scalebox{0.7}{$\overline{10}$}}
\newcommand{\ten}{\scalebox{0.7}{$10\phantom{\overline{1}}\hspace{-1ex}$}}
\newcommand{\tentenbar}{\ten\ms\tenbar}
\newcommand{\tenbarten}{\tenbar\,\ten}
\newcommand{\twensev}{27\,27}

\newcommand{\Rp}{R^{\ms \prime} \bs}
\newcommand{\Rpp}{R^{\ms \prime\prime} \bs}

\newcommand{\Rbar}{\overline{R}}
\newcommand{\Rpbar}{\overline{R}{}^{\ms \prime} \bs}
\newcommand{\Rppbar}{\overline{R}{}^{\ms \prime\prime} \bs}

\newcommand{\pr}[2]{{}^{#1}\bs #2}      % with small backspace
\newcommand{\prb}[2]{{}^{#1}\! #2}      % with big backspace
\newcommand{\prn}[2]{{}^{#1} #2}        % without backspace

\newcommand{\Dp}[2]{p_{\Delta #1\Delta #2}}
\newcommand{\delp}[2]{p_{\delta #1\delta #2}}

\graphicspath{{figures/}}

\newcommand{\rev}[1]{#1}

\begin{document}

\begin{flushright}
DESY-22-206 \\
IPARCOS-UCM-23-026 \\
arXiv:2212.11843 [hep-ph]
\end{flushright}

\begin{center}
\vspace{4\baselineskip}
\textbf{\Large Two-loop evolution kernels for colour dependent \\[0.2em]
   double parton distributions} \\
\vspace{3\baselineskip}
Markus~Diehl$^{\ms 1}$, Florian Fabry$^{\ms 1}$ and Alexey Vladimirov$^{\ms 2}$
\end{center}

\vspace{2\baselineskip}

\noindent
${}^{1}$ Deutsches Elektronen-Synchrotron DESY, Notkestr.~85, 22607 Hamburg, Germany\\
${}^{2}$ Departamento de F\'isica Te\'orica and IPARCOS, Universidad Complutense de Madrid, \\
\hphantom{${}^{2}$} Plaza de Ciencias 1, 28040 Madrid, Spain

\vspace{3\baselineskip}

\parbox{0.9\textwidth}{
}

A key ingredient in the description of double parton distributions is their scale dependence.  If the colour of each individual parton is summed over, the distributions evolve with the same DGLAP kernels as ordinary parton distributions.  This is no longer true if the two partons are colour correlated.  We compute the relevant kernels for this case at next-to-leading order in the strong coupling, \rev{for unpolarised or longitudinally polarised partons and for transversely polarised quarks}.

\vfill

\newpage

\tableofcontents

\begin{center}
\rule{0.6\textwidth}{0.3pt}
\end{center}

%%%%%%%%%%%%%%%%%%%%%%%%%%%%%%%%%%%%%%%%%%%%%%%%%%%%%%%%%%%%%%%%%%%%%%%%%%%%%%%%

\section{Introduction}
\label{sec:introduction}

Double parton scattering is the mechanism by which two pairs of partons undergo a hard interaction in the same proton-proton collision.  A broad range of experimental studies has been carried out at the Tevatron and the LHC (see \cite{Abe:1997xk, Abazov:2015nnn, Aaij:2016bqq, Aaboud:2018tiq, CMS:2022pio} and references therein), and data from the forthcoming LHC runs will hopefully yield even more detailed insight.  Building on foundations laid in the 1980s \cite{Paver:1982yp, Mekhfi:1983az, Sjostrand:1986ep}, considerable progress towards a systematic theory of double parton scattering has been made in the last decade
\cite{Blok:2010ge, Gaunt:2011xd, Ryskin:2011kk, Blok:2011bu,
Diehl:2011yj, Manohar:2012jr, Manohar:2012pe, Ryskin:2012qx, Gaunt:2012dd, Blok:2013bpa, Diehl:2017kgu, Cabouat:2019gtm, Cabouat:2020ssr}.  There is a sustained interest in various theoretical aspects, as illustrated for instance by the recent work in \cite{Fedkevych:2020cmd, Ceccopieri:2021luf, Blok:2022mtv, Golec-Biernat:2022wkx}.  An overview of the subject can be found in \cite{Bartalini:2017jkk}.

To compute double parton scattering cross sections, one needs to know double parton distributions (DPDs), which quantify the joint probability for extracting two specified partons from a proton.  Beyond their practical importance, they are of interest for studying hadron structure, since they describe correlations between partons that are not accessible in single-parton distributions.

The scale dependence of DPDs has long been a point of theoretical interest, mainly because of an inhomogeneous term that appears in the evolution equation when the distributions are integrated over the transverse distance $y$ between the two partons \cite{Kirschner:1979im, Shelest:1982dg, Snigirev:2003cq, Gaunt:2009re, Ceccopieri:2010kg}.  As shown in \cite{Diehl:2011yj, Diehl:2017kgu, Buffing:2017mqm}, this term is however absent if $y$ is kept fixed.  We concentrate on DPDs at fixed~$y$ in the present work.

Most theory investigations consider the case in which the colour of each individual parton is summed over; we will call the corresponding distributions ``colour singlet DPDs'' for reasons explained later.  In this case, the evolution equations involve the same kernels that appear in the DGLAP equations for ordinary parton distributions (PDFs).  These kernels are known to high perturbative orders \cite{Moch:2004pa, Vogt:2004mw, Moch:2014sna, Blumlein:2021enk, Blumlein:2021ryt, Vogt:2018miu}.
In general, however, two partons inside a proton are correlated in their colour as well as in their polarisation \cite{Mekhfi:1985dv}.  Colour correlations between the partons are suppressed by Sudakov logarithms \cite{Mekhfi:1988kj, Manohar:2012jr}, but it is not obvious that this suppression is so strong that the correlations can be neglected altogether.  A recent study suggests that their effect may be large enough to be observable \cite{Blok:2022mtv}.  Moreover, it is clear that the Sudakov suppression becomes weaker as the scale of the hard scattering decreases.  Colour correlations in double parton scattering could hence play a role in the production of additional jets with moderate transverse momentum, which contribute to the underlying event if they are not explicitly identified.

These considerations motivate us to investigate the scale evolution for DPDs with general colour structure.  At leading order (LO) in $\alpha_s$ the corresponding kernels were already calculated in \cite{Diehl:2011yj, Buffing:2017mqm}.  Since LO calculations are often insufficient for a realistic analysis of proton-proton collisions, we compute these kernels at next-to-leading order (NLO) in the present work.  We do this both for unpolarised and for polarised partons.

We use two independent methods.  One of them modifies the calculation of the ordinary DGLAP kernels that is reported in \cite{Curci:1980uw, Ellis:1996nn, Vogelsang:1996im, Vogelsang:1997ak}.  The other one adapts the two-loop calculation \cite{Echevarria:2016scs, Gutierrez-Reyes:2018iod} of transverse-momentum dependent parton distributions (TMDs) in the short-distance limit.  The second method heavily relies on a detailed analysis of the renormalisation properties of DPDs and TMDs.

This paper is organised as follows.  In \sect{\ref{sec:evolution}} we recall some basics of the evolution and renormalisation of DPDs with nontrivial colour dependence, followed by a detailed analysis of the renormalisation factor at higher orders.  The two methods we use to compute the evolution kernels are presented in \sects{\ref{sec:first-method}} and \ref{sec:second-method}.  In \sect{\ref{sec:channels}} we analyse the mixing between different parton combinations in the presence of colour correlations, and we revisit the scheme transformation that is commonly applied to polarised evolution kernels computed in the \msbar scheme.  We present our results in \sect{\ref{sec:results}} and summarise them in \sect{\ref{sec:summary}}.  Some relations for the renormalisation of TMDs are collected in \app{\ref{sec:tmd-renorm}}.

\section{Evolution and renormalisation of colour dependent DPDs}
\label{sec:evolution}

A collinear (i.e.\ transverse-momentum integrated) DPD $\pr{R_1 R_2}{F}_{a_1 a_2}(x_1,x_2,y,\mu_1,\mu_2,\zeta)$ depends on the momentum fractions $x_1, x_2$ of the two observed partons, on their distance $y$ in the transverse plane, and on three scales.  These are the renormalisation scales $\mu_1, \mu_2$ associated with each parton and a rapidity parameter $\zeta$ with the dimension of a squared mass.  The labels $a_1, a_2$ specify both the type and the polarisation of the two partons.  We write $q, \bar{q}, g$ for an unpolarised parton, $\Delta q, \Delta\bar{q}, \Delta g$ for a longitudinally polarised one, and $\delta q, \delta\bar{q}$ for transversely polarised quarks.  A complete description of the gluon sector also requires linear gluon polarisation, which is not considered in the present work.

Finally, $R_1$ and $R_2$ are irreducible representations of the colour group and specify the state of the pair of fields associated with each parton in the definition of a DPD (see below).  The possible values for a quark or antiquark are $1$ for the colour singlet and $8$ for the octet.  For gluons we have the colour singlet ($1$), the antisymmetric octet ($A$), the symmetric octet ($S$), the decuplet and antidecuplet ($10$ and $\overline{10}$), and the representation of dimension $27$.  We write $\Rbar$ for the conjugate of a representation $R$, bearing in mind that for the cases just enumerated $\Rbar$ differs from $R$ only for the decuplet and antidecuplet.

Following earlier work in \cite{Diehl:2011yj} and \cite{Manohar:2012jr}, the evolution equations for DPDs in general colour representations have been derived in \cite{Buffing:2017mqm}.  The dependence on the scale $\mu_1$ is given by a DGLAP equation
\begin{align}
\label{eq:DGLAP-DPD}
   &\frac{d}{d\ln\mu_1} \pr{R_1 R_2}{F}_{a_1 a_2}(x_1,x_2,y,\mu_1,\mu_2,\zeta)
   \notag \\
   &\qquad =
   2\sum_{b,\Rp} \pr{R_1\Rpbar}{P}_{a_1b}(x_1^\prime,\mu_1^{},x_1^2\zeta)
   \conv{x_1}
   \pr{\Rp R_2}{F}_{b a_2}(x_1^\prime,x_2^{},y,\mu_1^{},\mu_2^{},\zeta)
   \,,
\end{align}
and there is an analogous equation for the dependence on $\mu_2$.  The kernel $\pr{R\Rpbar}{P}_{a b}$ describes transitions between representations $\Rp$ and $R$ of equal dimension, and the colour singlet kernels $\pr{11}{P}_{a b}$ are identical to the DGLAP kernels for ordinary PDFs.  We use the term ``colour de\-pendent DGLAP kernels'' to denote the set of all $\pr{R\Rpbar}{P}_{a b}$.

In physical terms, the parameter $\zeta$ specifies a rapidity cutoff and is defined with respect to a reference momentum $k$.  Technically, this can be written as $\zeta \sim (k^+)^2$ for a particle moving fast in the positive $z$ direction.\footnote{%
We use light-cone coordinates $v^{\pm} = (v^0 \pm v^3) / \sqrt{2}$ for any four-vector and take boldface to denote its transverse part $\tvec{v} = (v^1, v^2)$.  The full four-vector in light-cone components is written as $(v^+, v^-, \tvec{v})$.}
For the reasons given in \app{I.2} of \cite{Buffing:2017mqm}, the rapidity
parameter $\zeta$ in the DPD is defined w.r.t.\ the plus-momentum $p^+$ of the incoming proton.  The evolution kernel $P$ cannot depend on the proton momentum but only on the parton momenta before and after the splitting.  Our convention is to use the latter in the rapidity argument of $P$, so that $\zeta$ is scaled by a factor $x_1^2$ in \eqref{eq:DGLAP-DPD}.  We define a generalisation of the ordinary Mellin convolution as
\begin{align}
   A\bigl( x^\prime, s_A(x,x^\prime) \, \zeta \bigr)
   \conv{x} B\bigl( x^\prime, s_B(x,x^\prime) \, \zeta \bigr)
   &=
   \int_x^1 \frac{dx^\prime}{x^\prime}
   A\bigl( x^\prime, s_A(x,x^\prime) \, \zeta \bigr)
   \, B\biggl( \frac{x}{x^\prime},
      s_B\Bigl(x, \frac{x}{x^\prime} \Bigr) \, \zeta \biggr)
   \,,
\label{eqn:convolution_rescaling_definition}
\end{align}
where $s_A$ and $s_B$ are scaling factors of $\zeta$ that can depend on the integration variable $x'$ and on the argument $x$ of the convolution.  The convolution product thus defined is commutative and satisfies
\begin{align}
   \delta(1-x') \conv{x}
   B\bigl( x^\prime, s_B(x,x^\prime) \, \zeta \bigr)
   &= B\bigl( x, s_B(x,x) \, \zeta \bigr)
   \,.
\end{align}
We will see in \eqn{\eqref{eqn:Mellin_conv_associative}} that the convolution product is \emph{not} associative if a scaling factor in \eqref{eqn:convolution_rescaling_definition} depends on $x$.

Two gluons in the antisymmetric octet $A$ transform like a single gluon under charge conjugation, whereas two gluons in the symmetric octet transform with an extra minus sign.  Charge conjugation invariance therefore forbids transitions between gluons in the $A$ and $S$ representations, and we have
\begin{align}
   \label{eq:no-AS}
   \pr{\AS}{P}_{g g} &= \pr{\SA}{P}_{g g} = 0
   \,.
\end{align}
Another consequence of charge conjugation invariance is
\begin{align}
   \label{eq:10-eq-10bar}
   \pr{\tentenbar}{P}_{g g} &= \pr{\tenbarten}{P}_{g g}
   \,,
\end{align}
so that we only need to give results for the first of the two combinations.  Analogues of \eqref{eq:no-AS} and \eqref{eq:10-eq-10bar} hold for polarised gluons.

The rapidity dependence of the DPDs is given by a Collins-Soper equation
\begin{align}
\label{eq:CS-DPD}
   \frac{\partial}{\partial\ln\zeta}
   \pr{R_1 R_2}{F}_{a_1 a_2}(x_1,x_2,y,\mu_1,\mu_2,\zeta)
   &= \frac{1}{2} \, \pr{R_1}{J}(y, \mu_1,\mu_2) \,
     \pr{R_1 R_2}{F}_{a_1 a_2}(x_1,x_2,y,\mu_1,\mu_2,\zeta)
\end{align}
that has the same form as the Collins-Soper equation for single-parton TMDs \cite{Collins:2011zzd} but involves a different kernel $\pr{R}{J}$.  This kernel is zero for the colour singlet, and it is equal for representations with the same dimension:
\begin{align}
   \pr{1}{J} &= 0
   \,,
   &
   \pr{8}{J} &= \pr{A}{J} = \pr{S}{J} \,,
   &
   \pr{\ten}{J} &= \pr{\tenbar}{J}
   \,.
\end{align}
Its scale dependence is given by
\begin{align}
\label{eq:CS-RGE}
   \frac{d}{d\ln\mu_1} \pr{R_1}{J}(y, \mu_1,\mu_2)
   &=
   {}- \prn{R}{\gamma}_J(\mu_1)
\end{align}
and its analogue for $\mu_2$, where $\prn{R}{\gamma}_J$ plays the role of a cusp anomalous dimension.  The consistency of \eqref{eq:DGLAP-DPD} with \eqref{eq:CS-DPD} and \eqref{eq:CS-RGE} requires
\begin{align}
   \frac{\partial}{\partial\ln\zeta} \pr{R\Rp}{P}_{ab}(x,\mu,\zeta)
   = -\frac{1}{4} \delta_{R\Rpbar} \ms \delta_{a b} \delta(1-x)
   \, \prn{R}{\gamma}_J(\mu)
   \,,
\label{eqn:P_zeta_derivative}
\end{align}
so that we can write
\begin{align}
   \pr{R\Rp}{P}_{ab}(x,\mu,\zeta)
   &= \pr{R\Rp}{\widehat{P}}_{ab}(x,\mu)
   - \frac{1}{4} \delta_{R\Rpbar} \ms
   \delta_{a b} \delta(1-x) \, \prn{R}{\gamma}_J(\mu) \ln\frac{\zeta}{\mu^2}
   \,.
\label{eqn:P_zeta_dependence}
\end{align}
The reduced kernel $\pr{R\Rp}{\widehat{P}}_{ab}$ depends on $\mu$ only via $\alpha_s(\mu)$.

%%%%%%%%%%%%%%%%%%%%%%%%%%%%%%%%%%%%%%%%%%%%%%%%%%%%%%%%%%%%%%%%%%%%%%%%%%%%%%%%

\subsection{Renormalisation of DPDs}
\label{sec:renorm-dpds}

The equations just given follow from the renormalisation and the handling of rapidity divergences in the definition of DPDs.  A detailed derivation was given in \cite{Buffing:2017mqm} and is quite lengthy.  In the following, we will only sketch those aspects that are important for understanding the methods of calculation used in the present work.

\paragraph{Operators and colour structure.}

DPDs with nontrivial colour structure are defined from a generalisation of the familiar twist-two operators in \rev{which} colour indices are left open.  For unpolarised quarks, the generalised operator reads
\begin{align}
\label{eq:quark-ops}
   \mathcal{O}_{q}^{i i'}(x, \tvec{y}, \tvec{z})
   = 2p^+ \int \frac{dz^-}{2\pi}\, dy^-\; e^{i x p^+ z^-}
   &
   \bar{q}_{B \bs,\ms j'}^{} \bigl( y - \half z \bigr)\,
   W_{j' i'}^\dagger \bigl(y-\half z, v_L \bigr) \,
   \half \gamma^+ \,
   \notag \\
   &\times
   W_{i j} \bigl(y+\half z, v_L \bigr) \,
   q_{B \bs,\ms j}^{}\bigl( y + \half z \bigr)
   \Big|_{z^+ = y^+_{} = 0}
   \,,
\end{align}
where $i$ and $i'$ are indices in the fundamental representation of the colour group.
Given the integrations over the minus-components of the field positions, this operator corresponds to a parton with plus-momentum $x p^+$.  The Wilson line
\begin{align}
\label{WL-def}
   W_{i j}(\xi, v) &= \operatorname{P} \exp\biggl[\ms i g_B^{} \ms t^a_{i j}
       \int_{-\infty}^0 \!\! d s\; v A_B^a(\xi + s v) \ms\biggr]
   \,,
\end{align}
is in the fundamental representation and runs along a path specified by the vector $v$, which is set to $v_L = (0, 1, \tvec{0})$ in the matrix element.  At this point, we use bare (i.e.\ unrenormalised) field operators $q_B$ and $A_B$ and the bare coupling $g_B$.  Analogous definitions for polarised quarks and for gluons are given in \sect{3.1} of \cite{Buffing:2017mqm}.

To define an ordinary PDF for parton $a$, one takes the matrix element of
$\mathcal{O}_{a}^{r r'}(x, \tvec{0}, \tvec{0})$ in a proton state with momentum $p$ and contracts the open colour indices with $\delta_{r r'}$ (with indices $r$ and $r'$ in the fundamental or adjoint representation as appropriate).  Single-parton TMDs are defined from \rev{the} same matrix element with finite $\tvec{z}$.

Distributions of two partons are defined from the operator product
\begin{align}
\label{eq:parton-op-product}
   \mathcal{O}_{a_1}^{r_1^{} r_1'}(x_1, \tvec{y}, \tvec{z}_1) \,
   \mathcal{O}_{a_2}^{r_2^{} r_2'}(x_2, \tvec{0}, \tvec{z}_2)
   \,.
\end{align}
Specifically, a bare and rapidity-unsubtracted DPD $(F_{B \ms \text{us}, a_1 a_2})^{r_1^{} r_1' \ms r_2^{} r_2'}$ with open colour indices is obtained from the proton matrix element of this operator product, as specified in \eqn{(3.11)} of \cite{Buffing:2017mqm}.
For collinear DPDs, the transverse distances $\tvec{z}_1$ and $\tvec{z}_2$ between parton fields are set to zero, as in the PDF case.  The result is then projected onto a definite colour combination by coupling the index pairs $(r_1^{} r_1')$ and $(r_2^{}\ms r_2')$ to irreducible representations $R_1$ and $R_2$ of the colour group, such that one obtains an overall colour singlet.  The appropriate normalisation factors are given in \eqn{(I.11)} of \cite{Buffing:2017mqm}.

The second element in the construction is a soft factor, defined in terms of the operator
\begin{align}
   \label{eq:soft-op}
   & \bigl[ \mathcal{O}_{S,a}(\tvec{y},\tvec{z}) \bigr]^{r r' \!,\ms s s'}
   \nonumber \\
   & \qquad
      = \bigl[ W^{}(\tvec{y}+\half \tvec{z},v_L)\,
      W^\dagger(\tvec{y}+\half \tvec{z},v_R) \bigr]_{r s} \,
      \bigl[ W^{}(\tvec{y}-\half \tvec{z},v_R)\,
      W^\dagger(\tvec{y}-\half \tvec{z},v_L) \bigr]_{s' r'}
   \; ,
\end{align}
where $z^\pm = y^\pm = 0$.
The vector $v_R = (1,0,\tvec{0})$ points in the direction opposite to $v_L$, and the Wilson lines are in the fundamental (adjoint) representation for quarks (gluons).
As shown in \sect{4.4} of \cite{Buffing:2017mqm}, important simplifications arise if one sets $\tvec{z} = \tvec{0}$ in this operator: $(i)$ the projection of the index  pairs $(r r')$ and $(s s')$ onto representations $R_r$ and $R_s$ gives zero unless $R_r = \Rbar_s$, and $(ii)$ the projection of both index pairs on conjugate representations depends only on their dimension (and is hence equal for $R_r = 8$, $A$, and $S$).  Because $W^\dagger W = 1$, one obtains the identity operator if both index pairs are projected on the colour singlet.

The bare soft factor $S_B$ for DPDs is obtained from the vacuum expectation value of the product
\begin{align}
\label{eq:soft-op-product}
   \mathcal{O}_{S,a_1}^{r_1^{} r_1' ,\, s_1^{} s_1'}(\tvec{y},\tvec{z}_1) \,
   \mathcal{O}_{S,a_2}^{r_2^{} r_2' ,\, s_2^{} s_2'}(\tvec{0},\tvec{z}_2)
\end{align}
with appropriate projections of the colour indices on an overall singlet, as specified in \eqn{(I.15)} of \cite{Buffing:2017mqm}.
For collinear DPDs one sets $\tvec{z}_1 = \tvec{z}_2 = \tvec{0}$ in the soft factor, just as in the rapidity-unsubtracted DPD.  With the simplifications just stated, one has different soft factors characterised by $(r_1^{} r_1')$ being in a colour octet, decuplet, or the 27 representation, whereas the soft factor for the colour singlet is unity.

For non-singlet representations, the construction just outlined gives rapidity divergences and must be modified.  A rapidity regulator is introduced to this end; two regulators that have been used in DPD studies are the Collins regulator with Wilson lines off the light-cone \cite{Collins:2011zzd} and the so-called $\delta$ regulator in the form of \cite{Echevarria:2015byo, Echevarria:2016scs}.\footnote{%
An overview of these and other regulators used in the TMD literature can be found in \app{B} of \cite{Ebert:2019okf}.
}
A bare rapidity-subtracted DPD $\pr{R_1 R_2}{F}_{B}$ is then obtained from $\pr{R_1 R_2}{F}_{B \ms \text{us}} / \sqrt{\pr{R_1}{S}_B}$ by removing the rapidity regulator, which leaves a dependence on the rapidity parameter $\zeta$.  Explicit expressions for how $\zeta$ is introduced in this procedure are given in \sect{2.2} of \cite{Diehl:2021wpp} for both regulators just mentioned.  We follow the procedure outlined in \cite{Collins:2011zzd} for TMDs and remove the rapidity regulator before the ultraviolet regulator is removed, i.e.\ before setting $\epsilon=0$ in dimensional regularisation.

%...............................................................................

\paragraph{Renormalisation.}

For $\tvec{z} = \tvec{0}$, the operators in \eqref{eq:quark-ops} and \eqref{eq:soft-op} have ultraviolet divergences beyond those associated with the bare fields, because they involve products of fields separated by a light-like distance.  The projections of the bare parton operators \rev{$\mathcal{O}^{r r'}_a(x, \tvec{y}, \tvec{0})$} on definite colour representations are renormalised by  convolution with functions $\pr{R \Rp}{Z}_{\text{us}, a b}(x)$.  Their index pairs $(R \Rp)$ and $(a b)$ are needed to describe the mixing between quark and gluon operators under renormalisation.  The bare soft operators \rev{$\mathcal{O}_{\smash{S}\vphantom{s}}^{r r' \!,\ms s s'}(\tvec{y}, \tvec{0})$}, with $(r r')$ projected on $R$ and $(s s')$ projected on $\Rbar$ are renormalised by multiplication with a factor $\pr{R}{Z}_S$.
The renormalisation factor $\pr{R \Rp}{Z}_{a b}(x)$ for the rapidity-subtracted DPD is then obtained from $\pr{R \Rp}{Z}_{\text{us}, a b}(x) / \sqrt{\pr{R}{Z}_S}$ by removing the rapidity regulator in the same way as for the DPD.  The renormalised DPD is obtained as
\begin{align}
   \pr{R_1 R_2}{F}_{a_1 a_2}(x_1,x_2,y,\mu_1,\mu_2,\zeta)
   &= \sum_{b_1,\Rp}
   \pr{R_1\Rpbar}{Z}_{a_1b_1}(x_1^\prime,\mu_1^{},x_1^2\zeta)
   \,\conv{x_1}\,
   \sum_{b_2,\Rpp}
   \pr{R_2\Rppbar}{Z}_{a_2b_2}(x_2^\prime,\mu_2^{},x_2^2\zeta)
   \notag\\
   &\qquad\quad
   \conv{x_2} \pr{\Rp\Rpp}{F}_{B \bs,\ms b_1b_2}(x_1^\prime,x_2^\prime,
      y,\mu_1^{},\mu_2^{},\zeta)
   \,,
\label{eqn:DPD_renormalisation}
\end{align}
where it is understood that after convolution with the $Z$ factors one can set $\epsilon=0$ in dimensional regularisation.
Since the bare distributions are $\mu$ independent by construction, one obtains the DGLAP equation \eqref{eq:DGLAP-DPD} schematically from
\begin{align}
\label{eq:RGE-DPD-derive}
   \frac{d}{d\ln\mu_1} F(\mu_1, \mu_2)
   &= \frac{d}{d\ln\mu_1}
   \bigl[ Z(\mu_1) \conv{x_1} Z(\mu_2) \conv{x_2} F_B \bigr]
   = \biggl[ \frac{d}{d\ln\mu_1} Z(\mu_1) \biggr]
      \conv{x_1} Z(\mu_2) \conv{x_2} F_B
   \notag \\[0.1em]
   & \underset{\text{def}}{=}\,
   \bigl[ P(\mu_1) \conv{x_1'} Z(\mu_1) \bigr]
      \conv{x_1} Z(\mu_2) \conv{x_2} F_B
   =  P(\mu_1) \conv{x_1} \bigl[ Z(\mu_1) \conv{x_1'} Z(\mu_2)
      \conv{x_2} F_B \bigr]
   \notag \\
   &= P(\mu_1) \conv{x_1} F(\mu_1, \mu_2)
   \,,
\end{align}
where we omitted all indices and arguments that are not essential for the manipulations.  Whilst this is a standard argument when manipulating a renormalisation group equation (RGE), we need to be careful about the rescaling of $\zeta$.  The renormalisation factors in \eqref{eqn:DPD_renormalisation} are defined such that they refer to the momentum of the first or second parton in the DPD, which explains the rescaling by $x_1^2$ or $x_2^2$ of $\zeta$ (given that the argument $\zeta$ of the DPD refers to the proton rapidity).  In the second line of \eqref{eq:RGE-DPD-derive} we need to use the modified law of associativity
\begin{align}
\label{eqn:Mellin_conv_associative}
   &
   \Bigl[A(x^{\prime\prime},x^2 \zeta) \conv{x^\prime}
   B(x^{\prime\prime},
      x^{\prime\prime\, 2\ms} x^2 \zeta / x^{\prime\, 2}) \Bigr]
   \conv{x} C(x^\prime,\zeta)
   \notag \\
   &\qquad\quad
   = A(x^{\prime},x^2\zeta) \conv{x}
      \Bigl[ B(x^{\prime\prime},x^{\prime \, 2} \zeta)
      \conv{x^{\prime}} C(x^{\prime\prime},\zeta) \Bigr]
   \,,
\end{align}
which is readily verified by writing out the convolution integrals and making appropriate substitutions of the integration variables.  The correct relation between the renormalisation factors and the associated DGLAP kernels is therefore
\begin{align}
   \frac{d}{d\ln\mu} \pr{R\Rp}{Z}_{ab}(x,\mu,\zeta)
   &= 2 \sum_{c,\Rpp}\pr{R\Rppbar} {P}_{ac}(x^\prime,\mu,\zeta)
   \conv{x} \pr{\Rpp\Rp}{Z}_{c b}
   \bigl(x^\prime,\mu,x^{\prime\ms 2} \zeta/ x^2 \bigr)
   \,.
\label{eqn:RGE_analysis_Z_derivative}
\end{align}

%...............................................................................

\paragraph{Rapidity dependence.}

The $\zeta$ dependence of the bare DPD $F_B$ is given by the
analogue of \eqref{eq:CS-DPD} with a bare Collins-Soper kernel $J_B$.  This kernel is renormalised additively as
\begin{align}
   \pr{R}{J}(y, \mu_1,\mu_2)
   &=
   \prn{R}{\Lambda}(\mu_1) + \prn{R}{\Lambda}(\mu_2) + \pr{R}{J}_{B}(y)
   \,,
\end{align}
where $\Lambda$ describes the $\zeta$ dependence of the renormalisation factor $Z$ via
\begin{align}
   \frac{\partial}{\partial\ln\zeta} \, \pr{R\Rp}{Z}_{a b}(x,\mu,\zeta)
   &=
   \frac{1}{2} \ms \prn{R}{\Lambda}(\mu)\, \pr{R\Rp}{Z}_{a b}(x,\mu,\zeta)
   \,.
\label{eqn:Lambda_def}
\end{align}
This leads to the Collins-Soper equation \eqref{eq:CS-DPD} for the renormalised DPD.  The anomalous dimension associated with $\Lambda$ is
\begin{align}
   \prn{R}{\gamma}_J(\mu)
   &= {} - \frac{d}{d\ln\mu}\prn{R}{\Lambda}(\mu)
   \,,
\end{align}
which gives the RGE \eqref{eq:CS-RGE} for the renormalised Collins-Soper kernel.

%%%%%%%%%%%%%%%%%%%%%%%%%%%%%%%%%%%%%%%%%%%%%%%%%%%%%%%%%%%%%%%%%%%%%%%%%%%%%%%%

\subsection{Explicit form of the renormalisation factor}
\label{sec:Z-factor}

The relation \eqref{eqn:RGE_analysis_Z_derivative} is the basis for extracting the colour dependent DGLAP kernels from the ultraviolet divergences of appropriately chosen graphs.  We now derive the explicit relation between $\pr{R\Rp}{P}$ and the poles of $\pr{R\Rp}{Z}$ in dimensional regularisation, order by order in perturbation theory.

We expand all quantities in
\begin{align}
   a_s &= \frac{\alpha_s}{2\pi}
\end{align}
and use the following convention for the $\beta$ function:
\begin{align}
   \frac{d a_s(\mu)}{d\ln\mu}
   &= \frac{\beta\bigl( a_s(\mu) \bigr)}{\pi}
   = {}- \sum_{n=1}^\infty \beta_{n-1} \, a_s^{n+1}(\mu)
   \,.
\end{align}
Evolution kernels and anomalous dimensions start at order $a_s$, and we write
\begin{align}
   \pr{R\Rp}{P}_{a b}(x,\mu,\zeta)
   &= \sum_{n=1}^\infty a_s^{n}(\mu) \,
      \pr{R\Rp}{P}_{a b}^{(n-1)}(x,\zeta/\mu^2)
   \,,
   \\
   \prn{R}{\gamma}_J(\mu)
   &= \sum_{n=1}^\infty a_s^{n}(\mu) \, \prn{R}{\gamma}_J^{(n-1)}
   \,,
\end{align}
where the expansion coefficients $P^{(n-1)}$ depend on the ratio $\zeta/\mu^2$ for dimensional reasons.  The renormalisation factor $Z$ has a tree-level term and reads
\begin{align}
   \label{eq:expand-Z-DPD}
   \pr{R\Rp}{Z}_{a b}(x,\mu,\zeta,\epsilon)
   &= \delta_{R\Rpbar} \ms \delta_{a b} \delta(1-x)
      + \sum_{n=1}^\infty a_s^n(\mu) \,
      \pr{R\Rp}{Z}_{a b}^{(n)}(x,\zeta/\mu^2,\epsilon)
   \,,
\intertext{whereas $\Lambda$ starts at order $a_s$,}
   \prn{R}{\Lambda}(\mu,\epsilon)
   &= \sum_{n=1}^\infty a_s^n(\mu)\ \prn{R}{\Lambda}^{(n)}(\epsilon)
   \,,
\end{align}
as one readily sees from \eqref{eqn:Lambda_def}.

Throughout this work, we assume that one uses \msbar renormalisation implemented in such a way that all counterterms are pure poles in $\epsilon$.  More detail will be given in \eqref{eq:as-renorm} and below.  We thus have
\begin{align}
\label{eq:Z-Lambda-epsilon}
   \pr{R\Rp}{Z}_{a b}^{(n)}(\epsilon)
   &= \sum_{i=1}^\infty \frac{1}{\epsilon^i} \, \pr{R\Rp}{Z}_{a b}^{(n,i)}
   \,,
   &
   \prn{R}{\Lambda}^{(n)}(\epsilon)
   &= \sum_{i=1}^\infty\frac{1}{\epsilon^i} \, \prn{R}{\Lambda}^{(n,i)}
   \,,
\end{align}
where the highest order of the poles for given $n$ will be determined below.
Using the RGE derivative in $4 - 2\epsilon$ dimensions,
\begin{align}
\label{eq:rge-derivative}
   \frac{d}{d\ln \mu}
   &=
   \frac{\partial}{\partial\ln\mu}
   + \biggl(\frac{d a_s(\mu)}{d\ln\mu} - 2 \epsilon a_s(\mu) \biggl) \ms
      \frac{\partial}{\partial a_s}
   \,,
\end{align}
one can show that $\pr{R\Rp}{P}$ defined by \eqref{eqn:RGE_analysis_Z_derivative} does not depend on $\epsilon$.  The proof is easy and proceeds by recursion in the perturbative order.

%...............................................................................

\paragraph{Poles of $\Lambda$.}
Using \eqref{eq:rge-derivative} together with the fact that $\partial\, \prn{R}{\Lambda} / (\partial\ln\mu) = 0$, we get
\begin{align}
   \prn{R}{\gamma}_J
   &=
   \biggl( \sum_{m=1}^\infty
      \beta_{m-1} \, a_s^{m+1} + 2 \epsilon a_s\biggr)
   \biggl( \sum_{n=1}^\infty n \ms a_s^{n-1}\ \prn{R}{\Lambda}^{(n)}(\epsilon) \biggr)
   \,,
\end{align}
where we indicate explicitly which expressions depend on $\epsilon$.  Inserting the Laurent series \eqref{eq:Z-Lambda-epsilon} and comparing the terms of $\epsilon^0$ on both sides, we find that
\begin{align}
   \prn{R}{\Lambda}^{(N,1)}
   &=
   \frac{1}{2N}\prn{R}{\gamma}_J^{(N-1)} \,,
\end{align}
whilst the absence of poles in $\epsilon$ implies that
\begin{align}
   \prn{R}{\Lambda}^{(N,m)}
   &=
   {}- \frac{1}{2N}\sum_{n=1}^{N-1} (N-n) \beta_{n-1}
   \prn{R}{\Lambda}^{(N-n,m-1)}
   &
   \text{ for } m>1.
\end{align}
By recursion in the perturbative order, one can show that
\begin{align}
\label{eq:Lambda-pole-constraint}
   \prn{R}{\Lambda}^{(N,m)} &= 0
   &
   \text{ for } m>N,
\end{align}
so that the highest pole of $\Lambda$ at order $a_s^N$ is $\epsilon^{-N}$.

%...............................................................................

\paragraph{Poles of $Z$.}

The RGE analysis of $Z$ is more involved, and we omit parton labels and the associated sums for the time being.
If in \eqref{eqn:RGE_analysis_Z_derivative} we insert the perturbative expansion of $Z$ in its convolution with $P$, we can isolate a term $P$ and write
\begin{align}
   \pr{R\Rp}{P}(a_s)
   &=
   \frac{1}{2} \frac{\partial}{\partial\ln\mu}
   \pr{R\Rp}{Z}(a_s,\epsilon)
   - \frac{1}{2} \biggl( \sum_{m=1}^\infty
   \beta_{m-1} \, a_s^{m+1} + 2 \epsilon a_s\biggr)
   \biggl(\sum_{n=1}^\infty n \ms a_s^{n-1}
   \ \pr{R\Rp}{Z}^{(n)}(\epsilon)\biggr)
   \notag\\
   & \quad
   {}- \sum_{\Rpp} \pr{R\Rppbar}{P}(a_s)
   \otimes \biggl(\sum_{n=1}^{\infty}a_s^{n}
   \ \pr{\Rpp\Rp}{Z}^{(n)}(\epsilon)\biggr)
   \,,
\label{eqn:RGE_analysis_general_ansatz}
\end{align}
where we have also used the form \eqref{eq:rge-derivative} of $d/(d\ln\mu)$ and explicitly performed the derivative of $Z$ w.r.t.\ $a_s$.  According to \eqref{eq:expand-Z-DPD}, the explicit $\mu$ dependence of $Z$ is via the ratio $\zeta/\mu^2$, so that we can rewrite
\begin{align}
\label{eqn:RGE_analysis_Z_mu_derivative}
   \frac{1}{2} \frac{\partial}{\partial\ln\mu} \, \pr{R\Rp}{Z}(\zeta/\mu^2)
   &=
   {}- \frac{\partial}{\partial\ln\zeta}\,
   \pr{R\Rp}{Z}(\zeta/\mu^2)
   \,,
\end{align}
where the r.h.s.\ is given by \eqref{eqn:Lambda_def}.  Using this in \eqref{eqn:RGE_analysis_general_ansatz} and extracting the term of order $a_s^N$ gives
\begin{align}
   \pr{R\Rp}{P}^{(N-1)}
   &=
   {}- \epsilon N \ms \prb{R\Rp}{Z}^{(N)}(\epsilon)
   - \frac{1}{2} \delta_{R\Rpbar} \ms \delta(1-x) \,
      \prn{R}{\Lambda}^{(N)}(\epsilon)
   - \sum_{n=1}^{N-1}
   \Biggl( \frac{1}{2} (N-n) \beta_{n-1} \prb{R\Rp}{Z}^{(N-n)}(\epsilon)
   \notag \\
   &\quad
   + \sum_{\Rpp} \pr{R\Rppbar}{P}^{(n-1)}
      \otimes \pr{\Rpp\Rp}{Z}^{(N-n)}(\epsilon)
   + \frac{1}{2} \prn{R}{\Lambda}^{(n)}(\epsilon) \,
      \prb{R\Rp}{Z}^{(N-n)}(\epsilon)
   \Biggr)
   \,.
\label{eqn:RGE_analysis_colour_singlet_arbitrary_order}
\end{align}
The only finite contribution at $\epsilon=0$ on the r.h.s.\ is due to the first term, so that we get
\begin{align}
   \label{eq:P-from-Z-residue}
   \pr{R\Rp}{P}^{(N-1)}
   &=
   {}-N \, \pr{R\Rp}{Z}^{(N,1)}
   \,,
\end{align}
whilst the absence of poles implies
\begin{align}
   \pr{R\Rp}{Z}^{(1,m)}
   &=
   {}- \frac{1}{2} \delta_{R\Rpbar} \ms \delta(1-x)
   \, \prn{R}{\Lambda}^{(1,m-1)}
   &
   \text{ for } m>1
\label{eqn:RGE_analysis_RZ_LO_coefficients_preliminary}
\end{align}
and
\begin{align}
   \pr{R\Rp}{Z}^{(N,m)}
   = {}- \frac{1}{N}
   &
   \Biggl\{
   \frac{1}{2} \delta_{R\Rpbar} \ms \delta(1-x)
   \, \prn{R}{\Lambda}^{(N,m-1)}
   + \sum_{n=1}^{N-1}
   \biggl( \frac{1}{2} (N-n) \beta_{n-1}
   \pr{R\Rp}{Z}^{(N-n,m-1)}
   \notag\\
   &
   + \sum_{\Rpp} \pr{R\Rppbar}{P}^{(n-1)}
   \otimes \pr{\Rpp\Rp}{Z}^{(N-n,m-1)}
   + \frac{1}{2} \sum_{i=1}^{m-2}
   \prn{R}{\Lambda}^{(n,i)} \; \pr{R\Rp}{Z}^{(N-n,m-i-1)}
   \biggr)
   \Biggr\}
   \notag \\[0.3em]
   & \hspace{18em}
   \text{ for } N>1 \text{ and } m>1.
\label{eqn:RGE_analysis_RZ_NLO_coefficients_preliminary}
\end{align}
Using the constraint \eqref{eq:Lambda-pole-constraint} for the poles of $\Lambda$, one can show by recursion over the perturbative order that
\begin{align}
   \pr{R\Rp}{Z}^{(N,m)} &= 0
   &
   \text{ for } m > 2 N,
\end{align}
such that the highest pole of $Z$ at order $a_s^{N}$ is $\epsilon^{- 2 N}$.  In the colour singlet case, where $\Lambda = 0$, the order highest pole is $\epsilon^{-N}$ instead.

%...............................................................................

\paragraph{Explicit results at LO and NLO.}
Restoring parton labels and function arguments, we finally obtain the explicit LO and NLO renormalisation factors:
\begin{align}
   \pr{R\Rp}{Z}_{ab}^{(1,1)}(x,\zeta/\mu^2)
   &=
   {}- \pr{R\Rp}{P}_{ab}^{(0)}(x,\zeta/\mu^2)
   \,,
   \\[0.2em]
   \pr{R\Rp}{Z}_{ab}^{(1,2)}(x)
   &=
   {}- \frac{1}{4} \delta_{R\Rpbar} \ms
   \delta_{a b} \delta(1-x) \prn{R}{\gamma}_J^{(0)}
   \intertext{and}
   \pr{R\Rp}{Z}_{ab}^{(2,1)}(x,\zeta/\mu^2)
   &=
   {}- \frac{1}{2} \, \pr{R\Rp}{P}_{ab}^{(1)}(x,\zeta/\mu^2)
   \,,
   \\[0.2em]
   \pr{R\Rp}{Z}_{ab}^{(2,2)}(x,\zeta/\mu^2)
   &=
   \frac{1}{2} \ms \biggl\{
   \sum_{c,\Rpp} \pr{R\Rppbar}{P}_{ac}^{(0)}(x^\prime,\zeta/\mu^2) \conv{x}
   \pr{\Rpp\Rp}{P}_{c b}^{(0)}\bigl(x^\prime,
      x^{\prime\, 2} \zeta/(x^2 \mu)^2 \ms\bigr)
   \notag\\
   &\qquad\
   + \frac{1}{2} \ms \beta_0 \pr{R\Rp}{P}^{(0)}_{ab}(x,\zeta/\mu^2)
   - \frac{1}{8} \delta_{R\Rpbar} \ms \delta_{a b} \delta(1-x)
   \prn{R}{\gamma}_J^{(1)} \biggr\} \notag\\
   &=
   \frac{1}{2} \ms \sum_{c,\Rpp} \pr{R\Rppbar}{\widehat{P}}_{ac}^{(0)}(x^\prime)
   \conv{x} \pr{\Rpp\Rp}{\widehat{P}}_{c b}^{(0)}(x^\prime)
   \notag\\
   &\quad
   + \frac{1}{4} \ms \biggl[ \beta_0 - \prn{R}{\gamma}_J^{(0)}
   \Bigl( \ln\frac{\zeta}{\mu^2} - \ln x \Bigr) \biggr]
   \pr{R\Rp}{\widehat{P}}_{ab}^{(0)}(x) \notag\\
   &\quad
   + \frac{1}{16} \delta_{R\Rpbar} \ms \delta_{a b} \delta(1-x) \notag\\
   &\qquad
   \times \biggl[ \frac{1}{2} \Bigr( \prn{R}{\gamma}_J^{(0)}
   \ln \frac{\zeta}{\mu^2} \Bigl)^2 - \beta_0 \ms \prn{R}{\gamma}_J^{(0)}
   \ln\frac{\zeta}{\mu^2} - \prn{R}{\gamma}_J^{(1)}\biggr]
   \,,
   \label{eq:Z22-final}
   \\[0.2em]
   \pr{R\Rp}{Z}_{ab}^{(2,3)}(x,\zeta/\mu^2)
   &=
   \frac{1}{4} \ms \prn{R}{\gamma}_J^{(0)}
   \biggl\{ \pr{R\Rp}{P}^{(0)}_{ab}(x,\zeta/\mu^2)
   + \frac{3}{8} \delta_{R\Rpbar} \ms \delta_{a b} \ms \delta(1-x) \ms \beta_0
   \biggr\}
   \,,
   \\[0.2em]
   \pr{R\Rp}{Z}_{ab}^{(2,4)}(x)
   &=
   \frac{1}{32} \delta_{R\Rpbar} \ms \delta_{a b} \ms \delta(1-x) \ms
   \Bigl( \prn{R}{\gamma}_J^{(0)} \Bigr)^2
   \,.
\end{align}
In the second expression for $\pr{R\Rp}{Z}_{ab}^{(2,2)}$, we have made the $\zeta$ dependence of $\pr{R\Rp}{P}_{ab}^{(0)}$ explicit by using \eqref{eqn:P_zeta_dependence}.  Notice the $\ln x$ term in the forth line of \eqref{eq:Z22-final}, which is a consequence of the rescaling of $\zeta$.

\section{First method: modified splitting kernel calculation}
\label{sec:first-method}

To calculate the colour dependent DGLAP kernels at NLO, it is natural to make use of one of the many computations of the ordinary DGLAP kernels at that order.  We do so by taking the calculations in \cite{Curci:1980uw, Ellis:1996nn, Vogelsang:1996im}, where detailed results are given for the contributions of individual Feynman graphs.  This allows us to extract the colour dependent kernels, with the limitations described below.

\begin{figure}
\begin{center}
   \subfigure[graph (h)]{\includegraphics[height=0.24\textwidth]{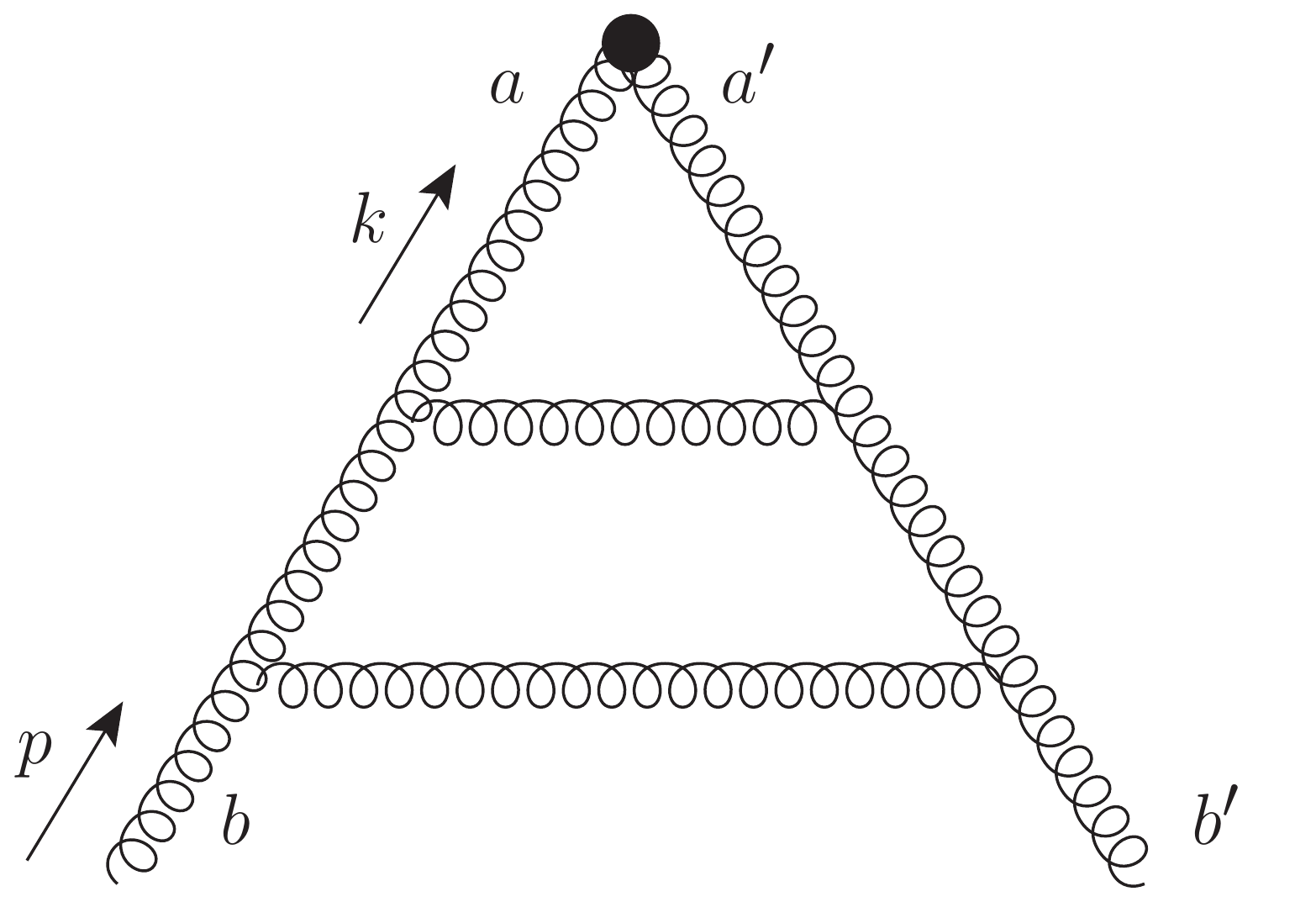}}
   \caption{\label{fig:ladder-graph} Graph contributing to the two-loop splitting kernel for gluons.  At the top of the graph is the vertex with the twist-two operator, and $k$ denotes the momentum flowing into this vertex.  $a, a'$ and $b, b'$ are open colour indices.  The label (h) for the graph corresponds to the nomenclature in \protect\cite{Curci:1980uw, Ellis:1996nn, Vogelsang:1996im}.}
\end{center}
\end{figure}

The method used in the above papers is to extract the DGLAP kernels from the $1/\epsilon$ poles of appropriate graphs.  In dimensional regularisation, a  quantity with mass dimension is required to obtain nonzero loop integrals, which is achieved by putting a cutoff on the transverse momentum $\tvec{k}$ of the parton lines that couple to the twist-two operator (see \fig{\ref{fig:ladder-graph}}).

According to \eqref{eq:P-from-Z-residue}, the relation between the DGLAP kernel and the $1/\epsilon$ pole of the renormalisation factor $Z(x)$ remains true in colour non-singlet channels, so that we can extract $\pr{R\Rp}{P}(x)$ by projecting the colour indices at the top of the graph onto $R$ and those at the bottom onto $\Rp$.  The corresponding tensors in colour space are given in \eqn{(5.19)} of \cite{Diehl:2011yj}.

For each graph that does not contain a four-gluon vertex, the colour projection gives a global factor depending on $R\Rp$, with the $x$ dependence remaining unchanged.
For unpolarised partons, the latter is given for individual graphs in \tab{1} and \eqn{(4.51)} of \cite{Curci:1980uw}, and in \tabs{1} to 4 and \eqn{(57)} of \cite{Ellis:1996nn}.  For longitudinal polarisation, it can be found in \tabs{1} to 4 and \eqn{(45)} of \cite{Vogelsang:1996im}.  Transverse polarisation is discussed in \sect{\ref{sec:transverse-graphs}}.

A minor technical complication arises because some entries in the tables refer to the sum of two graphs with the same topology but different parton types in the loops, namely the pairs (cd) and (fg) in the notation of the quoted papers.  However, each of these pairs contributes to two colour structures in the tables, e.g.\ to $C_F^2$ and $C_F \ms C_A$ for the pair (cd) in $\pr{11}{P}_{q q}$, or to $C_A^2$ and $C_A \ms n_f$ for the pair (fg) in $\pr{11}{P}_{g g}$.  This allows us to extract the contribution of individual graphs in all cases.

In the calculation of the splitting kernels, a graph with double-ladder topology (h) such as the one in \fig{\ref{fig:ladder-graph}} requires subtracting the convolution of two LO kernels, which is labelled as graph (i) in the quoted work.  The overall colour factor of (h) and (i) is identical, such that a separation of the two contributions is not needed in this case.

%...............................................................................

\paragraph{Choice of gauge and $\delta(1-x)$ terms.}

The calculations in \cite{Curci:1980uw, Ellis:1996nn, Vogelsang:1996im} are performed in light-cone gauge, with the principal value prescription for the poles at $\ell^+$ of the gluon propagator for momentum $\ell$:
\begin{align}
   \frac{1}{\ell^+}
   &\to
   \frac{\ell^+}{(\ell^+)^2 + \delta^2 \ms (p^+)^2}
   \,.
\end{align}
Here the momentum $p$ of the incoming parton (see \fig{\ref{fig:ladder-graph}}) serves as a reference, and $\delta$ is an infinitesimal parameter that must be set to zero at the end of the calculation.  In graphs that contain rapidity divergences, this prescription leads to integrals
\begin{align}
\label{eq:PV-gauge-integrals}
   I_i &= \int_0^1 du \, \frac{u \, \ln^{\ms i} \! u}{u^2+\delta^2}
   &&
   \text{ with } i = 0,1
\end{align}
that diverge for $\delta \to 0$ and therefore must cancel in the sum over all graphs.  We find that this cancellation takes place in all colour channels, which provides a consistency check of our calculation.

We are not aware of a way to connect the principal value prescription in the axial-gauge propagator with a rapidity regulator and an associated variable $\zeta$.  Since the $\zeta$ dependence of the DGLAP kernels comes with a factor $\delta(1-x)$ according to \eqref{eqn:P_zeta_dependence}, we can still extract the $x$ dependent part of the colour dependent kernels with the method just described.

In fact, the $\delta(1-x)$ part of the colour singlet kernels was not determined from Feynman graphs in \cite{Ellis:1996nn}, but by using the number and momentum sum rules for unpolarised partons.  Since there are no such sum rules for other colour representations, we cannot use this trick to get the $\delta(1-x)$ terms in the kernels for other colour representations.  Instead, this will be achieved using the method described in \sect{\ref{sec:second-method}}.

%%%%%%%%%%%%%%%%%%%%%%%%%%%%%%%%%%%%%%%%%%%%%%%%%%%%%%%%%%%%%%%%%%%%%%%%%%%%%%%%

\subsection{Graphs with four-gluon vertices}

The Feynman rule for the four-gluon vertex does not factorise into a part depending on the colour and another depending on the polarisation and momentum of the gluons.  It is therefore not guaranteed that the projection on different colour representations at the top and the bottom of the graphs leads to a global colour factor times a colour independent structure depending on the momentum fraction.

We have therefore recomputed the graphs containing a four-gluon vertex, which are shown in \fig{\ref{fig:4-g-graphs}}.  The loop integrals can be readily performed using the methods described in the appendices of \cite{Ellis:1996nn}.

\begin{figure}
\begin{center}
   \subfigure[graph (j)\label{fig:graph-j}]{\includegraphics[height=0.24\textwidth]{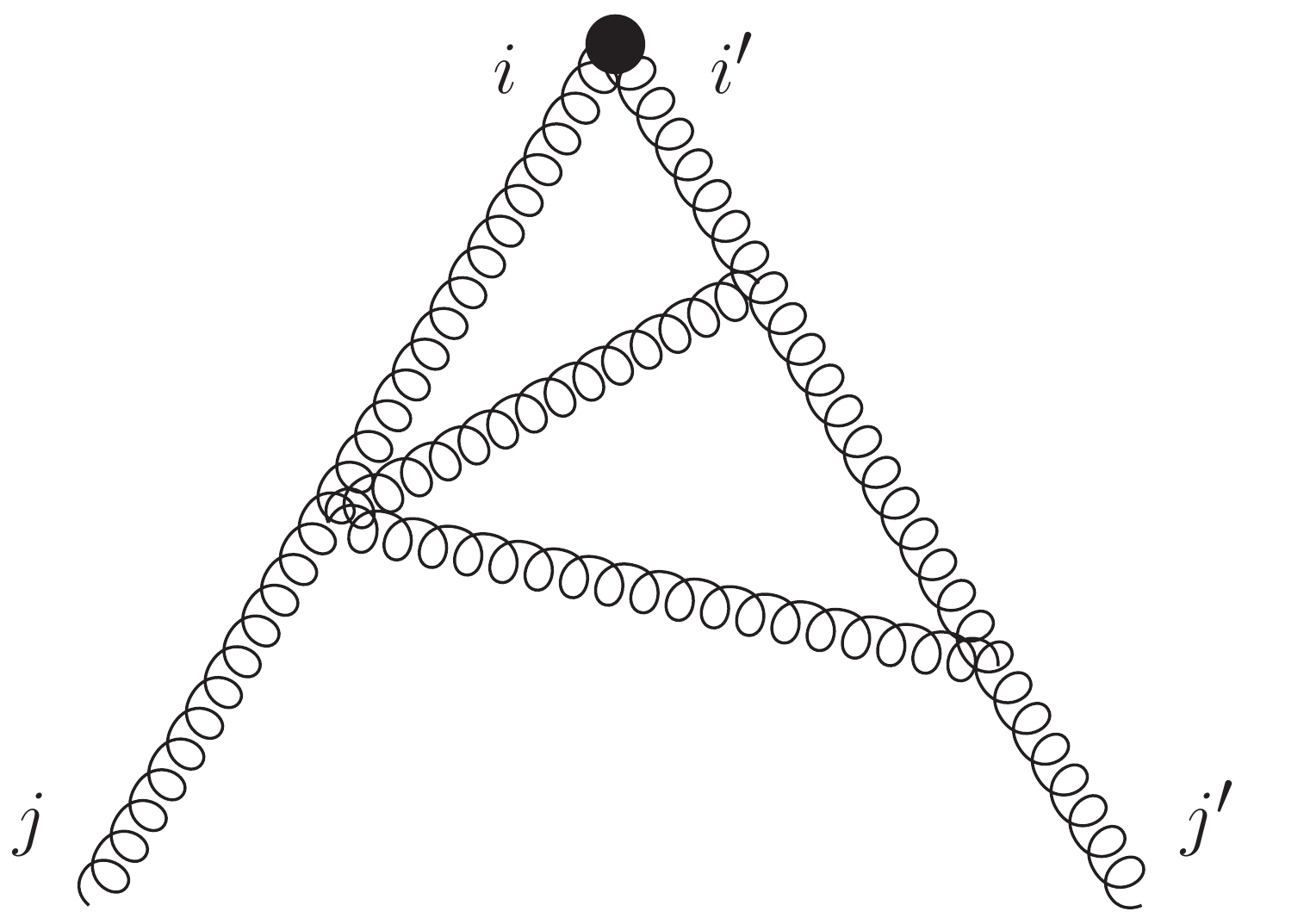}}
   \hspace{2.4em}
   \subfigure[graph (k)]{\includegraphics[height=0.24\textwidth]{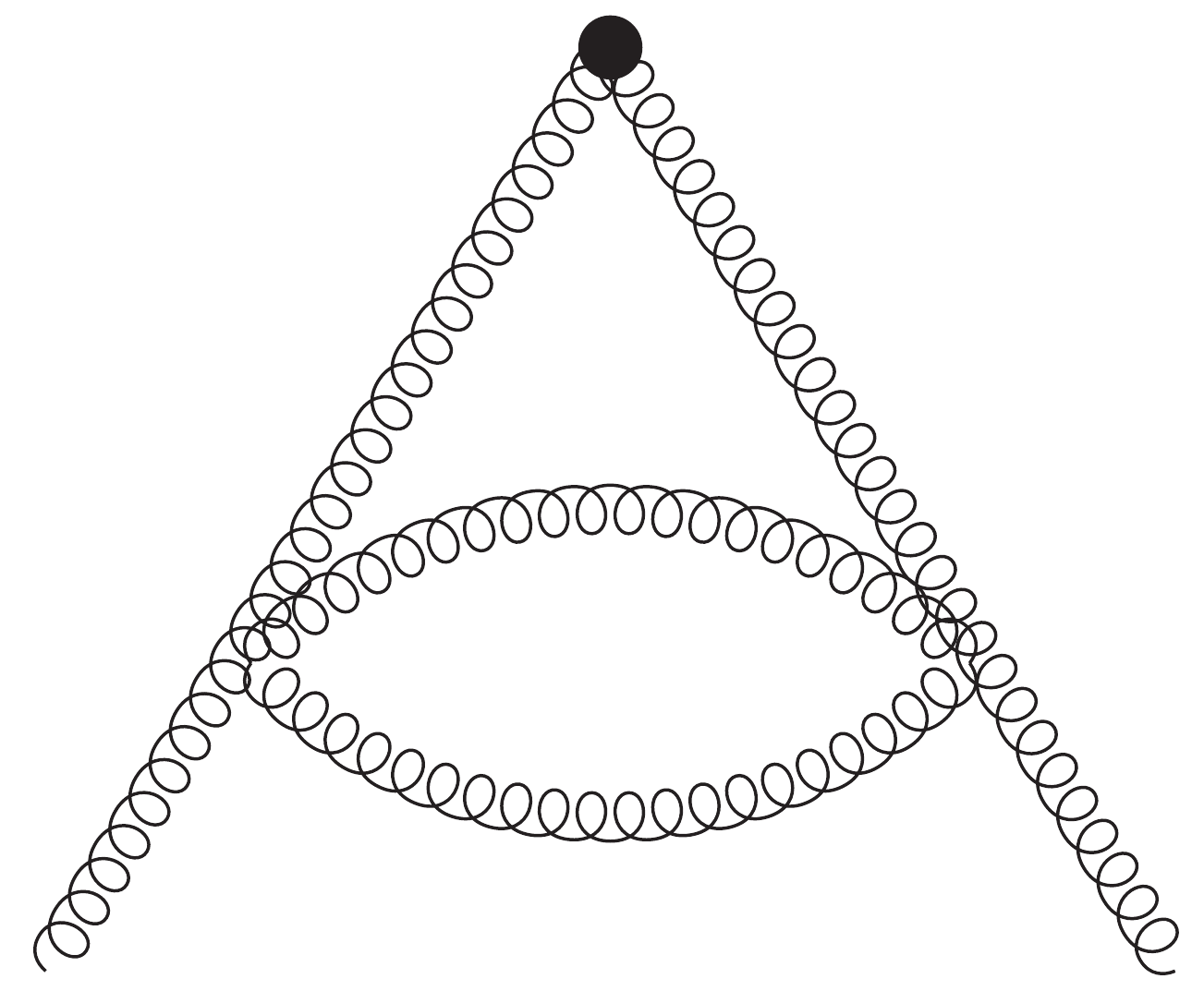}}
   \\[0.5em]
   \subfigure[graph (l)]{\includegraphics[height=0.24\textwidth]{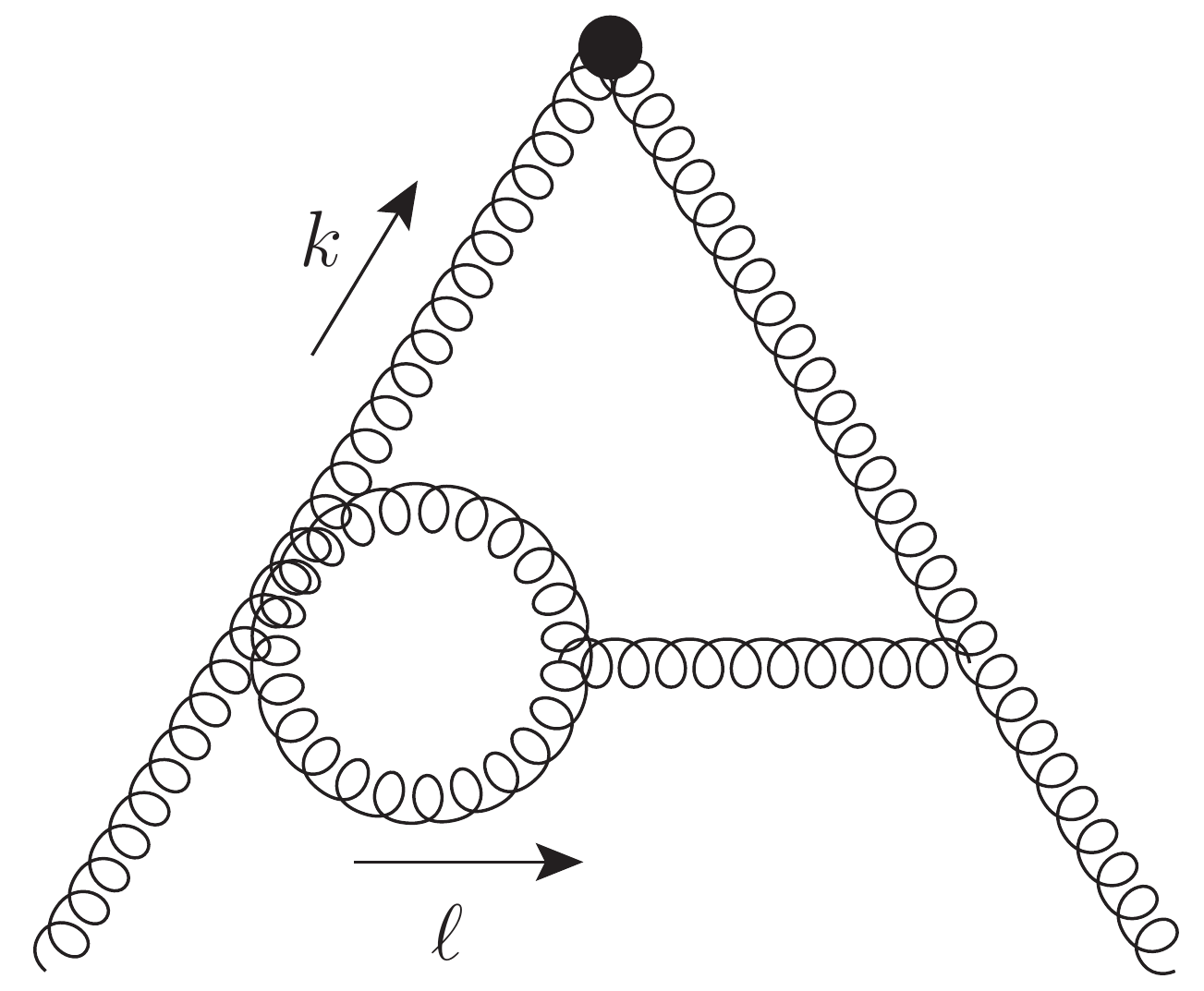}}
   \hspace{1.2em}
   \subfigure[graph (m)]{\includegraphics[height=0.24\textwidth]{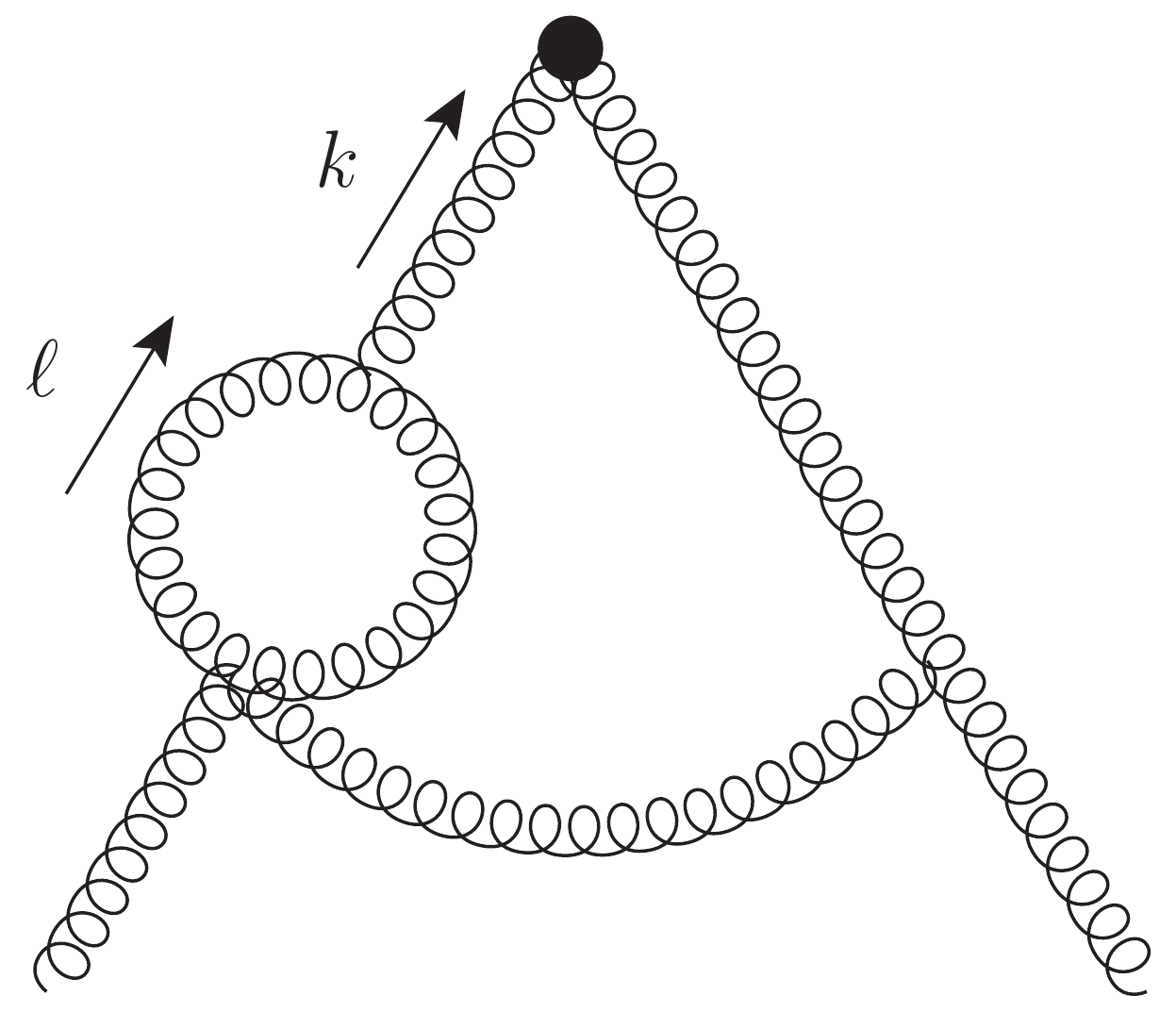}}
   \hspace{1.2em}
   \subfigure[graph (n)]{\includegraphics[height=0.24\textwidth]{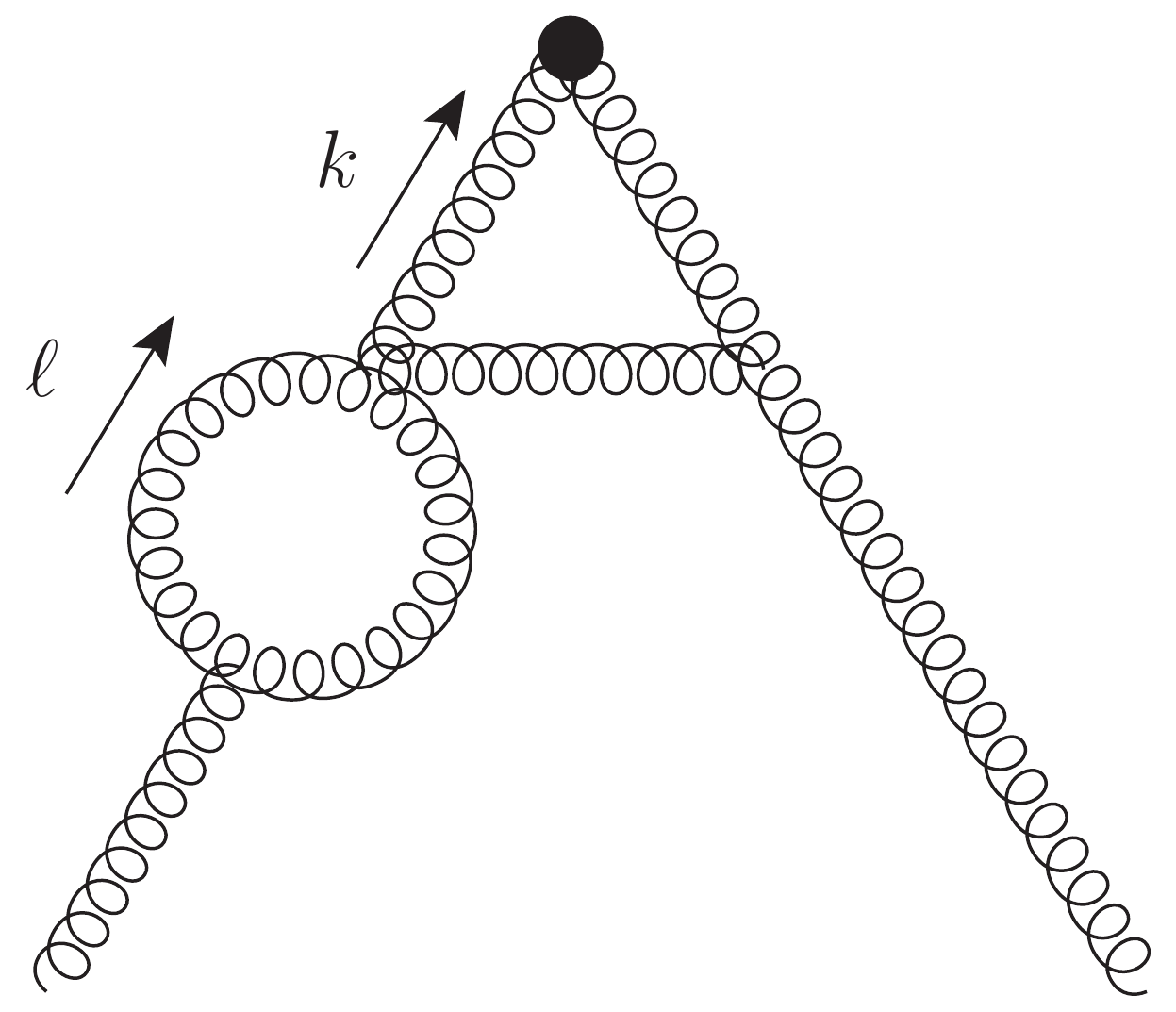}}
   \caption{\label{fig:4-g-graphs} Graphs that contribute to the two-loop DGLAP kernels and contain a four-gluon vertex.  \rev{The indices $i, i'$ and $j, j'$ specify gluon polarisation, and $k$ and $\ell$ are loop momenta.  Not shown are the complex conjugates of graphs (j), (l), (m), and (n).}}
\end{center}
\end{figure}

Before presenting the results of the calculation, we need to discuss the case of longitudinal polarisation.  Following the choice in \cite{Vogelsang:1996im}, we use the scheme of ’t Hooft, Veltman and Breitenlohner, Maison (HVBM)  \cite{tHooft:1972tcz, Breitenlohner:1977hr} to generalise the Levi-Civita tensor $\varepsilon$ to $D = 4 - 2\epsilon$ dimensions.  This prescription sets $\varepsilon_{\lambda\mu\nu\rho} = 0$ if any of the indices is not in the $4$ physical space-time dimensions, which breaks rotation invariance in the $2 - 2\epsilon$ transverse dimensions.  Since this tends to complicate calculations, we avoid the explicit use of the $\varepsilon$ tensor with the method explained in \sect{7.3.3} of \cite{Buffing:2017mqm}.  We first work with open transverse polarisation indices for the incoming gluons and the gluons in the twist-two operator, labelling these by $(j j')$ and $(i i')$ as shown in \fig{\ref{fig:graph-j}}.  Projecting the gluon operator on its asymmetric part in $i$ and $i'$ gives the contribution for longitudinal gluon polarisation.  The contribution to the polarised splitting kernel can then be obtained by contracting the graph with $(\delta_{i j} \delta_{i' j'} - \delta_{i j'} \delta_{i' j})$, which is the only structure in $2 - 2\epsilon$ dimensions that is invariant under rotations and antisymmetric in the pair $(i i')$.  For this method to work it is essential that the kernel does not depend on any external transverse direction.

The integration over minus-components of loop momenta can be used to put intermediate lines on shell in the graphs, which corresponds to fixing the final-state cut if one represents the kernel as the product of an amplitude and its complex conjugate.
After this step, we find that there is no contribution from \rev{the vertex correction graphs (l), (m), (n)} to the unpolarised or polarised kernels.\footnote{%
We note that \rev{these graphs were} not shown in \protect{\cite{Ellis:1996nn, Vogelsang:1996im}} because of their vanishing contribution [W.~Vogelsang, private communication].}
This is because \rev{their respective numerators are zero after integration over the angle of a suitably chosen linear combination of the transverse momenta $\tvec{k}$ and $\tvec{\ell}$ shown in \fig{\ref{fig:4-g-graphs}}.}

Adding the contributions of the other graphs with a four-gluon vertex, we obtain
\begin{align}
\label{eq:4g-vtx-general}
   \pr{AA}{\widetilde{P}}_{g g}^{(1)} \ms\Big|_{\text{4-g-vtx}}
   &= \pr{SS}{\widetilde{P}}_{g g}^{(1)} \ms\Big|_{\text{4-g-vtx}}
   \,,
   &
   \pr{\tentenbar}{\widetilde{P}}_{g g}^{(1)} \ms\Big|_{\text{4-g-vtx}}
   &= 0
\end{align}
and
\begin{align}
\label{eq:4g-vtx-specific}
   \pr{R\Rp}{\widetilde{P}}_{g g}^{(1)}(x) \ms\Big|_{\text{4-g-vtx}}
   &= \frac{1}{8} \ms \pr{R\Rp}{C}_\text{(k)} \ms (1-x)
   - \frac{1}{2} \ms \pr{R\Rp}{C}_\text{(j)} \,
      \Bigl(2(1+x)\ln(x) + 5 \ms (1-x)\Bigr)
   \,,
\end{align}
where $\widetilde{P}(x)$ denotes the part of the splitting kernels that is not proportional to $\delta(1-x)$.  The colour factors of graphs (k) and (j) are
\begin{align}
   \label{eq:Ck}
   \prn{11}{C}_\text{(k)} &= 9 N^2 \,,
   &
   \prn{AA}{C}_\text{(k)} &= 3 N^2 \,,
   &
   \pr{\twensev}{C}_\text{(k)} &= -3 \,,
   \\
   \label{eq:Cj}
   \prn{11}{C}_\text{(j)} &= 0 \,,
   &
   \prn{AA}{C}_\text{(j)} &= {}- N^2 / 4 \,,
   &
   \pr{\twensev}{C}_\text{(j)} &= 4 \,.
\end{align}
Analogues of \eqref{eq:4g-vtx-general} and \eqref{eq:4g-vtx-specific} hold for the polarised kernels $\pr{R\Rp}{\widetilde{P}}_{\Delta g \Delta g}$, with colour factors
\begin{align}
   \prn{11}{C}_{\Delta\text{(k)}} &= 3 N^2 \,,
   &
   \prn{AA}{C}_{\Delta\text{(k)}} &= 2 N^2 \,,
   &
   \pr{\twensev}{C}_{\Delta\text{(k)}} &= - 17 \,,
   \\\
   \prn{11}{C}_{\Delta\text{(j)}} &= {}- 3 N^2 \,,
   &
   \prn{AA}{C}_{\Delta\text{(j)}} &= {}- 3 N^2 / 4 \,,
   &
   \pr{\twensev}{C}_{\Delta\text{(j)}} &= - 3
\end{align}
instead of those in \eqref{eq:Ck} and \eqref{eq:Cj}.  Interestingly, we find that the dependence on the colour and on $x$ factorises for the individual graphs, although this is not obvious from the Feynman rule of the four-gluon vertex.  The colour factors given above are valid for a general number $N$ of colours, except for the 27 representation, where we have set $N=3$.

%%%%%%%%%%%%%%%%%%%%%%%%%%%%%%%%%%%%%%%%%%%%%%%%%%%%%%%%%%%%%%%%%%%%%%%%%%%%%%%%

\subsection{Transverse quark polarisation}
\label{sec:transverse-graphs}

The calculations in \cite{Curci:1980uw, Ellis:1996nn, Vogelsang:1996im} for unpolarised and longitudinally polarised partons were extended in \cite{Vogelsang:1997ak} to the case of transverse quark polarisation.
The contribution of individual graphs to the two-loop kernels is not given in that paper, but thankfully its author provided us with this information, which is collected in \tab{\ref{tab:transv-per-graph}}.

\begin{table}
   \centering
   \renewcommand*{\arraystretch}{1.4}
   \begin{tabular}{|c|c|c|c|c|c|c|c|}
      \hline
      Terms                                  &\multicolumn{7}{c|}{Graphs}\\
      \hline\hline
                                             &(b)   &(c)    &(d)   &(e)  &(f)      &(g)      &(hi)\\
      \hline
      $\delp{q}{q}(x) \ln^2(x)$              &$-1$  &1      &$-1$  &$-2$ &         &         &2\\
      $\delp{q}{q}(x) \ln^2(1-x)$            &      &       &$-2$  &     &         &$-1$     &\\
      $\delp{q}{q}(x) \ln(x)\ln(1-x)$        &      &2      &$-6$  &$-4$ &         &$-2$     &\\
      $\delp{q}{q}(x) \ln(x)$                &      &$-3/2$ &$3/2$ &     &$-2/3$   &$11/6$   &\\
      $\delp{q}{q}(x) \ln(1-x)$              &4     &$-3$   &$-5$  &3    &         &$-2$     &$-4$\\
      $\delp{q}{q}(x) \ms \pi^2/3$           &      &2      &$-3$  &$-2$ &         &$-1$     &\\
      $\delp{q}{q}(x) \ms\bigl[\ln(1-x)+\ln(x)\bigr] I_0$       &      &4    &$-8$     &$-4$     &         &$-2$      &\\
      $\delp{q}{q}(x) \ms I_1$               &      &$-4$   &8     &4    &         &2        &\\
      $\delp{q}{q}(x) \ms I_0$               &4     &       &$-8$  &     &         &$-2$     &$-4$\\
      $\delp{q}{q}(x)$                       &      &$-7$   &11    &7    &$-10/9$  &$103/18$ &\\
      $1-x$                                  &1     &       &      &     &         &         &\\
      \hline
   \end{tabular}
   \caption{\label{tab:transv-per-graph} Contributions of individual graphs to $\pr{RR}{P}^{V (1)}_{\delta q\delta q}(x)$.  The function $\delp{q}{q}(x)$ is given in \protect\eqref{eq:lo-p_dqdq}, and $I_0$ and $I_1$ are defined in \protect\eqref{eq:PV-gauge-integrals}.  More explanations are given in the text.  The sum of all contributions is published in \protect\cite{Vogelsang:1997ak}, and the individual terms shown here have been kindly provided by W.~Vogelsang.}
\end{table}

\begin{table}
   \centering
   \renewcommand*{\arraystretch}{1.4}
   \begin{tabular}{|c|c|c|c|c|c|c|c|}
      \hline
      R  &\multicolumn{7}{c|}{Graphs}\\
      \hline\hline
         &(b)                  &(c)               &(d)      &(e)              &(f)                  &(g)     &(hi)\\
      \hline
      1  & $- A / (4 N^2)$     & $- A / (4 N^2)$  & $- A/4$ & $A^2 / (4 N^2)$ & $n_f \ms A / (4 N)$ & $A/2$  & $A^2 / (4 N^2)$ \\
      8  & $(N^2+1) / (4 N^2)$ & $ 1/ (4 N^2)$    & $1/4$   & $- A / (4 N^2)$ & $- n_f / (4 N)$     & $-1/2$ & $1 / (4 N^2)$   \\
      \hline
   \end{tabular}
   \caption{\label{tab:colour-per-graph} The colour factor for each graph in $\pr{RR}{P}^{V (1)}_{\delta q\delta q}(x)$.  Here $n_f$ is the number of active quark flavours, $N$ is the number of colours, and $A = N^2 - 1$.}
\end{table}

To obtain the contribution of a graph to $\pr{RR}{P}^{V (1)}_{\delta q\delta q}(x)$, the entry in \tab{\ref{tab:transv-per-graph}} needs to be multiplied with the colour factor in \tab{\ref{tab:colour-per-graph}}.  The letters (b), (c), etc.\ refer to the graphs in \fig{3} of \cite{Ellis:1996nn}, and the entry (hi) gives the combined contributions from (h) and its subtraction term (i).

The kernel $\pr{11}{P}_{\delta q \delta\bar{q}}$ for the transition from a quark to an antiquark receives contributions from only one Feynman graph and is given in \eqn{(44)} of \cite{Vogelsang:1997ak}.  We thus have all ingredients needed to extract the evolution kernels for transverse quark and antiquark polarisation along the lines described above.

\section{Second method: modified TMD matching calculation}
\label{sec:second-method}

In general, the DGLAP kernels can be extracted from any quantity that obeys an evolution equation in which these kernels appear.  One such quantity is the matching kernel that connects TMDs at short distances with ordinary PDFs.
This method can be extended to the colour dependent case, and in this section we show how to extract the colour dependent NLO kernels by modifying the two-loop calculation of the TMD matching kernels in \cite{Echevarria:2016scs, Gutierrez-Reyes:2018iod}.  That calculation was done in Feynman gauge with the $\delta$ regulator for rapidity divergences.  This allows us to compute also the terms going with $\delta(1-x)$, along with providing an independent cross check for the $x$ dependent part of the kernels obtained with our first method.

To begin with, we recall that two-parton (or double) TMDs can be defined from
the opera\-tor products in \eqref{eq:parton-op-product} and
\eqref{eq:soft-op-product} at finite distances $\tvec{z}_1$ and $\tvec{z}_2$, as discussed at length in \cite{Buffing:2017mqm}.  If both distances are small, one can use the operator product expansion to match the double TMDs onto collinear ones.  This was done at LO in \cite{Buffing:2017mqm}, where it was observed that the colour dependent DGLAP kernels for DPDs can be extracted from the corresponding matching kernels of the same order in $a_s$.  We follow a modified version of that strategy at NLO in the present work.

The matching kernels can be computed from matrix elements of \emph{single} operators $\mathcal{O}_{a}(x, \tvec{y}, \tvec{z})$ and $\mathcal{O}_{S,a}(\tvec{y},\tvec{z})$ rather than their products \eqref{eq:parton-op-product} and
\eqref{eq:soft-op-product}.  To this end, one defines the matrix element of $\mathcal{O}_{a}$ between parton states and the vacuum expectation value of $\mathcal{O}_{S,a}$, with all colour indices being projected onto definite colour representations.  Details and normalisation factors are given in \eqn{(I.52)} of \cite{Buffing:2017mqm}.  After combining the matrix elements in the same way as described in \sect{\ref{sec:renorm-dpds}} for DPDs, the rapidity regulator can be removed.  One thus obtains ``TMD matrix elements'' $\pr{R\Rp}{\widehat{\mathcal{M}}}_{a b}(x, \tvec{z}, \mu, \zeta)$ for finite $\tvec{z}$ and ``PDF matrix elements'' $\pr{R\Rp}{\mathcal{M}}_{a b}(x, \mu, \zeta)$ for $\tvec{z} = \tvec{0}$, where $a$ denotes the parton in the operator and $b$ the target parton.\footnote{%
For ease of language, we use the terms TMD and PDF also for colour non-singlet channels here.}
The two types of matrix elements are connected by the matching equation
\begin{align}
   \pr{R\Rp}{\widehat{\mathcal{M}}}_{ab}(x,\tvec{z},\mu,\zeta)
   &=
   \sum_{c,\Rpp} \pr{R\Rppbar}{C}_{ac}(x^\prime,\tvec{z},\mu,x^2\zeta)
   \conv{x} \pr{\Rpp\Rp}{\mathcal{M}}_{c b}(x^\prime,\mu,\zeta)
   \,,
\label{eqn:matrix_elements_matching}
\end{align}
where $\pr{R\Rppbar}{C}_{ac}$ is the matching kernel that connects double TMDs with collinear DPDs.  The kernel $\prn{11}{C}_{ac}$ also describes the matching  between single-parton TMDs and PDFs.

Using the relation (I.65) in \cite{Buffing:2017mqm}, one could extract the colour dependent DGLAP kernels from the $\mu$ dependence of $\pr{R\Rppbar}{C}_{ac}$.  Instead, we extract them \rev{from the poles} of the unrenormalised TMD matrix elements.  This is simpler than computing the matching kernels, given that those require also the finite terms of the matrix elements, whose expressions are more involved.
The extraction from the poles requires an RGE analysis of the matching equation \eqref{eqn:matrix_elements_matching}, which is given in the following subsection.

%...............................................................................

\paragraph{Polarisation dependence and $\delta(1-x)$ terms.}
The TMD matching kernels of order $a_s^2$ were computed in \cite{Echevarria:2016scs} for unpolarised partons and in \cite{Gutierrez-Reyes:2018iod} for transversely polarised quarks.  The kernels for longitudinal polarisation have not been evaluated using these methods.  We can hence not directly calculate the $\delta(1-x)$ terms of the colour dependent DGLAP kernels.  This is, however, not necessary because these terms are polarisation independent.

To understand this, let us see which graphs can provide a $\delta(1-x)$ contribution to the bare matrix elements.  There are purely virtual graphs, which cannot depend on the parton polarisation since the loop corrections do not connect the two partons that enter the operator vertex.  In addition, there are contributions from mixed real-virtual or purely real graphs in the limit where the partons going across the final-state cut have zero plus-momentum.  More specifically, one has either one or two cut gluons, or a cut quark-antiquark pair that merges into a gluon on at least one side of the cut.  The connection between the two partons that enter the operator vertex --- which is the potential source of polarisation dependence --- thus always involves one or two gluons whose plus-momentum goes to zero.  In this limit, one can use the eikonal approximation for their couplings to parton lines with finite plus-momentum.  Since the eikonal coupling is spin independent, the $\delta(1-x)$ terms in the bare matrix elements cannot depend on the polarisation of the twist-two operator.  The same must therefore be true for the associated ultraviolet renormalisation factors, and hence for the full DGLAP kernels.

We explicitly verified this argument by direct computation of the $\delta(1-x)$ terms for unpolarised and transversely polarised quarks, which come out to be equal as it must be.

%%%%%%%%%%%%%%%%%%%%%%%%%%%%%%%%%%%%%%%%%%%%%%%%%%%%%%%%%%%%%%%%%%%%%%%%%%%%%%%%

\subsection{Renormalisation group analysis}
\label{sec:matching-RGE}

The PDF matrix element is renormalised by
\begin{align}
\label{eq:m-pdf-renorm}
   \pr{R\Rp}{\mathcal{M}}_{ab}(x,\mu,\zeta)
   &= Z_b^{-1}(\mu) \sum_{c,\Rpp} \pr{R\Rppbar}{Z}_{ac}(x^\prime,\mu,x^2\zeta)
   \conv{x} \pr{\Rpp\Rp}{\mathcal{M}}_{B \bs,\ms c b}(x^\prime,\zeta)
   \,,
\end{align}
where $\pr{R\Rppbar}{Z}_{ac}$ is the renormalisation factor for collinear DPDs that was analysed at length in \sects{\ref{sec:renorm-dpds}} and \ref{sec:Z-factor}.
The wave function renormalisation constant $Z_b$ connects bare and renormalised fields,
\begin{align}
   q_{B, i}^{} &= \sqrt{Z_q} \, q_i^{}
   \,,
   &
   A^{a, \mu}_B = \sqrt{Z_g} \ms A^{a, \mu}
   \,,
\end{align}
and enters in \eqref{eq:m-pdf-renorm} because the bare matrix element is taken with bare parton states (otherwise it would not be $\mu$ independent) whereas the renormalised matrix element requires renormalised parton states in order to be ultraviolet finite.

The TMD matrix element is renormalised multiplicatively as
\begin{align}
\label{eq:m-tmd-renorm}
   \pr{R\Rp}{\widehat{\mathcal{M}}}_{ab}(x,\tvec{z},\mu,\zeta)
   &= Z_b^{-1}(\mu) \, \widehat{Z}_a(\mu,x^2\zeta)
   \, \pr{R\Rp}{\widehat{\mathcal{M}}}_{B \bs,\ms a b}(x,\tvec{z},\zeta)
   \,.
\end{align}
The factor $\widehat{Z}_a(\mu,x^2 \zeta)$ renormalises the TMD operator $\mathcal{O}_{a}^{r r'}(x,\tvec{y},\tvec{z})$.  Unlike its analogue for the collinear operator, it is colour and spin independent and only differs between quarks and gluons.  This is because in the TMD case, ultraviolet divergences are associated only with each field operator and with the product of operators at transverse position $\tvec{y} + \tvec{z}/2$ or at transverse position $\tvec{y} - \tvec{z}/2$ in \eqref{eq:quark-ops} and \eqref{eq:soft-op}.  How these two operator products are coupled in their colour (or spin) indices is irrelevant for the renormalisation factor.

The renormalisation factor for TMDs is well studied and has in particular been used at two loops in the matching calculations we are modifying.  Expressions relevant for our present analysis are given in \app{\ref{sec:tmd-renorm}}.

For the following arguments, we need the perturbative expansions
\begin{align}
   \pr{R\Rp}{\widehat{\mathcal{M}}}_{ab}(x,\tvec{z},\mu,\zeta)
   &= \delta_{R\Rpbar} \ms \delta_{a b} \delta(1-x)
   + \sum_{n=1}^\infty a_s^n(\mu)
   \, \pr{R\Rp}{\widehat{\mathcal{M}}}_{ab}^{(n)}(x,\tvec{z},\mu,\zeta)
   \,,
   \\
   \pr{R\Rp}{\mathcal{M}}_{ab}(x,\mu,\zeta)
   &= \delta_{R\Rpbar} \ms \delta_{a b} \delta(1-x)
   + \sum_{n=1}^\infty a_s^n(\mu) \,
   \pr{R\Rp}{\mathcal{M}}_{ab}^{(n)}(x,\zeta/\mu^2)
   \,,
   \\
   \pr{R\Rp}{C}_{a b}(x,\tvec{z},\mu,\zeta)
   &= \delta_{R\Rpbar} \ms \delta_{a b} \delta(1-x)
   + \sum_{n=1}^\infty a_s^n(\mu) \,
   \pr{R\Rp}{C}_{a b}^{(n)}(x,\tvec{z},\mu,\zeta)
   \,,
   \label{eq:expand-matching-C}
   \\
   Z_b(\mu) &= 1 + \sum_{n=1}^\infty a_s^n(\mu) \, Z_b^{(n)}
   \,,
   \\
   \widehat{Z}_a(\mu,\zeta) &= 1
   + \sum_{n=1}^\infty a_s^n(\mu) \, \widehat{Z}_a^{(n)}(\zeta/\mu^2)
   \,,
   \label{eq:expand-Z-tmd}
\end{align}
and the corresponding expansion \eqref{eq:expand-Z-DPD} of $\pr{R\Rp}{Z}_{ab}(x,\mu,\zeta)$.

We compute the unrenormalised matrix elements $\widehat{\mathcal{M}}_{B}$ and $\mathcal{M}_B$ in bare perturbation theory, using the Lagrangian written in terms of bare fields and the bare coupling $g_B$.  The relation between \rev{the bare and renormalised versions of $a_s = \alpha_s / (2 \pi)$ is}
\begin{align}
   \label{eq:as-renorm}
   a_{s, B}
   &= \mu^{2\epsilon} a_s(\mu) \, Z_\alpha(\mu) / S_\epsilon
\end{align}
with
\begin{align}
   Z_\alpha(\mu) = 1 + \sum_{n=1}^\infty a_s^n(\mu) \, Z_\alpha^{(n)}
   &= 1 - a_s(\mu) \, \frac{1}{\epsilon} \, \frac{\beta_0}{2}
   + \mathcal{O}(a_s^2)
   \,.
\end{align}
The factor $S_\epsilon$ implements the \msbar scheme as specified in \sect{6.2} of \cite{Diehl:2018kgr}, such that all renormalisation counterterms are sums of pure poles in $\epsilon$.  We take the standard choice
\begin{align}
   \label{eq:S-eps-choice}
   S_\epsilon &= (4 \pi e^{- \gamma})^{\epsilon}
   \,,
\end{align}
where $\gamma$ is the Euler-Mascheroni constant.
The expansion of the TMD matrix element can then be written as
\begin{align}
   \pr{R\Rp}{\widehat{\mathcal{M}}}_{B \bs,\ms a b}(x,\tvec{z},\zeta)
   &= \delta_{R\Rpbar} \ms \delta_{a b} \delta(1-x)
   + \sum_{n=1}^\infty \bigl[ a_s(\mu)  \ms Z_\alpha(\mu) \bigr]^n \;
   \pr{R\Rp}{\widehat{\mathcal{M}}}_{B \bs,\ms a b}^{(n)}(x,\tvec{z},\mu,\zeta)
   \,,
\end{align}
where a factor of $(\mu^{2\epsilon} / S_\epsilon)^n$ is included in the expansion coefficients ${\widehat{\mathcal{M}}}_{B \bs,\ms a b}^{(n)}\ms$.

The PDF matrix element is given by its tree-level expression
\begin{align}
   \pr{R\Rp}{\mathcal{M}_{B \bs,\ms a b}}(x,\zeta)
   &= \delta_{R\Rpbar} \ms \delta_{a b} \delta(1-x)
   \,.
\end{align}
If one works with the Collins regulator for rapidity divergences, this readily follows because all loop integrals are scale-less and hence zero in dimensional regularisation.  The argument is slightly more involved for the $\delta$ regulator, which involves regulating parameters $\delta^+$ and $\delta^-$ with the dimension of a mass.  In the matrix element of $\mathcal{O}_{a}$ with parton states of momentum $p$, the rapidity regulator only appears in the form of the dimensionless combination $\delta^+ / p^+$ because of boost invariance, so that the transverse loop integrals are zero in $2 - 2\epsilon$ dimensions.  Individual graphs of the corresponding soft factor are nonzero because they depend on the dimensionful and boost invariant combination $\delta^+ \delta^-$.  However, since rapidity divergences must cancel between the partonic and soft matrix elements, the sum over all graphs for the soft matrix element must give zero.\footnote{%
See also the related discussion in \sect{II} of \protect\cite{Echevarria:2015byo}.}

Expanded in $a_s$, the renormalised matrix elements are related to the bare ones by
\begin{align}
   \pr{R\Rp}{\widehat{\mathcal{M}}}_{ab}^{(1)}(x,\zeta)
   &=
   \delta_{R\Rpbar} \ms \delta_{a b} \delta(1-x)
   \Bigl( \widehat{Z}_a^{(1)}(x^2\zeta) - Z_b^{(1)}\Bigr)
   + \pr{R\Rp}{\widehat{\mathcal{M}}}_{B \bs,\ms a b}^{(1)}(x,\zeta)
   \,,
   \\[0.2em]
   \pr{R\Rp}{\widehat{\mathcal{M}}}_{ab}^{(2)}(x,\zeta)
   &=
   \delta_{R\Rpbar} \ms \delta_{a b} \delta(1-x)
   \Bigl[ \bigl(Z_b^{(1)}\bigr)^2 - Z_b^{(2)} + \widehat{Z}_a^{(2)}(x^2\zeta)
   - Z_b^{(1)} \widehat{Z}_a^{(1)}(x^2\zeta) \Bigr] \notag\\
   &\quad
   + \Bigl( \widehat{Z}_a^{(1)}(x^2\zeta) - Z_b^{(1)} + Z_\alpha^{(1)} \Bigr)
   \pr{R\Rp}{\widehat{\mathcal{M}}}_{B \bs,\ms a b}^{(1)}(x,\zeta)
   + \pr{R\Rp}{\widehat{\mathcal{M}}}_{B \bs,\ms a b}^{(2)}(x,\zeta)
\end{align}
and
\begin{align}
   \pr{\Rpp\Rp}{\mathcal{M}}_{c b}^{(1)}(x^\prime,\zeta)
   &=
   {}- \delta_{\Rpp\Rpbar} \ms \delta_{c b} \delta(1-x^\prime) Z_b^{(1)}
   + \pr{\Rpp\Rp}{Z}_{c b}^{(1)}(x^\prime,x^{\prime \ms 2}\zeta)
   \,,
   \\[0.2em]
   \pr{\Rpp\Rp}{\mathcal{M}}_{c b}^{(2)}(x^\prime,\zeta)
   &=
   \delta_{\Rpp\Rpbar} \ms \delta_{c b} \delta(1-x^\prime)
   \Bigl[ \bigl(Z_b^{(1)}\bigr)^2 - Z_b^{(2)} \Bigr]
   + \pr{\Rpp\Rp}{Z}_{c b}^{(2)}(x^\prime,x^{\prime \ms 2}\zeta)
   \notag\\
   &\quad
   - Z_b^{(1)} \, \prb{\Rpp\Rp}{Z}_{c b}^{(1)}(x^\prime,x^{\prime \ms 2}\zeta)
   \,.
\end{align}
With the expansion \eqref{eq:expand-matching-C} of the matching kernel, the $a_s$ part of the matching equation \eqref{eqn:matrix_elements_matching} thus reads
\begin{align}
   \delta_{R\Rpbar} \ms
   & \delta_{ab} \delta(1-x)
   \Bigl( \widehat{Z}_a^{(1)}(x^2\zeta) - Z_b^{(1)}\Bigr)
   + \pr{R\Rp}{\widehat{\mathcal{M}}}_{B \bs,\ms a b}^{(1)}(x,\zeta)
   \notag\\
   &=
   {}- \delta_{R\Rpbar} \ms \delta_{a b} \delta(1-x) Z_b^{(1)}
   + \pr{R\Rp}{Z}_{ab}^{(1)}(x,x^2\zeta)
   + \pr{R\Rp}{C}_{ab}^{(1)}(x, x^2 \zeta)
   \,.
   \label{eq:matching-eq-LO}
\end{align}
Extracting the single pole in $\epsilon$ of this relation and using \eqref{eq:P-from-Z-residue}, we can extract the LO DGLAP splitting kernels as
\begin{align}
   \pr{R\Rp}{P}_{ab}^{(0)}(x,x^2\zeta)
   &=
   - \biggl[ \delta_{R\Rpbar} \ms \delta_{a b} \delta(1-x)
   \widehat{Z}_a^{(1)}(x^2\zeta)
   + \pr{R\Rp}{\widehat{\mathcal{M}}}_{B \bs,\ms a b}^{(1)}(x,\zeta) \biggr]_{-1}
   \,,
   \label{eqn:splitting_kernel_LO}
\end{align}
where we write
\begin{align}
   \label{eq:def-laurent-series-notation}
   Q(\epsilon) &= \sum_{k} \epsilon^{k} \, \bigl[ Q \bigr]_{k}
\end{align}
for the Laurent expansion of a quantity around $\epsilon=0$.

Using the modified TMD matching computations of \cite{Echevarria:2016scs, Gutierrez-Reyes:2018iod}, we correctly obtain the known LO DGLAP kernels for all colour channels.  We also verified that the double poles in $\epsilon$ are equal on both sides of \eqref{eq:matching-eq-LO}.

The $a_s^2$ part of the matching equation \eqref{eqn:matrix_elements_matching} reads
\begin{align}
   \delta_{R\Rpbar} \ms
   &
   \delta_{ab} \delta(1-x)
   \Bigl[ \bigl(Z_b^{(1)}\bigr)^2 - Z_b^{(2)}
   + \widehat{Z}_a^{(2)}(x^2\zeta)
   - Z_b^{(1)} \widehat{Z}_a^{(1)}(x^2\zeta) \Bigr]
   \notag\\
   &\quad
   + \Bigl( \widehat{Z}_a^{(1)}(x^2\zeta)
   - Z_b^{(1)} + Z_\alpha^{(1)} \Bigr)
   \pr{R\Rp}{\widehat{\mathcal{M}}}_{B \bs,\ms a b}^{(1)}(x,\zeta)
   + \pr{R\Rp}{\widehat{\mathcal{M}}}_{B \bs,\ms a b}^{(2)}(x,\zeta)
   \notag\\
   &=
   \delta_{R\Rpbar} \ms \delta_{a b} \delta(1-x)
   \Bigl[ \bigl(Z_b^{(1)}\bigr)^2 - Z_b^{(2)} \Bigr]
   + \pr{R\Rp}{Z}_{ab}^{(2)}\bigl( x, x^2 \zeta \bigr)
   - Z_b^{(1)} \, \pr{R\Rp}{Z}_{ab}^{(1)}\bigl(x, x^2 \zeta\bigr)
   \notag\\
   &\quad
   + \sum_{c,\Rpp} \pr{R\Rppbar}{C}_{ac}^{(1)}(x^\prime,x^2\zeta)
   \conv{x} \biggl\{
   - \delta_{\Rpp\Rpbar} \ms \delta_{c b} \delta(1-x^\prime) Z_b^{(1)}
   + \pr{\Rpp\Rp}{Z}_{c b}^{(1)}\bigl(x^\prime, x^{\prime \ms 2} \zeta \bigr)
   \biggr\}
   \notag\\
   &\quad + \pr{R\Rp}{C}_{ab}^{(2)}(x,x^2\zeta)
   \,.
   \label{eqn:matching_equation_at_NNLO}
\end{align}
The last term in this expression is finite and does not contribute to the poles in $\epsilon$.
We can eliminate $\pr{R\Rppbar}{C}_{ac}^{(1)}$ by using the LO relation \eqref{eq:matching-eq-LO}.  First replacing $x$ with $x'$ and then rescaling the variable $\zeta$, we get
\begin{align}
   &\pr{R\Rp}{C}_{ab}^{(1)}(x', x^2 \zeta)
   \notag\\
   &\qquad
   = \delta_{R\Rpbar} \ms \delta_{a b} \delta(1-x')
     \widehat{Z}_a^{(1)}(x^2 \zeta)
   - \pr{R\Rp}{Z}_{ab}^{(1)}(x', x^2 \zeta)
   + \pr{R\Rp}{\widehat{\mathcal{M}}}_{B \bs,\ms a b}^{(1)}(x',
        x^2\zeta /x^{\prime \ms 2})
   \,.
   \label{eqn:matching_kernel_NLO}
\end{align}
Using again the relation \eqref{eq:P-from-Z-residue} we obtain the NLO DGLAP kernels as
\begin{align}
   \pr{R\Rp}{P}_{ab}^{(1)}(x,x^2\zeta)
   = - 2 \,
   &
   \biggl[ \delta_{R\Rpbar} \ms \delta_{a b}
   \delta(1-x) \widehat{Z}_a^{(2)}(x^2\zeta)
   + \Big( \widehat{Z}_a^{(1)}(x^2\zeta) + Z_\alpha^{(1)} \Big)
   \pr{R\Rp}{\widehat{\mathcal{M}}}_{B \bs,\ms a b}^{(1)}(x,\zeta)
   \notag\\[0.3em]
   &
   - \sum_{c,\Rpp}
   \pr{R\Rppbar}{\widehat{\mathcal{M}}}^{(1)}_{B \bs,\ms a c}
   \bigl(x^\prime,x^2\zeta/x^{\prime \ms 2}\bigr)
   \conv{x} \pr{\Rpp\Rp}{Z}_{c b}^{(1)}(x^\prime, x^{\prime \ms 2}\zeta)
   \notag\\[-0.2em]
   &
   + \pr{R\Rp}{\widehat{\mathcal{M}}}_{B \bs,\ms a b}^{(2)}(x,\zeta) \biggr]_{-1}
   \,.
   \label{eqn:splitting_kernel_from_matching_NLO}
\end{align}
To evaluate the second line on the r.h.s., we need the following convolution of two plus-distributions:
\begin{align}
   \label{eq:convol-plus-distrib}
   \bigg[ \frac{1}{1-x^\prime} \bigg]_+
               \conv{x} \bigg[ \frac{1}{1-x^\prime} \bigg]_+
   &=
   2 \bigg[ \frac{\ln(1-x)}{1-x} \bigg]_+
               - \frac{\ln(x)}{1-x} - \frac{\pi^2}{6} \delta(1-x)
   \,,
\end{align}
where we define as usual
\begin{align}
   \label{eq:def-plus-distrib}
   \int_x^1 d x\, \biggl[ \frac{\ln^k(1-x)}{1-x} \biggr]_{+} \, \varphi(x)
   &= \int_x^1 d x\, \frac{\ln^k(1-x)}{1-x} \,
      \bigl[ \varphi(x) - \varphi(1) \bigr]
   \notag \\
   &\quad
   {}- \varphi(1) \int_0^x d x \, \frac{\ln^k(1-x)}{1-x}
   &
   \text{ with } k=0,1,\ldots
\end{align}
for a test function $\varphi(x)$.
The relation \eqref{eq:convol-plus-distrib} is easily verified by taking its Mellin transform, which turns the l.h.s.\ into a simple product.  The required transforms can for instance be found in entries number 1, 3, 9, and 10 of the appendix in \cite{Blumlein:1998if}.

For the higher-order poles in \eqref{eqn:matching_equation_at_NNLO}, we obtain
\begin{align}
   &
   \pr{R\Rp}{Z}_{ab}^{(2,m)}(x,x^2\zeta)
   = \biggl[ \delta_{R\Rpbar} \ms \delta_{a b} \delta(1-x)
   \widehat{Z}_a^{(2)}(x^2\zeta) - \widehat{Z}_a^{(1)}(x^2\zeta)
   \pr{R\Rp}{Z}_{ab}^{(1)}(x,x^2\zeta)
   \notag\\[0.5em]
   &\qquad
   + \sum_{c,\Rpp} \pr{R\Rppbar}{Z}^{(1)}_{ac}(x^\prime,x^2\zeta)
   \conv{x} \pr{\Rpp\Rp}{Z}_{c b}^{(1)}(x^\prime, x^{\prime \ms 2}\zeta)
   + \Big( \widehat{Z}_a^{(1)}(x^2\zeta) + Z_\alpha^{(1)} \Big)
   \pr{R\Rp}{\widehat{\mathcal{M}}}_{B \bs,\ms a b}^{(1)}(x,\zeta)
   \notag\\
   &\qquad
   - \sum_{c,\Rpp}
   \pr{R\Rppbar}{\widehat{\mathcal{M}}}^{(1)}_{B \bs,\ms a c}(x^\prime,x^2\zeta/x^{\prime \ms 2})
   \conv{x} \pr{\Rpp\Rp}{Z}_{c b}^{(1)}(x^\prime, x^{\prime \ms 2} \zeta)
   + \pr{R\Rp}{\widehat{\mathcal{M}}}_{B \bs,\ms a b}^{(2)}(x,\zeta) \biggr]_{-m}
   \notag \\
   & \hspace{28em}
   \text{ for } m = 2,3,4.
   \label{eqn:Z2_higher_poles}
\end{align}
Since the terms on the r.h.s.\ are all known, this serves as a cross check of our calculation.

Taking the calculations in \cite{Echevarria:2016scs} and \cite{Gutierrez-Reyes:2018iod} and changing the colour projection for all contributing graphs, we have computed the colour dependent DGLAP kernels at order $a_s^2$ from \eqref{eqn:splitting_kernel_from_matching_NLO}, both for unpolarised partons and for transversely polarised quarks.  The $x$ dependent terms obtained in this way fully agree with the results of our first method.
Let us note that we have \emph{not} applied the rescaling of the regulating parameter $\delta^+$ or $\delta^-$ with a momentum fraction, as prescribed in \eqs{(3.9)} and (3.10) of \cite{Echevarria:2016scs}.  That prescription was the result of missing the necessary rescaling between the $\zeta$ parameters defined with reference to the hadron or the parton momentum in \cite{Echevarria:2016scs}.

In the decuplet sector, we find that the colour factors for all graphs contributing to the bare matrix element are zero, so that $\pr{\tentenbar}{\widehat{\mathcal{M}}}_{B, g g}^{(n)} = 0$ for $n=1,2$.  This implies that
\begin{align}
   \label{eq:ten-only-Z-tmd}
   \pr{\tentenbar}{P}_{a b}^{(N-1)}(x,x^2\zeta)
   &=
   {}- N \Bigl[ \delta_{a b} \ms
   \delta(1-x) \widehat{Z}_a^{(N)}(x^2\zeta) \Bigr]_{-1}
   &&
   \text{ for } N=1,2.
\end{align}
We will come back to this in \sect{\ref{sec:results}}.

\section{Evolution channels and finite renormalisation}
\label{sec:channels}

%%%%%%%%%%%%%%%%%%%%%%%%%%%%%%%%%%%%%%%%%%%%%%%%%%%%%%%%%%%%%%%%%%%%%%%%%%%%%%%%

\subsection{Linear combinations of DPDs and evolution kernels}
\label{sec:evol-basis}

Before presenting our results, we need to discuss in some detail the structure of the DGLAP kernels regarding the parton labels.  This is well known in the colour singlet sector (see e.g.~\cite{Moch:2004pa, Vogt:2004mw}), but there are some important differences in colour non-singlet channels that we need to review.  For definiteness, we consider evolution in the scale $\mu_1$ of the first parton in the DPD.

It is useful to form linear combinations of distributions, in order to decouple the DGLAP equations between different parton channels as much as possible.
Combinations with definite charge conjugation for the first parton are
\begin{align}
   \pr{R_1R_2}{F}_{\smash{q_i^\pm} a_2}
   &=
   \pr{R_1R_2}{F}_{q_i a_2}
   \pm \pr{R_1R_2}{F}_{\bar{q}_i a_2}
   \,.
\end{align}
The DGLAP kernels for different parton transitions are related by charge conjugation invariance as follows:
\begin{align}
   \pr{RR}{P}_{q_i q_k}
   &= \pr{RR}{P}_{\ms \bar{q}_i\bar{q}_k}
   \,,
   \label{eqn:P_qiqj_charge_conjugation}
   \\
   \pr{RR}{P}_{q_i\bar{q}_k}
   &= \pr{RR}{P}_{\ms \bar{q}_i q_k}
   \,,
   \label{eqn:P_qbariqj_charge_conjugation}
   \\
   \pr{R\Rp}{P}_{q_i g}
   &= \eta(\Rp\ms)\, \pr{R\Rp}{P}_{\ms \bar{q}_i g}
   \,,
   \label{eqn:P_qg_charge_conjugation}
   \\
   \pr{R\Rp}{P}_{g\ms q_i}
   &= \eta(R)\, \pr{R\Rp}{P}_{g\ms \bar{q}_i}
   \,,
   \label{eqn:P_gq_charge_conjugation}
\end{align}
where $\eta(A) = -1$ and $\eta(R) = + 1$ for all other representations.  Due to this factor, the mixing pattern for gluons in the asymmetric octet differs from the well-known one in the colour singlet sector, as we shall see shortly.  Notice that in \eqref{eqn:P_qiqj_charge_conjugation} and \eqref{eqn:P_qbariqj_charge_conjugation} we simplified the representation indices, using that for quarks the only allowed representation pairs are $R\Rp = 11$ and $88$.

The flavour dependence for quark and antiquark kernels can be split into valence and pure singlet parts,
\begin{align}
   \pr{RR}{P}_{a_i b_k}
   &= \delta_{i k} \pr{RR}{P}_{ab}^V + \pr{RR}{P}_{ab}^S
   &
   \text{ with } a,b = q,\bar{q},
   \label{eq:quark-valence-sea-decomp}
\end{align}
where $i$ and $k$ denote the quark flavour.
As a result of the above relations, flavour non-singlet combinations of DPDs decouple from gluons and evolve as
\begin{align}
   \frac{d}{d\ln\mu_1}
      \pr{R_1R_2}{F}_{\smash{(q_i^\pm - q_k^\pm)}\ms a_2}
   &=
   2 \; \prb{R_1R_1}{P}^\pm \conv{x_1}
      \pr{R_1R_2}{F}_{\smash{(q_i^\pm - q_k^\pm)}\ms a_2}
   \label{eqn:non_singlet_evolution}
\end{align}
with kernels
\begin{align}
   \pr{RR}{P}^\pm
   &=
   \pr{RR}{P}^V_{q q} \pm \pr{RR}{P}^V_{q\bar{q}}
   \,,
\label{eqn:P_pm_def}
\end{align}
where we abbreviated
${F}_{\smash{(q_i^\pm - q_k^\pm)}\ms a_2} = {F}_{\smash{q_i^\pm} a_2} - {F}_{\smash{q_k^\pm} a_2} \ms$.

The flavour singlet combinations for the first parton are defined as
\begin{align}
   \pr{R_1R_2}{F}_{\Sigma^\pm a_2}
   &= \sum_i\pr{R_1R_2}{F}_{\smash{q_i^\pm} a_2}
\end{align}
and mix with gluons.  In the charge conjugation even sector, this results in the coupled system
\begin{align}
   \frac{d}{d\ln\mu_1}
   \begin{pmatrix}
   \pr{R R_2}{F}_{\Sigma^+a_2} \\[0.1em]
   \pr{\Rp R_2}{F}_{g\ms a_2} \
   \end{pmatrix}
   &= 2
   \begin{pmatrix}
   \pr{R R}{P}_{\Sigma^+\Sigma^+} &
   \pr{R \Rp}{P}_{\Sigma^+g} \\[0.1em]
   \pr{\Rp R}{P}_{g\ms\Sigma^+} &
   \pr{\Rp \Rp}{P}_{g g}
   \end{pmatrix}
   \conv{x_1}
   \begin{pmatrix}
   \pr{R R_2}{F}_{\Sigma^+a_2} \\[0.1em]
   \pr{\Rp R_2}{F}_{g\ms a_2} \
   \end{pmatrix}
   \notag \\[0.1em]
   &&
   \hspace{-6em} \text{ for } R \Rp = 11, 8S.
\label{eqn:Sigma_plus_evolution}
\end{align}
Due to the factors $\eta(R)$ in the relations \eqref{eqn:P_qg_charge_conjugation} and \eqref{eqn:P_gq_charge_conjugation}, there is also quark-gluon mixing in the charge conjugation odd sector:
\begin{align}
   \frac{d}{d\ln\mu_1}
   \begin{pmatrix}
      \pr{8 R_2}{F}_{\Sigma^-a_2} \\[0.1em]
      \pr{A R_2}{F}_{g\ms a_2}
   \end{pmatrix}
   &=
   2
   \begin{pmatrix}
      \pr{88}{P}_{\Sigma^-\Sigma^-} &
      \pr{8A}{P}_{\Sigma^-g} \\[0.1em]
      \pr{A8}{P}_{g\ms \Sigma^-} &
      \pr{AA}{P}_{g g}
   \end{pmatrix}
   \conv{x_1}
   \begin{pmatrix}
      \pr{8 R_2}{F}_{\Sigma^-a_2} \\[0.1em]
      \pr{A R_2}{F}_{g\ms a_2}
   \end{pmatrix}
   \,.
\label{eqn:Sigma_minus_octet_evolution}
\end{align}
The kernels appearing in these equations are given by
\begin{align}
   \pr{RR}{P}_{\Sigma^\pm\Sigma^\pm}
   &= \pr{RR}{P}^\pm + n_f \ms \Bigl[ \pr{RR}{P}_{q q}^S
   \pm \pr{RR}{P}_{q\bar{q}}^S \ms \Bigr]
   &
   \text{for } R = 1,8
   \label{eqn:P_Sigma_def}
\end{align}
and
\begin{align}
   \pr{R\Rp}{P}_{\Sigma^+ g}
   &= 2 n_f \, \pr{R\Rp}{P}_{q_i g}
   \,,
   &
   \pr{\Rp R}{P}_{g\Sigma^+}
   &= \pr{\Rp R}{P}_{g \ms q_i}
   &
   \text{for } R\Rp = 11,8S,
   \\
   \pr{8A}{P}_{\Sigma^- g}
   &= 2 n_f \, \pr{8A}{P}_{q_i g}
   \,,
   &
   \pr{A8}{P}_{g\Sigma^-}
   &= \pr{A8}{P}_{g \ms q_i}
   \,,
\end{align}
where $n_f$ is the number of active quark flavours.
Quark-gluon mixing kernels for the remaining combinations of $R \Rp$ and $\Sigma^\pm$ are zero.

In the colour singlet channel, the flavour singlet combination with odd charge conjugation parity evolves by itself:
\begin{align}
   \frac{d}{d\ln\mu_1} \pr{1 R_2}{F}_{\Sigma^-a_2}
   &= 2 \; \pr{11}{P}_{\Sigma^-\Sigma^-} \otimes \pr{1 R_2}{F}_{\Sigma^-a_2}
   \,.
\label{eqn:Sigma_minus_singlet_evolution}
\end{align}
The gluon distributions for higher colour representations, $\pr{\tentenbar}{F}_{g g}$, $\pr{\tenbarten}{F}_{g g}$, and $\pr{\twensev}{F}_{g g}$ evolve without mixing as well.

The preceding discussion applies to polarised channels in complete analogy.  For the polarised valence kernels, we use the notation
\begin{align}
   \pr{RR}{P}^\pm_\Delta
   &=
   \pr{RR}{P}^V_{\Delta q\Delta q}
   \pm \pr{RR}{P}^V_{\Delta q\Delta\bar{q}}
   \,,
   &
   \pr{RR}{P}^\pm_\delta
   &=
   \pr{RR}{P}^V_{\delta q\delta q}
   \pm \pr{RR}{P}^V_{\delta q\delta\bar{q}}
   \,.
\label{eqn:P_Delta_pm_def}
\end{align}
Transversely polarised quarks do not mix with gluons under evolution, so that the corresponding kernels are all zero.  The same holds for the pure singlet combinations,
\begin{align}
   \pr{RR}{P}^S_{\delta q\delta q}
   &= \pr{RR}{P}^S_{\delta q\delta\bar{q}}
   = 0
   \,.
\end{align}

%%%%%%%%%%%%%%%%%%%%%%%%%%%%%%%%%%%%%%%%%%%%%%%%%%%%%%%%%%%%%%%%%%%%%%%%%%%%%%%%

\subsection{Finite renormalisation for longitudinal polarisation}
\label{sec:finite-renorm}

It is known that \msbar renormalisation of spin dependent quantities computed with the HVBM scheme for $\gamma_5$ and the $\varepsilon$ tensor have a number of undesirable features.  These can be removed by an additional finite renormalisation in $4$ space-time dimensions, after one has performed \msbar subtraction and removed the dimensional regulator.

The corresponding scheme change for longitudinally polarised PDFs is analysed at length in \sect{3.1} of \cite{Vogelsang:1996im}, see also \sect{2} of \cite{Moch:2014sna}.  It restores the relation
\begin{align}
   \label{eq:hel-conservation}
   \pr{11}{P}^{V}_{\Delta q \Delta q}
   &= \pr{11}{P}^{V}_{q q}
   \,,
\end{align}
which corresponds to helicity conservation in the valence sector and is broken at order $a_s^2$ in the \msbar scheme.  Restoration of the equality \eqref{eq:hel-conservation} also ensures the scale independence of the flavour non-singlet axial current, which was for instance discussed in \cite{Larin:1993tq}.

We shall review the necessary scheme transformation here, because we need to extend it to the colour octet sector.  The finite renormalisation can be implemented at the level of the twist-two operators $\prn{R}{\mathcal{O}}_{a}^{r r'}$ that are projected on a definite colour representation.  It reads
\begin{align}
   \label{eq:scheme-trf-q}
   \prn{R}{\mathcal{O}}_{\Delta q_i}
   &= \pr{R}{\widetilde{Z}} \conv{x}
   \prn{R}{\mathcal{O}}_{\Delta q_i}^{\,\msbars}
   \,,
   \\
   \label{eq:scheme-trf-qbar}
   \prn{R}{\mathcal{O}}_{\Delta\bar{q}_i}
   &= \pr{R}{\widetilde{Z}} \conv{x}
   \prn{R}{\mathcal{O}}_{\Delta\bar{q}_i}^{\,\msbars}
   \,,
\end{align}
where for brevity we have omitted the colour indices of the operators.
One obtains the same transformation for the operator combinations associated with the distributions $F_{\smash{\Delta q_i^\pm} a_2}$ and $F_{\Delta \Sigma^\pm \ms a_2}$.  The \msbar renormalised gluon operators are left untouched.
Notice that the renormalisation factor $\pr{R}{\smash{\widetilde{Z}}}(x,\mu)$ does not depend on $\zeta$, so that the Collins-Soper equation for DPDs remains unchanged and polarisation independent.

Let us (re)derive the change of the evolution kernels induced by this finite renormalisation.
For a flavour non-singlet combination, the change in operator renormalisation specified by \eqref{eq:scheme-trf-q} and \eqref{eq:scheme-trf-qbar}  implies
\begin{align}
   \pr{R_1R_2}{F}_{\smash{(\Delta q_{i}^\pm - \Delta q_{k}^\pm)}\ms  a_2}
   &=
   \pr{R_1}{\widetilde{Z}} \conv{x_1}
   \pr{R_1R_2}{F}_{\smash{(\Delta q_{i}^\pm - \Delta q_{k}^\pm)}\ms
      a_2}^{\,\msbars}
   \,,
\end{align}
where it is understood that the scheme transformation is only performed for the first parton in the DPD.
Taking the RGE derivative in $\mu_1$ on both sides gives
\begin{align}
\label{eq:scheme-trf-1}
   2\, \pr{R_1R_1}{P}^\pm_\Delta
   &
   \,\conv{x_1}\, \Bigl[
   \pr{R_1}{\widetilde{Z}} \conv{x^\prime_1}
   \pr{R_1R_2}{F}_{(\Delta q_{i}^\pm - \Delta q_{k}^\pm)\ms a_2}^{\,\msbars}
   \Bigr]
   = \Bigl(\frac{d}{d\ln\mu_1} \pr{R_1}{\widetilde{Z}}\Bigr)
   \,\conv{x_1}\,
   \pr{R_1R_2}{F}_{(\Delta q_{i}^\pm - \Delta q_{k}^\pm)\ms
     a_2}^{\,\msbars}
   \notag\\[0.2em]
   &
   + 2\, \pr{R_1}{\widetilde{Z}} \conv{x_1}
   \Bigl[
   \pr{R_1R_1}{P}^\pm_{\Delta,\msbars}
   \,\conv{x^\prime_1}\,
   \pr{R_1R_2}{F}_{(\Delta q_{i}^\pm - \Delta q_{k}^\pm)\ms
     a_2}^{\,\msbars}
   \Bigl]
   \,.
\end{align}
With the modified law of associativity \eqref{eqn:Mellin_conv_associative} for triple convolution products, this implies
\begin{align}
   \pr{RR}{P}^\pm_\Delta(x^\prime, \zeta) \conv{x}
   \pr{R}{\widetilde{Z}}(x^\prime)
   &= \frac{1}{2} \frac{d}{d\ln\mu} \pr{R}{\widetilde{Z}}(x)
   + \pr{R}{\widetilde{Z}}(x^\prime) \conv{x}
   \pr{RR}{P}^\pm_{\Delta,\msbars}\bigl( x^\prime,
      x^{\prime\ms 2} \zeta / x^2 \bigr)
   \,.
\label{eqn:P_pm_Delta_shifting_formula}
\end{align}
Taking the expansion
\begin{align}
   \pr{R}{\widetilde{Z}}(x, \mu)
   &=
   \delta(1-x)
   + \sum_{n=1}^{\infty} a_s^{n}(\mu) \, \pr{R}{\widetilde{Z}}^{(n)}(x)
\end{align}
and the form \eqref{eq:rge-derivative} of the RGE derivative for $\epsilon = 0$, we obtain
\begin{align}
   \pr{RR}{P}^{\pm (0)}_\Delta(x, \zeta)
   &= \pr{RR}{P}^{\pm (0)}_{\Delta,\msbars}(x, \zeta)
   \,,
   \\
   \pr{RR}{P}^{\pm (1)}_\Delta(x, \zeta)
   &= \pr{RR}{P}^{\pm (1)}_{\Delta,\msbars}(x, \zeta)
   + \frac{1}{2} \Bigl( \prn{R}{\gamma}_J^{(0)} \ln(x) - \beta_0 \Bigr) \ms
   \pr{R}{\widetilde{Z}}^{(1)}(x)
   \,,
\label{eqn:P_pm_NLO_shift}
\end{align}
where in the second relation we used the commutativity of the convolution product and
\begin{align}
   \pr{RR}{P}^{\pm}_{\Delta,\msbars}\bigl( x^\prime,
         x^{\prime\ms 2} \zeta / x^2 \bigr)
   - \pr{RR}{P}^{\pm}_{\Delta,\msbars}(x^\prime, \zeta)
   &=
   \frac{1}{2} \delta(1 - x^\prime) \, \prn{R}{\gamma}_J \ms \ln(x)
   \,,
\end{align}
which follows from \eqref{eqn:P_Delta_pm_def}, the polarised analogue of \eqref{eq:quark-valence-sea-decomp}, and \eqref{eqn:P_zeta_dependence}.
The equal shift of $\pr{RR}{P}_{\Delta}^{+ (1)}$ and $\pr{RR}{P}_{\Delta}^{- (1)}$ in \eqref{eqn:P_pm_NLO_shift} leads to the same shift for
\begin{align}
   \pr{RR}{P}^V_{\Delta q\Delta q}
   &= \Bigl(\pr{RR}{P}^+_\Delta + \pr{RR}{P}^-_\Delta\Bigl) \Big/ 2
   \,,
\end{align}
whereas
\begin{align}
   \pr{RR}{P}^V_{\Delta q\Delta \bar{q}}
   &= \Bigl(\pr{RR}{P}^+_\Delta - \pr{RR}{P}^-_\Delta\Bigl) \Big/ 2
\label{eqn:P_Delta_qbar_in_terms_of_P_pm}
\end{align}
remains unchanged.

The scheme transformation in the flavour singlet sector can be derived in the same way.  We obtain
\begin{align}
   &\begin{pmatrix}
      \pr{RR}{P}_{\Delta\Sigma^+\Delta\Sigma^+} &
      \pr{R\Rp}{P}_{\Delta\Sigma^+ \Delta g}
      \\[0.2em]
      \pr{\Rp R}{P}_{\Delta g \ms \Delta\Sigma^+} &
      \pr{\Rp\Rp}{P}_{\Delta g\Delta g}
   \end{pmatrix}
   \conv{x}
   \begin{pmatrix}
      \pr{R}{\widetilde{Z}}(x') & 0 \\[0.2em]
      0 & \delta(1-x')
   \end{pmatrix}
   =
   \frac{1}{2}
   \begin{pmatrix}
      d \ms \pr{R}{\widetilde{Z}}(x) / (d\ln\mu) & 0 \\[0.2em]
      0 & 0
   \end{pmatrix}
   \notag\\[0.3em]
   & \qquad
   + \begin{pmatrix}
      \pr{R}{\widetilde{Z}}(x') & 0 \\[0.2em]
      0 & \delta(1-x')
   \end{pmatrix}
   \conv{x}
   \begin{pmatrix}
      \pr{RR}{P}_{\Delta\Sigma^+\Delta\Sigma^+,\msbars} &
      \pr{R\Rp}{P}_{\Delta\Sigma^+ \Delta g,\msbars}
      \\[0.2em]
      \pr{\Rp R}{P}_{\Delta g \ms \Delta \Sigma^+,\msbars} &
      \pr{\Rp\Rp}{P}_{\Delta g\Delta g,\msbars}
   \end{pmatrix}
   \notag \\[0.1em]
   &
   \hspace{26em} \text{ with } R\Rp = 11, 8S,
\label{eqn:P_Singlet_shifting_formula}
\end{align}
where the arguments of the DGLAP kernels are as in \eqref{eqn:P_pm_Delta_shifting_formula}.  An analogous relation holds for $R\Rp = 8A$ with $\Sigma^-$ instead of $\Sigma^+$.  The identity for the upper left component in \eqref{eqn:P_Singlet_shifting_formula} has the same form as the one for $\pr{RR}{P}^\pm_\Delta$ in \eqref{eqn:P_pm_Delta_shifting_formula}.  One also obtains the same type of identity for $\pr{11}{P}_{\Delta\Sigma^- \Delta\Sigma^-}$, which appears in the polarised version of the evolution equation \eqref{eqn:Sigma_minus_singlet_evolution}.  In summary, we find that the four combinations
\begin{align}
   \pr{RR}{P}^{(1)}_{\Delta\Sigma^+ \Delta\Sigma^+} \,,
   \pr{RR}{P}^{(1)}_{\Delta\Sigma^- \Delta\Sigma^-}
   &&
   \text{ with } R=1,8
\end{align}
transform in the same manner as $\pr{RR}{P}^{\pm (1)}_{\Delta}$ in \eqref{eqn:P_pm_NLO_shift}, whereas their LO analogues are invariant.
Given \eqref{eqn:P_Sigma_def} this implies that the pure singlet kernels $P_{\Delta q\Delta q}^{S}$ and $P_{\Delta q\Delta\bar{q}}^{S}$ remain unchanged at both LO and NLO.

The off-diagonal elements of the matrix equation \eqref{eqn:P_Singlet_shifting_formula} and its analogue with $\Sigma^-$ lead to the scheme change
\begin{align}
   \pr{R\Rp}{P}^{(1)}_{\Delta \Sigma^\pm \Delta g}
   &= \pr{R\Rp}{P}^{(1)}_{\Delta \Sigma^\pm\Delta g,\msbars}
   \,+\,
   \pr{R}{\widetilde{Z}}^{(1)}
   \otimes \pr{R\Rp}{P}^{(0)}_{\Delta \Sigma^\pm \Delta g,\msbars}
   \,,
\label{eqn:qg_shift}
   \\
   \pr{\Rp R}{P}^{(1)}_{\Delta g  \ms \Delta \Sigma^\pm}
   &= \pr{\Rp R}{P}^{(1)}_{\Delta g \ms \Delta \Sigma^\pm,\msbars}
   \,-\,
   \pr{\Rp R}{P}^{(0)}_{\Delta g \ms \Delta \Sigma^\pm,\msbars}
   \otimes \pr{R}{\widetilde{Z}}^{(1)}
\label{eqn:gq_shift}
\end{align}
for the relevant colour combinations, i.e.\ $R\Rp = 11, 8S$ for $\Sigma^+$ and $R\Rp = 8A$ for $\Sigma^-$.  At LO the quark-gluon mixing kernels remain unchanged.  We recall that these mixing kernels are $\zeta$ independent at all orders.  Finally, the evolution kernel $\pr{R\Rp}{P}_{\Delta g \Delta g}$ for gluons is invariant under the scheme change in all colour channels.

For $R=1$ we adopt the renormalisation factor $\pr{R}{\widetilde{Z}}$ taken in \cite{Vogelsang:1997ak}.  Our choice for $R=8$ is presented in \sect{\ref{sec:longit-results}}.

\section{Results}
\label{sec:results}

In this section, we present the DGLAP kernels for all colour channels at LO and NLO accuracy.  Our results are for the standard \msbar scheme specified in \eqref{eq:as-renorm} and \eqref{eq:S-eps-choice}, followed by the finite renormalisation for longitudinally polarised partons that was described in \sect{\ref{sec:finite-renorm}}.  We recall the values
\begin{align}
   C_A &= N \,,
   &
   C_F &= \frac{N^2-1}{2N} \,,
   &
   \beta_0 &= \frac{11}{3} C_A - \frac23 n_f
\end{align}
for the colour group SU($N$).  We have set $T_F = 1/2$ throughout.  For the colour singlet, octet, and decuplet channels, we give results for general $N$, whereas for the 27 representation we always take $N=3$ for the number of colours.\footnote{%
The reason for this specialisation is explained in footnote 4 of \protect\cite{Buffing:2017mqm}.}
At both LO and NLO, we find the following structure of the DGLAP kernels:
\begin{align}
   \label{eq:P-master-form}
   \pr{R\Rp}{P}^{(n)}_{a b}(x, \zeta/\mu^2)
   &=
   \pr{R\Rp}{\widetilde{P}}^{(n)}_{a b}(x)
   \notag\\
   &\quad
   + \frac{1}{2} \delta_{R\Rpbar}\, \delta_{a b}\, \delta(1-x)\,
     \biggl[
      d_a^{(n)} + \pr{R}{c}^{(n)}
      - \frac{1}{2} \prn{R}{\gamma}_J^{(n)} \ln\frac{\zeta}{\mu^2}
      \biggr]
   \,,
\end{align}
where $d_a$ depends on the parton species but not on the polarisation,
\begin{align}
   d_{q \vphantom{\Delta}}^{(n)} &= d_{\Delta q}^{(n)} = d_{\delta q}^{(n)}
   \,,
   &
   d_{g \vphantom{\Delta}}^{(n)} &= d_{\Delta g}^{(n)}
   \,,
\end{align}
whilst $\pr{R}{c}$ depends only on the multiplicity of the colour representation $R$ and is zero for the colour singlet:
\begin{align}
   \pr{1}{c}^{(n)} &= 0
   \,,
   &
   \pr{8}{c}^{(n)} &= \pr{A}{c}^{(n)} = \pr{S}{c}^{(n)}
   \,,
   &
   \pr{\ten}{c}^{(n)} &= \pr{\tenbar}{c}^{(n)}
   \,.
\end{align}
The terms going with $\delta(1-x)$ thus have a remarkably simple dependence on the different quantum numbers.

The coefficients $d_{a}$ are well known from the ordinary DGLAP kernels.  They read
\begin{align}
   d_q^{(0)} &= 3 C_F \,,
   &
   d_g^{(0)} &= \beta_0
\end{align}
and
\begin{align}
   d_q^{(1)} &=
   C_F^2 \biggl( \frac{3}{4} - \pi^2 + 12\zeta_3 \biggr)
   + C_F \ms C_A \biggl( \frac{17}{12} + \frac{11}{9} \pi^2 - 6\zeta_3 \biggr)
   - C_F \ms n_f \biggl( \frac{1}{6} + \frac{2}{9} \pi^2 \biggr)
   \,,
   \\
   d_g^{(1)} &=
   C_A^2 \biggl( \frac{16}{3} + 6\zeta_3 \biggr)
   - \frac{4}{3} C_A \ms n_f
   - C_F \ms n_f
\end{align}
and can be determined from the kernels $\pr{11}{\widetilde{P}}_{a b}(x)$ together with the number and momentum sum rules for parton distributions.
Here $\zeta_n$ denotes the Riemann $\zeta$ function at argument $n$.

At LO, we have colour dependent coefficients
\begin{align}
   \prn{R}{c}^{(0)} &= 0
\end{align}
and
\begin{align}
   \prn{8}{\gamma}_J^{(0)}
   &=
   2 C_A
   \,,
\end{align}
whilst at NLO we find
\begin{align}
   \prn{8}{c}^{(1)}
   &= C_A^2 \biggl( \frac{101}{27} - \frac{11}{72} \pi^2
   - \frac{7}{2} \zeta_3 \biggr)
   + C_A \ms n_f \biggl( - \frac{14}{27} + \frac{1}{36} \pi^2 \biggr)
   \,,
   \\
   \label{eq:gammaJ-NLO}
   \prn{8}{\gamma}_J^{(1)}
   &=
   C_A^2 \biggl( \frac{67}{9} - \frac{1}{3} \pi^2 \biggr)
   - \frac{10}{9} C_A \ms n_f
   \,.
\end{align}
\rev{As shown in \sect{7.2.1} of \cite{Buffing:2017mqm}, the cusp anomalous dimensions for colour octet DPDs and for single-gluon TMDs are related as}
\begin{align}
   \label{eq:gamma-J-K}
   \prn{8}{\gamma_J} &= \frac{1}{2} \ms \gamma_{K, g}
\end{align}
\rev{at all orders in the strong coupling.  Comparing \eqref{eq:gammaJ-NLO} with \eqref{eq:tmd-g-cusp}, we see that this relation is fulfilled by our calculation, which constitutes another cross check.}

Remarkably, we find that the colour dependent coefficients for higher representations all satisfy Casimir scaling up to NLO:
\begin{align}
   \label{eq:casimir-10}
   \frac{\prn{\ten}{c}^{(1)}}{\pr{8}{c}^{(1)}}
   &=
   \frac{\prn{\ten}{\gamma}_J^{(0)}}{\prn{8}{\gamma}_J^{(0)}}
   =
   \frac{\prn{\ten}{\gamma}_J^{(1)}}{\prn{8}{\gamma}_J^{(1)}}
   = \frac{C_{10}}{C_A}
   = 2
   \,,
   \\[0.4em]
   \frac{\prn{27}{c}^{(1)}}{\prn{8}{c}^{(1)}}
   &=
   \frac{\prn{27}{\gamma}_J^{(0)}}{\prn{8}{\gamma}_J^{(0)}}
   =
   \frac{\prn{27}{\gamma}_J^{(1)}}{\prn{8}{\gamma}_J^{(1)}}
   = \frac{C_{27}}{C_A}
   = \frac{8}{3}
   \,,
\end{align}
where $C_{10}$ and $C_{27}$ are the eigenvalues of the quadratic Casimir operators for the $10$ and $27$ representations \cite{Bali:2000un}.

%...............................................................................

\paragraph{Decuplet distributions.}

Combining \eqref{eq:ten-only-Z-tmd} and \eqref{eq:gamma-tmd-from-Z} we find for the decuplet kernels
\begin{align}
   \label{eq:full-kernel-10}
   \pr{\tentenbar}{P}^{(n)}_{a a}(x, \zeta/\mu^2)
   &= \frac{1}{2} \ms \delta(1-x) \, \gamma_{F,g}^{(n)}(\zeta/\mu^2)
   &
   \text{ for } n=0,1,
\end{align}
where $a = g$ or $\Delta g$.  At NLO accuracy, the decuplet distributions hence evolve as
\begin{align}
   \label{eq:DGLAP-10}
   \frac{d}{d\ln\mu_1}
   \pr{\tentenbar}{F}_{a_1 a_2}(x_1,x_2,y,\mu_1,\mu_2,\zeta)
   &= \gamma_{F,g}(\mu_1^{}, x_1^2 \zeta) \;
   \pr{\tentenbar}{F}_{a_1 a_2}(x_1,x_2,y,\mu_1,\mu_2,\zeta)
\end{align}
with $(a_1 a_2) = (g g)$ or $(\Delta g \Delta g)$.  This has the same form as the evolution equation for a \rev{single-gluon} TMD, and its analytic solution is well known.  Since the anomalous dimension on the r.h.s.\ \rev{of} \eqref{eq:DGLAP-10} depends on $x_1$, the momentum fraction dependence of the DPD changes under evolution.

The comparison of \eqref{eq:full-kernel-10} with \eqref{eq:P-master-form} implies
\begin{align}
   \label{eq:gammaJ-10}
   \pr{10}{\gamma}_J^{(n)}
   &= \gamma_{K,g}^{(n)}
   &
   \text{ for } n=0,1
\end{align}
and
\begin{align}
   d_{g}^{(0)} &= \gamma_g^{(0)} \,,
   &
   d_{g}^{(1)} + \prn{10}{c}^{(1)} &= \gamma_g^{(1)}
   \,,
\end{align}
where we used $\prn{R}{c}^{(0)} = 0$ \rev{and the relation \eqref{eqn:TMD_anom_dim_def} between the cusp and non-cusp parts $\gamma_{K, g}$ and $\gamma_g$ of the TMD anomalous dimension $\gamma_{F, g}$}.
Notice that \eqref{eq:gammaJ-10} is consistent with \eqref{eq:gamma-J-K} and \eqref{eq:casimir-10}.

%...............................................................................

\paragraph{Terms depending on $x$.}
At leading order, the $x$ dependent parts of the kernels for different colour channels are proportional to each other:
\begin{align}
   \label{eq:LO-PVqq}
   \pr{88}{\widetilde{P}}^{V (0)}_{q q}(x)
   &= c_{q q}(88) \, \pr{11}{\widetilde{P}}^{V (0)}_{q q}(x)
   \,,
   \\[0.2em]
   \pr{8A}{\widetilde{P}}^{(0)}_{\Sigma^- g}(x)
   &= c_{q g}(8A) \, \pr{11}{\widetilde{P}}^{(0)}_{\Sigma^+ g}(x)
   \,,
   &&
   \pr{A8}{\widetilde{P}}^{(0)}_{g \ms \Sigma^-}(x)
   = c_{g q}(A8) \, \pr{11}{\widetilde{P}}^{(0)}_{g \ms \Sigma^+}(x)
   \,,
   \\[0.2em]
   \pr{8S}{\widetilde{P}}^{(0)}_{\Sigma^+ g}(x)
   &= c_{q g}(8S) \, \pr{11}{\widetilde{P}}^{(0)}_{\Sigma^+ g}(x)
   \,,
   &&
   \pr{S8}{\widetilde{P}}^{(0)}_{g \ms \Sigma^+}(x)
   = c_{g q}(S8) \, \pr{11}{\widetilde{P}}^{(0)}_{g \ms \Sigma^+}(x)
   \,,
   \\[0.2em]
   \pr{R\Rbar \ms}{\widetilde{P}}^{(0)}_{g g}(x)
   &= c_{g g}(R\Rbar) \, \pr{11}{\widetilde{P}}^{(0)}_{g g}(x)
   \label{eq:LO-Pgg}
\end{align}
with colour factors
\begin{align}
   c_{q q}(88) &= {}- \frac{1}{N^2-1}
   \,,
   \\[0.2em]
   c_{q g}(8A) = c_{g q}(A8) &= \sqrt{\frac{N^2}{2(N^2-1)}}
   \,,
   \\
   c_{q g}(8S) = c_{g q}(S8) &= \sqrt{\frac{N^2-4}{2(N^2-1)}}
   \,,
   \\[0.4em]
   c_{g g}(AA) = c_{g g}(SS) &= \frac{1}{2}
   \,,
   \label{eqn:cggA/S}
   \\
   c_{g g}(10\ms\overline{10}) &= 0
   \,,
   \vphantom{\sqrt{\frac{1^2}{1^2}}}
   \\
   c_{g g}(\twensev) &= - \frac{1}{3}
   \,.
\end{align}
Analogues of \eqref{eq:LO-PVqq} to \eqref{eq:LO-Pgg} hold for the polarised kernels, with the same colour factors as in the unpolarised case.
For completeness, we list the LO kernels in the colour singlet representation:
\begin{align}
   \pr{11}{\widetilde{P}}^{V (0)}_{q q}(x)
   &= C_F \ms p_{q q}(x) \,,
   \\
   \pr{11}{\widetilde{P}}^{(0)}_{\Sigma^+g}(x)
   &= n_f \ms p_{q g}(x) \,,
   \\
   \pr{11}{\widetilde{P}}^{(0)}_{g\Sigma^+}(x)
   &= C_F \ms p_{g q}(x) \,,
   \\
   \pr{11}{\widetilde{P}}^{(0)}_{g g}(x)
   &= 2 C_A \, p_{g g}(x) \,,
   \intertext{and}
   \pr{11}{\widetilde{P}}^{V (0)}_{\Delta q\Delta q}(x)
   &= \pr{11}{\widetilde{P}}^{V (0)}_{q q}(x) \,,
   \\
   \pr{11}{\widetilde{P}}^{(0)}_{\Delta \Sigma^+ \Delta g}(x)
   &= n_f \ms \Dp{q}{g}(x) \,,
   \label{eqn:DeltaPqgSinglet}
   \\
   \pr{11}{\widetilde{P}}^{(0)}_{\Delta g\Delta \Sigma^+}(x)
   &= C_F \ms \Dp{g}{q}(x) \,,
   \label{eqn:DeltaPgqSinglet}
   \\
   \pr{11}{\widetilde{P}}^{(0)}_{\Delta g\Delta g}(x)
   &= 2 C_A \, \Dp{g}{g}(x) \,,
   \\[0.2em]
   \pr{11}{\widetilde{P}}^{V (0)}_{\delta q\delta q}(x)
   &= C_F \ms \delp{q}{q}(x)
   \,,
\end{align}
where
\begin{align}
   p_{q q}(x) &= \frac{1 + x^2}{\bigl[1-x\bigr]_+} \,,
   \\
   p_{q g}(x) &= x^2 + (1-x)^2 \,,
   \\
   p_{g q}(x) &= \frac{1+(1-x)^2}{x} \,,
   \\
   p_{g g}(x) &= \frac{1}{\bigl[1-x\bigr]_+} + \frac{1}{x} + x(1-x) - 2 \,,
   \\[0.2em]
   \Dp{q}{g}(x) &= x^2 - (1-x)^2 = 2x - 1 \,,
   \\
   \Dp{g}{q}(x) &= \frac{1-(1-x)^2}{x} = 2-x \,,
   \\
   \Dp{g}{g}(x) &= \frac{1}{\bigl[1-x\bigr]_+} - 2x + 1
   \,,
   \\[0.2em]
   \label{eq:lo-p_dqdq}
   \delp{q}{q}(x) &= \frac{2x}{[1-x]_+}
   \,.
\end{align}
The plus-distribution is defined in \eqref{eq:def-plus-distrib}.
We recall that at LO there are no quark-antiquark transitions and no pure singlet kernels, so that one has
\begin{align}
   \pr{RR}{P}^{V (0)}_{q \bar{q}}
   &= \pr{RR}{P}^{S (0)}_{q \bar{q}}
    = \pr{RR}{P}^{S (0)}_{q q}
    = 0
\end{align}
and likewise for the polarised cases.

In the following subsections, we list the $x$ dependent parts of the NLO kernels, including for definiteness the expressions of the colour singlet kernels from \cite{Ellis:1996nn, Vogelsang:1996im, Vogelsang:1997ak}.  Dilogarithms appear in the combination
\begin{align}
   S_2(x) &= \int\limits_{x/(1+x)}^{1/(1+x)}
   \frac{dz}{z} \ln\biggl(\frac{1-z}{z}\biggr)
   = -2 \ms \mathrm{Li}_2(-x) + \frac{1}{2} \ln^2(x)
   - 2 \ln(x) \ln(1+x) - \frac{\pi^2}{6}
   \,.
   \label{eqn:S2}
\end{align}

%%%%%%%%%%%%%%%%%%%%%%%%%%%%%%%%%%%%%%%%%%%%%%%%%%%%%%%%%%%%%%%%%%%%%%%%%%%%%%%%

\subsection{Unpolarised partons}
\label{sec:unpol-results}

\paragraph{Quark kernels.}
For the valence kernels, we obtain
\begin{align}
   \pr{11}{\widetilde{P}}^{V (1)}_{q q}(x)
   &= \frac{C_F}{6} \ms\Biggl\{ C_F\biggl[-3 \ms (1+x) \ln^2(x)
   - 12 p_{q q}(x) \ln(x)\ln(1-x) \notag\\
   &\quad\qquad\qquad - \Bigl(9p_{q q}(x) + 21x + 9\Bigr)\ln(x)
   - 30 \ms (1-x)\biggr] \notag\\
   &\quad\qquad + C_A \biggl[3p_{q q}(x) \ln^2(x)
   + \Bigl(11p_{q q}(x) + 6 \ms (1+x)\Bigr)\ln(x) \notag\\
   &\quad\qquad\qquad\quad + \Bigl(\frac{67}{3}
   - \pi^2\Bigr)p_{q q}(x) + 40(1-x)\biggr] \notag\\
   &\quad\qquad + n_f\biggl[ -2p_{q q}(x) \ln(x)
   - \frac{10}{3} p_{q q}(x) - 4(1-x)\biggr]\Biggr\}
   \,,
   \\
   \pr{88}{\widetilde{P}}^{V (1)}_{q q}(x)
   &= c_{q q}(88) \ms\Biggl\{ {\pr{11}{\widetilde{P}}^{V (1)}_{q q}(x)}
   - \frac{C_F \ms C_A}{4} \biggl[\Bigl(2p_{q q}(x) - (1+x)\Bigr)\ln^2(x) \notag\\
   &\quad\qquad\qquad\qquad\qquad\qquad\qquad\quad
   + (8 - 4x)\ln(x) + 6(1-x)\biggr]\Biggr\}
   \,,
   \\
   \pr{11}{\widetilde{P}}^{V (1)}_{q\bar{q}}(x)
   &= {}- C_F \ms (C_A - 2C_F)
   \ms\Biggl\{ S_2(x) \, p_{q q}(-x) + (1+x) \ln(x) + 2 \ms (1-x)\Biggr\}
   \,,
   \\[0.3em]
   \pr{88}{\widetilde{P}}^{V (1)}_{q\bar{q}}(x)
   &= c_{q q}(88) \, (N^2+1) \, \pr{11}{\widetilde{P}}^{V (1)}_{q\bar{q}}(x)
   \,,
   \label{eq:88-V-qqbar}
\intertext{and for the pure singlet kernels we get}
   \pr{11}{\widetilde{P}}^{S (1)}_{q q}(x)
   &= \frac{C_F}{2} \ms\Biggl\{ - (1+x) \ln^2(x)
   + \biggl( \frac{8}{3} x^2 + 5x + 1 \biggr) \ln(x)
   - \frac{56}{9} x^2 + 6x - 2 + \frac{20}{9x}\Biggr\}
   \,,
   \\
   \pr{88}{\widetilde{P}}^{S (1)}_{q q}(x)
   &= {}- c_{q q}(88) \, (N^2-2) \, \pr{11}{\widetilde{P}}^{S (1)}_{q q}(x)
   \,,
   \vphantom{\frac{1}{1}}
   \label{eq:88-S-qq}
   \\
   \pr{11}{\widetilde{P}}^{S (1)}_{q\bar{q}}(x)
   &= \pr{11}{\widetilde{P}}^{S (1)}_{q q}(x)
   \,,
   \vphantom{\frac{1}{1}}
   \\
   \pr{88}{\widetilde{P}}^{S (1)}_{q\bar{q}}(x)
   &= 2 c_{q q}(88) \, \pr{11}{\widetilde{P}}^{S (1)}_{q\bar{q}}(x)
   \,.
   \vphantom{\frac{1}{1}}
   \label{eq:88-S-qqbar}
\end{align}
Notice that in the colour singlet sector one has $\pr{11}{\widetilde{P}}^{S}_{q q} = \pr{11}{\widetilde{P}}^{S}_{q\bar{q}}$ at both LO and NLO, so that according to \eqref{eqn:P_Sigma_def} the parton combination $\Sigma^-$ evolves with the same kernel $P^{-}$ as the flavour non-singlet combinations $q_i^- - q_{\smash{k} \vphantom{i}}^-$.  A difference between the pure singlet kernels for $q q$ and $q \bar{q}$ transitions only arises at order $a_s^3$, due to graphs with three gluons in the $t$ channel \cite{Vogt:2004mw}.  By contrast, the pure singlet kernels for $q q$ and $q \bar{q}$ transitions in the colour octet sector differ already at NLO.

%...............................................................................

\paragraph{Quark-gluon mixing kernels.}
The kernels for quark-gluon transitions read
\begin{align}
   \pr{11}{\widetilde{P}}^{(1)}_{\Sigma^+ g}(x)
   &= C_F \ms n_f \ms\Biggl\{ \frac{1}{2} (2x-1)\ln^2(x)
   + p_{q g}(x)\ln^2\biggl(\frac{x}{1-x}\biggr)
   + \Bigl(2p_{q g}(x)+2x-\frac{1}{2}\Bigr)\ln(x) \notag\\
   &\quad\qquad \quad \ + 2\Bigl(1-p_{q g}(x)\Bigr)\ln(1-x)
   + \Bigl(5 - \frac{\pi^2}{3}\Bigr)p_{q g}(x)
   - \frac92x+2\Biggr\} \notag\\
   &\quad + C_A \ms n_f \ms\Biggl\{ - \Bigl(\frac{1}{2}p_{q g}(x)
   + 4 x + 1\Bigr)\ln^2(x) - p_{q g}(x) \ln^2(1-x) \notag\\
   &\qquad\qquad \quad \ + \frac{1}{3} \Bigl(22p_{q g}(x)
   + 68x - 19\Bigr)\ln(x) + 2\Bigl(p_{q g}(x)-1\Bigr) \ln(1-x)\notag\\
   &\qquad\qquad \quad \ + \Bigl(\frac{\pi^2}{6}
   - \frac{109}{9}\Bigr)p_{q g}(x)
   + S_2(x) \, p_{q g}(-x) + \frac79x
   + \frac{91}{9} + \frac{20}{9x}\Biggr\}
   \,,
   \\
   \pr{8A}{\widetilde{P}}^{(1)}_{\Sigma^- g}(x)
   &= c_{q g}(8A) \ms\Biggl\{ {\pr{11}{\widetilde{P}}^{(1)}_{\Sigma^+ g}(x)}
   + \frac{1}{2} C_A \ms n_f \biggl[\Bigl( - p_{q g}(x) + 3x
   + \frac{3}{2}\Bigr) \ln^2(x) \notag\\
   &\quad\qquad\qquad\qquad\qquad\qquad \qquad\quad \
   - \frac{1}{3} \Bigl( 22p_{q g}(x) + 89x - 4\Bigr) \ln(x)
   + \frac{109}{9}p_{q g}(x)
   \notag\\[0.2em]
   &\quad\qquad\qquad\qquad\qquad\qquad \qquad\quad \
   - 2 S_2(x) \, p_{q g}(-x) + \frac{83}{9}x
   - \frac{172}{9} - \frac{20}{9x}\biggr]\Biggr\}
   \,,
   \\
   \pr{8S}{\widetilde{P}}^{(1)}_{\Sigma^+ g}(x)
   &= \frac{c_{q g}(8S)}{c_{q g}(8A)} \;
   \pr{8A}{\widetilde{P}}^{(1)}_{\Sigma^- g}(x)
   \label{eq:8S-8A-qg}
\intertext{and}
   \pr{11}{\widetilde{P}}^{(1)}_{g \ms \Sigma^+}(x)
   &= \frac{C_F}{18} \ms\Biggl\{ C_F \biggl[(9x - 18)\ln^2(x)
   - 18 p_{g q}(x) \ln^2(1-x) + (63x+36)\ln(x) \notag\\
   &\quad\qquad\qquad
   - \Bigl( 54p_{g q}(x) + 36x \Bigr)\ln(1-x)
   - 63x - 45\biggr] \notag \\
   &\quad\qquad
   + C_A \biggl[\Bigl(9p_{g q}(x)+18x+72\Bigr)\ln^2(x)
   - 36p_{g q}(x)\ln(x)\ln(1-x) \notag\\
   &\quad\qquad\qquad\ \
   + 18p_{g q}(x)\ln^2(1-x)
   - \Bigl( 48x^2 + 90x + 216 \Bigr)\ln(x) \notag\\
   &\quad\qquad\qquad\ \
   + \Bigl( 66p_{g q}(x) + 36x \Bigr)\ln(1-x)
   + \bigl(9-3\pi^2\bigr)p_{g q}(x) \notag\\
   &\quad\qquad\qquad\ \
   + 18 S_2(x) \, p_{g q}(-x) + 88x^2 + 65x + 56\biggr] \notag\\
   &\quad\qquad + n_f \biggl[-12p_{g q}(x)\ln(1-x)
   - 12x - 20p_{g q}(x)\biggr]\Biggr\}
   \,,
   \\
   \pr{A8}{\widetilde{P}}^{(1)}_{g \ms \Sigma^-}(x)
   &= c_{g q}(A8) \ms\Biggl\{ {\pr{11}{\widetilde{P}}^{(1)}_{g \ms \Sigma^+}(x)}
   + \frac{C_F \ms C_A}{18} \biggl[-\Bigl( 9p_{g q}(x)
   + \frac{27}{2}x + 27\Bigr)\ln^2(x) \notag\\
   &\qquad\qquad\qquad\qquad\qquad\qquad\qquad
   + \Bigl( 24x^2 + 27x + 135 \Bigr)\ln(x) + 58p_{g q}(x) \notag\\
   &\qquad\qquad\qquad\qquad\qquad\qquad\qquad
   - 18 S_2(x) \, p_{g q}(-x) - 44x^2 - 58x + 44\biggr]\Biggr\}
   \,,
   \\
   \pr{S8}{\widetilde{P}}^{(1)}_{g \ms \Sigma^+}(x)
   &= \frac{c_{g q}(S8)}{c_{g q}(A8)} \;
   \pr{A8}{\widetilde{P}}^{(1)}_{g \ms \Sigma^-}(x)
   \,.
   \label{eq:S8-A8-gq}
\end{align}
Notice the simple scaling relations \eqref{eq:8S-8A-qg} and \eqref{eq:S8-A8-gq} between pairs of kernels with a gluon in the antisymmetric or the symmetric octet.  At LO these relations are trivial because only one real graph contributes to each kernel, but remarkably they remain valid at NLO.

%...............................................................................

\paragraph{Gluon kernels.}
The pure gluon kernels read
\begin{align}
   \pr{11}{\widetilde{P}}^{(1)}_{g g}(x)
   &= C_F \ms n_f \ms\Biggl\{ -(1+x)\ln^2(x) - (5x+3)\ln(x)
   + \frac{10}{3}x^2 + 4 x - 8 + \frac{2}{3x}\Biggr\} \notag\\
   &\quad + C_A^2 \ms\Biggl\{ \Bigl(p_{g g}(x) + 4 \ms (1+x)\Bigr)\ln^2(x)
   - 4 p_{g g}(x) \ln(x) \ln(1-x) \notag\\
   &\qquad\qquad - \frac{1}{3}\Bigl(44 x^2 - 11 x + 25\Bigr)\ln(x)
   + \frac{1}{3} \Bigl( \frac{67}{3} - \pi^2 \Bigr) p_{g g}(x)
   + 2 S_2(x) \, p_{g g}(-x) \notag\\
   &\qquad\qquad + \frac{67}{9}x^2 + \frac{27}{2}(1-x)
   - \frac{67}{9x}\Biggr\} \notag\\
   &\quad + C_A \ms n_f \ms\Biggl\{ -\frac23(1+x)\ln(x) - \frac{10}{9}p_{g g}(x)
   + \frac{13}{9}x^2 - x + 1 - \frac{13}{9x}\Biggr\}
   \,,
   \\
   \pr{AA}{\widetilde{P}}^{(1)}_{g g}(x)
   &= c_{g g}(AA) \ms\Biggl\{ C_A^2 \biggl[2 \ms (1+x)\ln^2(x)
   - 4p_{g g}(x) \ln(x) \ln(1-x) \notag\\
   &\qquad\qquad\qquad\quad
   - \frac{1}{3} \Bigl( 22 x^2 - 14 x + 4 \Bigr) \ln(x)
   \notag\\[0.3em]
   &\qquad\qquad\qquad\quad
   + \frac{1}{3} \Bigl( \frac{67}{3} - \pi^2 \Bigr) p_{g g}(x) + 6 (1-x) \biggr]
   \notag\\
   &\qquad\qquad
   + C_A \ms n_f \biggl[-\frac{1}{2} (1+x) \ln^2(x)
   - \frac{1}{6} (19 x + 13) \ln(x)
   - \frac{10}{9} p_{g g}(x) \notag\\
   &\qquad\qquad\qquad\qquad
   + \frac{28}{9}x^2 + x - 3 - \frac{10}{9x}\biggr]\Biggr\}
   \,,
   \\
   \pr{SS}{\widetilde{P}}^{(1)}_{g g}(x)
   &= \frac{c_{g g}(SS)}{c_{g g}(AA)} \, \pr{AA}{\widetilde{P}}^{(1)}_{g g}(x) \notag\\
   & + 2 c_{g g}(SS) \ms (C_A - 2 C_F) \ms n_f
   \ms\Biggl\{ (1+x) \ln^2(x) + (5x + 3) \ln(x)
   \notag\\[-0.2em]
   &\qquad\qquad\qquad\qquad\qquad\qquad
   - \frac{10}{3} x^2 - 4 x + 8 - \frac{2}{3x}\Biggr\}
   \label{eqn:SPgg(1)}
   \,,
   \\
   \pr{\tentenbar}{\widetilde{P}}^{(1)}_{g g}(x)
   &= 0
   \,,
   \\
   \pr{\twensev}{\widetilde{P}}^{(1)}_{g g}(x)
   &= c_{g g}(\twensev) \ms\Biggl\{ - \Bigl( 15 p_{g g}(x)
   + 12 \ms (1+x)\Bigr) \ln^2(x) - 36 p_{g g}(x) \ln(x) \ln(1-x) \notag\\
   &\quad\qquad\qquad\ \ \,
   + \Bigl(44 x^2 + 57 x + 93 \Bigr) \ln(x)
   + 3 \ms \Bigl( \frac{67}{3} - \pi^2\Bigr) p_{g g}(x)
   \notag\\[0.4em]
   &\quad\qquad\qquad\ \ \,
   - 30 S_2(x) \, p_{g g}(-x) - \frac{335}{3}x^2
   - \frac{117}{2} (1-x) + \frac{335}{3x} \notag\\
   &\quad\qquad\qquad\ \ \,
   + n_f \biggl[ -2 \ms (1+x) \ln(x) - \frac{10}{3} p_{g g}(x)
   + \frac{13}{3} x^2 + 3 (1-x) - \frac{13}{3x} \biggr]\Biggr\}
   \,.
\end{align}
Note that according to \eqref{eqn:cggA/S} \rev{one} has $c_{g g}(SS)/c_{g g}(AA) = 1$.  We write out this ratio in \eqn{\eqref{eqn:SPgg(1)}} in order to make the proportionality $\pr{RR}{\widetilde{P}}_{g g} \propto c_{g g}(RR)$ explicit.
Whereas at LO the kernels $\pr{RR}{\widetilde{P}}_{g g}$ for $R=A$ and $R=S$ are equal, this is no longer the case at NLO.

%%%%%%%%%%%%%%%%%%%%%%%%%%%%%%%%%%%%%%%%%%%%%%%%%%%%%%%%%%%%%%%%%%%%%%%%%%%%%%%%

\subsection{Longitudinally polarised partons}
\label{sec:longit-results}

\paragraph{Quark kernels.}
Using the HVBM prescription in the \msbar scheme, one obtains
\begin{align}
   \pr{11}{\widetilde{P}}^{V (1)}_{\Delta q\Delta q,\msbars}(x)
   &= \pr{11}{\widetilde{P}}^{V (1)}_{q q}(x) - 2 C_F \beta_0(1-x)
   \,,
   \label{eqn:DeltaPqqV_singlet_MSbar}
   \\
   \pr{88}{\widetilde{P}}^{V (1)}_{\Delta q\Delta q,\msbars}(x)
   &= \pr{88}{\widetilde{P}}^{V (1)}_{q q}(x)
      + c_{q q}(88) \ms\Biggl\{ 4 C_F \ms C_A (1-x) \ln(x)
            - 2C_F\beta_0 \ms (1-x) \Biggr\}
   \label{eqn:DeltaPqqV_octet_MSbar}
\end{align}
for the quark-to-quark valence kernels.  In the colour singlet case, the relation \eqref{eq:hel-conservation} for helicity conservation in the valence sector can be restored by
\begin{align}
   \pr{1}{\widetilde{Z}}^{(1)}(x)
   &= -4C_F (1-x) \,,
   \label{eqn:1zqq}
\end{align}
as observed in \cite{Vogelsang:1996im}.  Remarkably, the same is achieved in the colour octet case with the choice
\begin{align}
   \pr{8}{\widetilde{Z}}^{(1)}(x)
   &= c_{q q}(88) \; \pr{1}{\widetilde{Z}}^{(1)}(x)
   \,,
\end{align}
given the transformation \eqref{eqn:P_pm_NLO_shift} with $\prn{8}{\gamma}_J^{(0)} = 2 C_A$.
With the finite renormalisation thus specified, the polarised quark kernels are
\begin{align}
   \label{eq:11-V-DeltaqDeltaq}
   \pr{11}{\widetilde{P}}^{V (1)}_{\Delta q\Delta q}(x)
   &= \pr{11}{\widetilde{P}}^{V (1)}_{q q}(x)
   \,,
   \\
   \pr{88}{\widetilde{P}}^{V (1)}_{\Delta q\Delta q}(x)
   &= \pr{88}{\widetilde{P}}^{V (1)}_{q q}(x)
   \,,
   \\[0.2em]
   \pr{11}{\widetilde{P}}^{V (1)}_{\Delta q\Delta \bar{q}}(x)
   &= {}- \pr{11}{\widetilde{P}}^{V (1)}_{q\bar{q}}(x)
   \,,
   \\
   \label{eq:88-V-DeltaqbarDeltaqbar}
   \pr{88}{\widetilde{P}}^{V (1)}_{\Delta q\Delta \bar{q}}(x)
   &= {}- \pr{88}{\widetilde{P}}^{V (1)}_{q\bar{q}}(x)
\intertext{and}
   \pr{11}{\widetilde{P}}^{S (1)}_{\Delta q\Delta q}(x)
   &= \frac{C_F}{2} \ms\Biggl\{ -(1+x)\ln^2(x) + (3x-1)\ln(x) + 1 - x \Biggr\}
   \,,
   \\
   \pr{88}{\widetilde{P}}^{S (1)}_{\Delta q\Delta q}(x)
   &= {}- c_{q q}(88) \, (N^2-2) \,
   \pr{11}{\widetilde{P}}^{S (1)}_{\Delta q\Delta q}(x)
   \,,
   \vphantom{\frac{1}{1}}
   \label{eq:88-S-DeltaqDeltaq}
   \\
   \pr{11}{\widetilde{P}}^{S (1)}_{\Delta q\Delta \bar{q}}(x)
   &= \pr{11}{\widetilde{P}}^{S (1)}_{\Delta q\Delta q}(x)
   \,,
   \vphantom{\frac{1}{1}}
   \\
   \pr{88}{\widetilde{P}}^{S (1)}_{\Delta q\Delta \bar{q}}(x)
   &= 2 c_{q q}(88) \, \pr{11}{\widetilde{P}}^{S (1)}_{\Delta q\Delta\bar{q}}(x)
   \,.
   \vphantom{\frac{1}{1}}
   \label{eq:88-S-DeltaqDeltaqbar}
\end{align}
Notice that \eqref{eq:88-S-DeltaqDeltaq} to \eqref{eq:88-S-DeltaqDeltaqbar} have the same signs and colour factors on the r.h.s.\ as their unpolarised counterparts \eqref{eq:88-S-qq} to \eqref{eq:88-S-qqbar}.

%...............................................................................

\paragraph{Quark-gluon mixing kernels.}
The polarised quark-gluon mixing kernels read
\begin{align}
   \pr{11}{\widetilde{P}}^{(1)}_{\Delta \Sigma^+ \Delta g}(x)
   &= C_F \ms n_f \ms\Biggl\{ \frac{1}{2}\Dp{q}{g}(x)\ln^2(x)
   - 2\Dp{q}{g}(x)\ln(x)\ln(1-x) + \Dp{q}{g}(x)\ln^2(1-x) \notag\\
   &\quad\qquad \quad \ - \frac{9}{2}\ln(x) + 4(1-x)\ln(1-x)
   - \frac{\pi^2}{3}\Dp{q}{g}(x) + \frac{27}{2}x - 11\Biggr\} \notag\\
   &\quad + C_A \ms n_f \ms\Biggl\{ -\frac{3}{2} (2x+1) \ln^2(x)
   - \Dp{q}{g}(x)\ln^2(1-x) \notag\\
   &\qquad\qquad \quad \ + (8x+1) \ln(x) - 4 \ms (1-x) \ln(1-x) \notag\\
   &\qquad\qquad \quad \ + \frac{\pi^2}{6}\Dp{q}{g}(x)
   - S_2(x) \, \Dp{q}{g}(-x) - 11x + 12\Biggr\}
   \,,
   \\
   \pr{8A}{\widetilde{P}}^{(1)}_{\Delta \Sigma^- \Delta g}(x)
   &= c_{q g}(8A) \ms\Biggl\{ \pr{11}{\widetilde{P}}^{(1)}_{\Delta \Sigma^+ \Delta g}(x) \notag\\
   &\qquad\qquad\quad
   + C_A \ms n_f \ms \biggl[ \frac{1}{4} \Bigl( \Dp{q}{g}(x) + 6\Bigr)\ln^2(x)
    - \frac{1}{4} \Bigl(7\Dp{q}{g}(x) + 11\Bigr) \ln(x) \notag\\
   &\qquad\qquad\qquad\qquad\quad
   + S_2(x) \, \Dp{q}{g}(-x) - \frac{9}{2} (1-x)\biggr]\Biggr\}
   \,,
   \\
   \pr{8S}{\widetilde{P}}^{(1)}_{\Delta \Sigma^+ \Delta g}(x)
   &= \frac{c_{q g}(8S)}{c_{q g}(8A)} \;
   \pr{8A}{\widetilde{P}}^{(1)}_{\Delta \Sigma^- \Delta g}(x)
\intertext{and}
   \pr{11}{\widetilde{P}}^{(1)}_{\Delta g \ms \Delta \Sigma^+}(x)
   &= \frac{C_F}{18} \ms\Biggl\{ C_F\biggl[9\Dp{g}{q}(x)\ln^2(x)
   - 18\Dp{g}{q}(x)\ln^2(1-x) + 9 \ms (x-4)\ln(x) \notag\\
   &\qquad\qquad\quad\ - 18 \ms (x+2)\ln(1-x) + 72x - 153\biggr] \notag\\
   &\qquad\quad + C_A\biggl[27 \ms (x+2) \ln^2(x) - 36\Dp{g}{q}(x)\ln(x)\ln(1-x) \notag\\
   &\qquad\qquad\qquad + 18\Dp{g}{q}(x)\ln^2(1-x) + (72-234\ms x) \ln(x)
   \notag\\[0.6em]
   &\qquad\qquad\qquad + 6 \ms (x+10) \ln(1-x) - 3\pi^2\Dp{g}{q}(x) \notag\\
   &\qquad\qquad\qquad - 18 S_2(x) \, \Dp{g}{q}(-x) + 70x + 82\biggr]
   \notag\\[0.3em]
   &\qquad\quad + n_f\biggl[-12\Dp{g}{q}(x) \ln(1-x) - 4x - 16\biggr] \Biggr\}
   \,,
   \\
   \pr{A8}{\widetilde{P}}^{(1)}_{\Delta g \ms \Delta \Sigma^-}(x)
   &= c_{g q}(A8) \ms\Biggl\{ \pr{11}{\widetilde{P}}^{(1)}_{\Delta g \ms \Delta \Sigma^+}(x)
   + \frac{C_F \ms C_A}{4} \biggl[-(x+10) \ln^2(x) + 10 \ms (2x + 1) \ln(x) \notag\\
   &\qquad\qquad\qquad\qquad\qquad\qquad\qquad\quad
   + 4 S_2(x) \, \Dp{g}{q}(-x) + 30(1-x)\biggr]\Biggr\}
   \,,
   \\
   \pr{S8}{\widetilde{P}}^{(1)}_{\Delta g \ms \Delta \Sigma^+}(x)
   &= \frac{c_{g q}(S8)}{c_{g q}(A8)} \;
   \pr{A8}{\widetilde{P}}^{(1)}_{\Delta g \ms \Delta \Sigma^-}(x)
   \,.
\end{align}

%...............................................................................

\paragraph{Gluon kernels.}
For the polarised pure gluon kernels, we have
\begin{align}
   \pr{11}{\widetilde{P}}^{(1)}_{\Delta g\Delta g}(x)
   &= C_F \ms n_f \ms\Biggl\{ -(1+x)\ln^2(x) + (x-5)\ln(x) - 5 \ms (1-x)\Biggr\} \notag\\
   &\quad + C_A^2 \ms\Biggl\{ \Bigl(\Dp{g}{g}(x)+4 \ms (1+x)\Bigr)\ln^2(x)
   - 4\Dp{g}{g}(x)\ln(x)\ln(1-x)
   \notag\\
   &\qquad\qquad + \frac{1}{3} (29-67x) \ln(x)
   + \frac{1}{3} \Bigl( \frac{67}{3} - \pi^2 \Bigr) \Dp{g}{g}(x)
   \notag\\
   &\qquad\qquad - 2 S_2(x) \, \Dp{g}{g}(-x)
   - \frac{19}{2}(1-x)\Biggr\}
   \notag\\
   &\quad + C_A \ms n_f \ms\Biggl\{ -\frac23(1+x)\ln(x)
   - \frac{10}{9}\Dp{g}{g}(x) - 2 \ms (1-x)\Biggr\}
   \,,
   \\
   \pr{AA}{\widetilde{P}}^{(1)}_{\Delta g\Delta g}(x)
   &= c_{g g}(AA) \ms\Biggl\{ C_A^2\biggl[2 \ms (1+x)\ln^2(x)
   - 4\Dp{g}{g}(x)\ln(x)\ln(1-x)
   \notag\\
   &\qquad\qquad\qquad\ \ + \frac{1}{3} (32-40x) \ln(x)
   + \frac{1}{3} \Bigl( \frac{67}{3} - \pi^2 \Bigr) \Dp{g}{g}(x) \biggr]
   \notag\\
   &\qquad\qquad\quad + C_A \ms n_f \biggl[-\frac{1}{2}(1+x)\ln^2(x)
   - \frac{1}{6}(x+19)\ln(x)
   \notag\\
   &\qquad\qquad\qquad\qquad\ \ \, - \frac{10}{9}\Dp{g}{g}(x)
   - \frac92(1-x)\biggr]\Biggr\}
   \,,
   \\
   \pr{SS}{\widetilde{P}}^{(1)}_{\Delta g\Delta g}(x)
   &= \frac{c_{g g}(SS)}{c_{g g}(AA)}\,
   \pr{AA}{\widetilde{P}}^{(1)}_{\Delta g\Delta g}(x) \notag\\
   &\quad + 2 c_{g g}(SS) \ms (C_A-2C_F) \ms n_f
   \ms\Biggl\{ (1+x)\ln^2(x) + (5-x)\ln(x)
   \notag \\
   &\qquad\qquad\qquad\qquad\qquad\qquad \ \,
   + 5 \ms (1-x)\Biggl\}
   \label{eqn:DeltaSPgg(1)}
   \,,
   \\
   \pr{\tentenbar}{\widetilde{P}}^{(1)}_{\Delta g\Delta g}(x)
   &= 0
   \,,
   \\
   \pr{\twensev}{\widetilde{P}}^{(1)}_{\Delta g\Delta g}(x)
   &= c_{g g}(\twensev) \ms\Biggl\{ -\Bigl(15\Dp{g}{g}(x)
   + 12 \ms (1+x)\Bigr)\ln^2(x)
   \notag\\[0.2em]
   &\qquad\qquad\qquad \, {}- 36\Dp{g}{g}(x)\ln(x)\ln(1-x)
   \notag\\[0.6em]
   &\qquad\qquad\qquad \, + (15x+111)\ln(x)
   + 3 \ms \Bigl( \frac{67}{3} - \pi^2 \Bigr) \Dp{g}{g}(x)
   \notag\\[0.2em]
   &\qquad\qquad\qquad \, + 30 S_2(x) \, \Dp{g}{g}(-x)
   + \frac{285}{2} (1-x)
   \notag\\
   &\qquad\qquad\qquad \, + n_f \biggl[ - 2 (1+x)\ln(x)
   - \frac{10}{3} \Dp{g}{g}(x) - 6 \ms (1-x) \biggr]\Biggr\}
   \,.
\end{align}
As in the unpolarised case, the kernels $\pr{RR}{\widetilde{P}}_{\Delta g\Delta g}$ for $R = A$ and $R = S$ are equal at LO but different at NLO.

%%%%%%%%%%%%%%%%%%%%%%%%%%%%%%%%%%%%%%%%%%%%%%%%%%%%%%%%%%%%%%%%%%%%%%%%%%%%%%%%

\subsection{Transversely polarised quarks}
\label{sec:transv-results}

For transverse polarisation, we only have valence kernels.  They read
\begin{align}
   \pr{11}{\widetilde{P}}^{V (1)}_{\delta q\delta q}(x)
   &= C_F \ms\Biggl\{ C_F \biggl[ - \Bigl( 2\ln(x)\ln(1-x)
   + \frac32\ln(x) \Bigr) \delp{q}{q}(x) + 1 - x \biggr] \notag\\
   &\qquad\quad
   + \frac{1}{2} C_A \biggl[
      \Bigl( \ln^2(x) + \frac{11}{3}\ln(x) \Bigl) \delp{q}{q}(x)
      + \frac{1}{3} \Bigl( \frac{67}{3} - \pi^2 \Bigr) \delp{q}{q}(x)
      - (1 - x) \biggr] \notag\\
   &\qquad\quad {}- \frac{1}{9} n_f \Bigl(3\ln(x) + 5\Bigr)
   \delp{q}{q}(x) \Biggr\}
   \,,
   \\
   \pr{88}{\widetilde{P}}^{V (1)}_{\delta q\delta q}(x)
   &= c_{q q}(88) \ms\Biggl\{ \pr{11}{\widetilde{P}}^{V (1)}_{\delta q\delta q}(x)
   - \frac{1}{2} C_F \ms C_A \biggl[ \ln^2(x) \, \delp{q}{q}(x) + 1 - x \biggr]\Biggr\}
   \,,
   \\
   \pr{11}{\widetilde{P}}^{V (1)}_{\delta q\delta \bar{q}}(x)
   &= - \frac{1}{2} C_F \ms (C_A - 2C_F)
   \ms\Biggl\{ 2 S_2(x) \, \delp{q}{q}(-x) - (1 - x) \Biggr\}
   \,,
   \\[0.2em]
   \pr{88}{\widetilde{P}}^{V (1)}_{\delta q\delta \bar{q}}(x)
   &= c_{q q}(88) \, (N^2+1)
   \, \pr{11}{\widetilde{P}}^{V (1)}_{\delta q\delta \bar{q}}(x)
   \,.
   \vphantom{\frac{1}{1}}
   \label{eq:11-V-deltaqbardeltaqbar}
\end{align}

%%%%%%%%%%%%%%%%%%%%%%%%%%%%%%%%%%%%%%%%%%%%%%%%%%%%%%%%%%%%%%%%%%%%%%%%%%%%%%%%

\subsection{Limiting behaviour}
\label{sec:kin-limits}

Let us finally examine the behaviour of our results for $x\to 1$ and $x\to 0$.  In the following, we give the leading behaviour of the kernels in these limits and investigate its dependence on the colour channel.  For the sake of brevity, we will omit the argument $x$ in $\widetilde{P}(x)$.

%...............................................................................

\paragraph{Large $x$.}  For all NLO kernels, we find that the leading term for $x\to 1$ is proportional to the value in the colour singlet channel.  Specifically, we have
\begin{align}
   \pr{88}{\widetilde{P}}_{q q}^{V (1)}
   &\approx
   c_{q q}(88) \, \pr{11}{\widetilde{P}}_{q q}^{V (1)}
   \,,
   \\
   \pr{8A}{\widetilde{P}}_{\Sigma^- g}^{(1)}
   &\approx c_{q g}(8A) \, \pr{11}{\widetilde{P}}_{\Sigma^+ g}^{(1)}
   \,,
   &
   \pr{A8}{\widetilde{P}}_{g \ms \Sigma^-}^{(1)}
   &\approx c_{g q}(A8) \, \pr{11}{\widetilde{P}}_{g \ms \Sigma^+}^{(1)}
   \,,
   \\
   \pr{8S}{\widetilde{P}}_{\Sigma^+ g}^{(1)}
   &\approx c_{q g}(8S) \, \pr{11}{\widetilde{P}}_{\Sigma^+ g}^{(1)}
   \,,
   &
   \pr{S8}{\widetilde{P}}_{g \ms \Sigma^+}^{(1)}
   &\approx c_{g q}(S8) \, \pr{11}{\widetilde{P}}_{g \ms \Sigma^+}^{(1)}
   \,,
   \\
   \pr{RR}{\widetilde{P}}_{g g}^{(1)}
   &\approx
   c_{g g}(RR) \, \pr{11}{\widetilde{P}}_{g g}^{(1)}
   \,,
\end{align}
with the same factors of proportionality as in the exact LO relations \eqref{eq:LO-PVqq} to \eqref{eq:LO-Pgg}.  The flavour diagonal kernels behave like $1/ [1-x]_+\ms$,
\begin{align}
   \pr{11}{\widetilde{P}}_{q q \vphantom{\Delta}}^{V (1)}
   \approx
   \pr{11}{\widetilde{P}}_{\Delta q\Delta q}^{V (1)}
   \approx
   \pr{11}{\widetilde{P}}_{\delta q\delta q}^{V (1)}
   &\approx
   \frac{1}{9}\ms C_F \Bigl( C_A \ms \bigl( 67 - 3\pi^2 \bigr)
   - 10 \ms n_f  \Bigr) \ms
   \frac{1}{[1-x]_+}
   \,,
   \\
   \pr{11}{\widetilde{P}}_{g g \vphantom{\Delta}}^{(1)}
   \approx
   \pr{11}{\widetilde{P}}_{\Delta g\Delta g}^{(1)}
   &\approx
   \frac{C_A}{C_F} \; \pr{11}{\widetilde{P}}_{q q \vphantom{\Delta}}^{V (1)}
   \,,
\intertext{whilst the quark-gluon mixing kernels have a double logarithmic enhancement}
   \pr{11}{\widetilde{P}}_{\Sigma^+ g}^{(1)}
   \approx
   \pr{11}{\widetilde{P}}_{\Delta\Sigma^+ \Delta g}^{(1)}
   &\approx
   - (C_A - C_F) \ms n_f \ln^2(1 - x)
   \,,
   \\
   \pr{11}{\widetilde{P}}_{g \ms \Sigma^+}^{(1)}
   \approx
   \pr{11}{\widetilde{P}}_{\Delta g \ms \Delta\Sigma^+}^{(1)}
   &\approx
   C_F (C_A - C_F) \ms \ln^2(1 - x)
   \,.
\end{align}
Notice that the leading behaviour is polarisation independent, in extension of our statement about the $\delta(1-x)$ terms in \sect{\ref{sec:second-method}}.
The kernels that are zero at LO vanish in the $x\to 1$ limit at NLO, i.e.\ we have
\begin{align}
   \pr{RR}{\widetilde{P}}_{q \bar{q}}^{V (1)} \,,
   \pr{RR}{\widetilde{P}}_{q q}^{S (1)} \,,
   \pr{RR}{\widetilde{P}}_{q \bar{q}}^{S (1)} \,,
   \pr{RR}{\widetilde{P}}_{\Delta q \Delta \bar{q}}^{V (1)} \,,
   \pr{RR}{\widetilde{P}}_{\Delta q \Delta q}^{S (1)} \,,
   \pr{RR}{\widetilde{P}}_{\Delta q \Delta \bar{q}}^{S (1)} \,,
   \pr{RR}{\widetilde{P}}_{\delta q \delta \bar{q}}^{V (1)}
   \; \underset{x\to 1}{\to} \; 0
\end{align}
for both $R=1$ and $R=8$.

%...............................................................................

\paragraph{Small $x$.}
In the limit $x\to 0$, the functional form of the kernels for given partons and polarisation is independent of the colour channel, but we see no simple overall pattern in the prefactors.
For unpolarised partons, we find
\begin{align}
    \pr{11}{\widetilde{P}}_{q q}^{V (1)}
    &\approx
    \frac{1}{2} \ms C_F (C_A - C_F) \ln^2(x)
    \,,
    \\[0.2em]
    \frac{\pr{88}{\widetilde{P}}_{q q}^{V (1)}}{c_{q q}(88)}
    &\approx
    \frac{1}{4} \ms C_F (C_A - 2C_F) \ln^2(x)
    \,,
    \\[0.2em]
    \pr{11}{\widetilde{P}}_{q\bar{q}}^{V (1)}
    &\approx
    - \frac{1}{2} \ms  C_F (C_A - 2C_F) \ln^2(x)
    \,,
    \\[0.3em]
    \pr{11}{\widetilde{P}}_{q q}^{S (1)} = \pr{11}{\widetilde{P}}_{q\bar{q}}^{S (1)}
    &\approx \frac{10}{9}  \ms C_F \, \frac{1}{x}
\end{align}
in the quark sector, with the colour octet kernels $\pr{88}{\widetilde{P}}_{q \smash{\bar{q}}}^{V (1)}$, $\pr{88}{\widetilde{P}}_{q q}^{S (1)}$, and $\pr{88}{\widetilde{P}}_{q \smash{\bar{q}}}^{S (1)}$ being proportional to their colour singlet counterparts according to \eqref{eq:88-V-qqbar}, \eqref{eq:88-S-qq}, and \eqref{eq:88-S-qqbar}.
All kernels involving one or two gluons behave like $1/x$, with coefficients
\begin{align}
    \pr{11}{\widetilde{P}}_{\Sigma^+g}^{(1)}
    &\approx
    \frac{20}{9} \ms C_A \ms n_f \ms \frac{1}{x}
    \,, \\[0.2em]
    \frac{\pr{8A}{\widetilde{P}}_{\Sigma^- g}^{(1)}}{c_{q g}(8A)}
    = \frac{\pr{8S}{\widetilde{P}}_{\Sigma^+ g}^{(1)}}{c_{q g}(8S)}
    &\approx
    \frac{10}{9} \ms C_A \ms n_f \ms \frac{1}{x}
    \,,
    \\[0.2em]
    \pr{11}{\widetilde{P}}_{g\Sigma^+}^{(1)}
    &\approx
    \frac{1}{9} \ms C_F \ms (9 C_A - 20 n_f) \, \frac{1}{x}
    \,,
    \\[0.2em]
    \frac{\pr{A8}{\widetilde{P}}_{g \ms \Sigma^-}^{(1)}}{c_{g q}(A8)}
    = \frac{\pr{S8}{\widetilde{P}}_{g \ms \Sigma^+}^{(1)}}{c_{g q}(S8)}
    &\approx
    \frac{1}{9} \ms C_F \Big(C_A \ms \bigl( 67 - 3\pi^2 \bigr)
      - 20 n_f\Big) \, \frac{1}{x}
\end{align}
and
\begin{align}
    \pr{11}{\widetilde{P}}_{g g}^{(1)}
    &\approx
    \frac{1}{9} \ms \Big[ 6 C_F \ms n_f - 23 C_A \ms n_f \Big] \, \frac{1}{x}
    \,,
    \\[0.2em]
    \frac{\pr{AA}{\widetilde{P}}_{g g}^{(1)}}{c_{g g}(AA)}
    &\approx
    \frac{1}{9} \ms
    \Big[ C_A^2 \ms \big( 67 - 3\pi^2 \big) - 20 C_A \ms n_f \Big] \, \frac{1}{x}
    \,,
    \\[0.2em]
    \frac{\pr{SS}{\widetilde{P}}_{g g}^{(1)}}{c_{g g}(SS)}
    &\approx
    \frac{1}{9} \ms \Big[ C_A^2 \ms \big( 67 - 3\pi^2 \big)
        + 24 C_F \ms n_f - 32 C_A \ms n_f \Big] \, \frac{1}{x}
    \,,
    \\[0.2em]
    \frac{\pr{\twensev}{\widetilde{P}}_{g g}^{(1)}}{c_{g g}(\twensev)}
    &\approx
    \frac{1}{3} \ms
    \Big[ \ms 8 \ms \big( 67 - 3\pi^2 \big) - 23 n_f \Big] \,
    \frac{1}{x}
    \,.
\end{align}
For longitudinally polarised partons, all kernels behave like $\ln^2(x)$ in the small $x$ limit.  The valence kernels are equal or opposite to their unpolarised analogues in all colour channels (see \eqref{eq:11-V-DeltaqDeltaq} to \eqref{eq:88-V-DeltaqbarDeltaqbar}), whereas in the pure singlet case we have
\begin{align}
    \pr{11}{\widetilde{P}}_{\Delta q \Delta q}^{S (1)}
    = \pr{11}{\widetilde{P}}_{\Delta q \Delta\bar{q}}^{S (1)}
    &\approx
    - \frac{1}{2} \ms C_F \ms \ln^2(x)
    \,,
\end{align}
with $\pr{88}{\widetilde{P}}_{\Delta q \Delta q}^{S (1)}$ and $\pr{88}{\widetilde{P}}_{\Delta q \Delta \bar{q}}^{S (1)}$ being proportional to their colour singlet counterparts as specified in \eqs{\eqref{eq:88-S-DeltaqDeltaq}} and \eqref{eq:88-S-DeltaqDeltaqbar}.
For kernels involving gluons, we get
\begin{align}
    \pr{11}{\widetilde{P}}_{\Delta \Sigma^+\Delta g}^{(1)}
    &\approx
    - \frac{1}{2} \ms (C_F + 2 C_A) \ms n_f \ln^2(x)
    \,,
    \\[0.2em]
    \frac{\pr{8A}{\widetilde{P}}_{\Delta \Sigma^- \Delta g}^{(1)}}{c_{q g}(8A)}
    = \frac{\pr{8S}{\widetilde{P}}_{\Delta \Sigma^+ \Delta g}^{(1)}}{c_{q g}(8S)}
    &\approx
    - \frac{1}{4} \ms (2C_F + C_A) \ms n_f \ln^2(x)
    \,,
    \\[0.2em]
    \pr{11}{\widetilde{P}}_{\Delta g \ms \Delta \Sigma^+}^{(1)}
    &\approx
    C_F (C_F + 2 C_A) \ln^2(x)
    \vphantom{\frac{1}{1}}
    \,,
    \\[0.2em]
    \frac{\pr{A8}{\widetilde{P}}_{\Delta g \ms \Delta \Sigma^-}^{(1)}}{c_{g q}(A8)}
    = \frac{\pr{S8}{\widetilde{P}}_{\Delta g \ms \Delta \Sigma^+}^{(1)}}{c_{g q}(S8)}
    &\approx
    \frac{1}{2} \ms C_F (2C_F + C_A) \ln^2(x)
\end{align}
and
\begin{align}
    \pr{11}{\widetilde{P}}_{\Delta g\Delta g}^{(1)}
    &\approx
    \bigl(4 C_A^2 - C_F \ms n_f \bigr) \ln^2(x)
    \,,
    \vphantom{\frac{1}{1}}
    \\[0.2em]
    \frac{\pr{AA}{\widetilde{P}}_{\Delta g\Delta g}^{(1)}}{c_{g g}(AA)}
    &\approx
    \frac{1}{2} \ms \bigl(4 C_A^2 - C_A \ms n_f\bigr) \ln^2(x)
    \,,
    \\[0.2em]
    \frac{\pr{SS}{\widetilde{P}}_{\Delta g\Delta g}^{(1)}}{c_{g g}(SS)}
    &\approx
    \frac{1}{2} \ms \bigl(4 C_A^2 - 8 C_F \ms n_f + 3 C_A \ms n_f\bigr) \ln^2(x)
    \,,
    \\[0.2em]
    \frac{\pr{\twensev}{\widetilde{P}}_{\Delta g\Delta g}^{(1)}}{c_{g g}(\twensev)}
    &\approx
    - 12 \ln^2(x)
    \,.
    \vphantom{\frac{1}{1}}
\end{align}
The kernels for transverse polarisation remain finite in the small $x$ limit, with
\begin{align}
    \pr{11}{\widetilde{P}}_{\delta q \delta q}^{V (1)}
    &\approx
    - \frac{1}{2} \ms C_F (C_A - 2 C_F)
    \,,
    \\[0.2em]
    \frac{\pr{88}{\widetilde{P}}_{\delta q \delta q}^{V (1)}}{c_{q q}(88)}
    &\approx
    - C_F (C_A - C_F)
    \,,
    \\[0.2em]
    \pr{11}{\widetilde{P}}_{\delta q \delta\bar{q}}^{V (1)}
    &\approx
    \frac{1}{2} \ms  C_F (C_A - 2 C_F)
\end{align}
and $\pr{88}{\widetilde{P}}_{\delta q \delta\bar{q}}^{V (1)}$ being proportional to $\pr{11}{\widetilde{P}}_{\delta q \delta\bar{q}}^{V (1)}$ as specified in \eqn{\eqref{eq:11-V-deltaqbardeltaqbar}}.

\section{Summary}
\label{sec:summary}

We have computed the DGLAP kernels for colour dependent DPDs at two-loop order, for unpolarised or longitudinally polarised partons and for transversely polarised quarks.  We use two independent methods, adapting existing calculations for the ordinary two-loop DGLAP kernels \cite{Curci:1980uw, Ellis:1996nn, Vogelsang:1996im, Vogelsang:1997ak} or for the matching of TMDs on PDFs \cite{Echevarria:2016scs, Gutierrez-Reyes:2018iod}.  The first calculation is done in light-cone gauge with a principal value regulator, whereas the second one uses Feynman gauge and the $\delta$ regulator for rapidity divergences.
\rev{With the first method, we obtain the $x$ dependent part of the kernels for all polarisations specified above, and with the second method we compute the full kernels for unpolarised partons and for transverse quark polarisation.  Combining the methods and using that the $\delta(1-x)$ terms are polarisation independent, we obtain the full kernels for all polarisations.}

We find full agreement between the results \rev{that can be obtained with both methods}.  We also confirm the polarisation independence of the terms going with $\delta(1-x)$ by explicit calculation for unpolarised or transversely polarised quarks.  Further cross checks validate our calculations: $(i)$ rapidity divergences of individual graphs cancel in the final results for both methods, $(ii)$ in the calculation with the second method, double and triple poles in $\epsilon$ have the structure required by an RGE analysis, and $(iii)$ our result for the $\zeta$ dependent part satisfies the all-order relation \eqref{eq:gamma-J-K} between the cusp anomalous dimensions for colour octet DPDs and single-gluon TMDs.

For the terms proportional to $\delta(1-x)$ in the splitting kernels, we find the simple structure in \eqn{\eqref{eq:P-master-form}} up to two loops.  In addition to the terms $d_a$ already present in the colour singlet case (which are spin independent and differ only between quarks and gluons), there are terms \mbox{$\prn{R}{c} - \prn{R}{\gamma}_{J} \, \log\sqrt{\zeta}/\mu$} with coefficients that depend only on the dimension of the colour representation $R$.  They are zero for the singlet, and their values for the higher representations $10$ and $27$ obey Casimir scaling relative to the octet.

For the complete kernels $\pr{R\Rp}{P}_{a b}(x)$, we find a number of noteworthy features.
\begin{enumerate}
\item At LO, the $x$ dependent part of each kernel factorises into a colour dependent prefactor and an $x$ dependent function.  This no longer holds at NLO, except for $\pr{RR}{P}_{q \bar{q}}^{V}$, $\pr{RR}{P}_{q q}^{S}$, and $\pr{RR}{P}_{q \bar{q}}^{S}\ms$.  These kernels receive contributions only from one graph, or from one graph and a subtraction term with identical colour structure.
\item The pure singlet kernels $\pr{RR}{P}_{q q}^{S}$ and $\pr{RR}{P}_{q \bar{q}}^{S}$ differ already at two-loop order for $R=8$, whereas for $R=1$ their difference starts only at three loops.
\item The colour octet kernels $\pr{AA}{P}_{g g}$ and $\pr{SS}{P}_{g g}$ are equal at LO but differ at NLO.
\item In the colour octet sector, the kernels describing quark-gluon mixing satisfy the simple relation $\pr{8A}{P}_{\Sigma^- g} : \pr{8S}{P}_{\Sigma^+ g} =
\pr{A8}{P}_{g \Sigma^-} : \pr{S8}{P}_{g \Sigma^+} = N : \sqrt{N^2 -4}$, both at LO and at NLO accuracy.
\item The $x$ dependent part of the kernels in the decuplet sector is zero, and the corresponding two-gluon DPD evolves like a single-gluon TMD, as specified in \eqn{\eqref{eq:DGLAP-10}}.
\item Both for $x\to 1$ and for $x\to 0$, the NLO kernels in colour non-singlet channels become proportional to their colour singlet counterparts.  For large $x$, the factor of proportionality is the same as at LO, whereas for small $x$ it depends on both $N$ and $n_f$ and does not exhibit any simple pattern.
\end{enumerate}
All statements just made also apply to the kernels for polarised partons.  We furthermore find that with a suitably chosen renormalisation scheme transformation, one can achieve quark helicity conservation for the NLO valence kernels not only in the colour singlet sector (which is well known) but also in the colour octet channel.

The evolution kernels we have computed result from the renormalisation of the soft operator \eqref{eq:soft-op} combined with the twist-two operator \eqref{eq:quark-ops} or with its analogues for gluons or polarised partons.  In this sense, the kernels are not specific to the evolution of colour dependent DPDs and can possibly be used in other contexts.  An example are the twist-three operators for TMDs discussed in \cite{Vladimirov:2021hdn}.  We leave it to future work to investigate this in more detail.

Finally, it will be interesting to study the quantitative impact of the NLO kernels on the scale dependence of DPDs with colour correlations.  This is left to future work.

\section*{Acknowledgements}

We are indebted to Werner Vogelsang for discussions and especially for providing us with the results in \tab{\ref{tab:transv-per-graph}}.  Our thanks also go to Peter Pl{\"o}{\ss}l for his careful reading of the manuscript.
This work is in part supported by the Deutsche Forschungsgemeinschaft (DFG, German Research Foundation) -- grant number 409651613 (Research Unit FOR 2926) and grant number 491245950.
A.V. is funded by the \emph{Atracci\'on de Talento Investigador} program of the Comunidad de Madrid (Spain) No.~2020-T1/TIC-20204, and by the Spanish Ministry grant PID2019-106080GB-C21.
The Feynman graphs in this manuscript were produced with JaxoDraw \cite{Binosi:2003yf, Binosi:2008ig}.  In our calculations we used FORM \cite{Vermaseren:2000nd}, as well as the ColorMath package \cite{Sjodahl:2012nk}.

\appendix
\section{Anomalous dimension and renormalisation factor for TMDs}
\label{sec:tmd-renorm}

In this appendix, we give the anomalous dimension and the renormalisation factor for single-parton TMDs up to second order in $a_s$.  These expressions are used in the computation described in \sect{\ref{sec:second-method}}.

The renormalisation factor and anomalous dimension of the twist-two TMD operator
are related in the standard manner by
\begin{align}
   \frac{d}{d\ln\mu} \widehat{Z}_a(\mu,\zeta)
   &= \gamma_{F,a}(\mu,\zeta) \, \widehat{Z}_a(\mu,\zeta)
   \,,
   \label{eqn:RGE_Z_hat}
\end{align}
where
\begin{align}
   \gamma_{F,a}(\mu,\zeta)
   &= \gamma_a(\mu) - \frac{1}{2} \gamma_{K,a}(\mu)\ln\frac{\zeta}{\mu^2}
   \,.
   \label{eqn:TMD_anom_dim_def}
\end{align}
As explained below \eqref{eq:m-tmd-renorm}, $\widehat{Z}_a$ is different for quarks and gluons but does not depend on the polarisation or the colour channel.

Our convention for the perturbative expansion of all anomalous dimensions is
\begin{align}
   \gamma &= \sum_{n=1}^\infty a_s^n \ms \gamma^{(n-1)}
   &
   \text{ for } \gamma = \gamma_{F, a}, \gamma_{a}, \gamma_{K, a}.
\end{align}
The LO coefficients then read
\begin{align}
   \gamma_q^{(0)} &= 3 C_F
   \,,
   &
   \gamma_g^{(0)} &= \beta_0
   \,,
   \\
   \gamma_{K,q}^{(0)} &= 4 C_F
   \,,
   &
   \gamma_{K,g}^{(0)} &= 4 C_A
   \,.
\end{align}
The NLO values can for instance be found in \app{D.2} of \cite{Echevarria:2016scs}.  The notation there is related to the one we are using by
\begin{align}
   \gamma_{K,q}^{(n)} &= 2^{\ms 2-n} \ms C_{F} \, \Gamma^{(n+1)}
   \,,
   &
   \gamma_{K,g}^{(n)} &= 2^{\ms 2-n} \ms C_{A} \, \Gamma^{(n+1)}
   \,,
   &
   \gamma_a^{(n)} &= - 2^{-(n+1)} \ms \gamma_V^{a(n+1)}
   \,,
\end{align}
which gives
\begin{align}
   \gamma_q^{(1)} &= C_F^2 \biggl( \frac34 - \pi^2 + 12 \zeta_3 \biggr)
   + C_F \ms C_A \biggl( \frac{961}{108} + \frac{11}{12} \pi^2 - 13 \zeta_3 \biggr)
   - C_F \ms n_f \Biggl( \frac{65}{54} + \frac16 \pi^2 \Biggr)
   \,,
   \\
   \gamma_g^{(1)} &= C_A^2 \biggl( \frac{346}{27} - \frac{11}{36} \pi^2
   - \zeta_3 \biggr)
   + C_A \ms n_f \biggl( -\frac{64}{27} + \frac{1}{18} \pi^2 \biggr)
   - C_F \ms n_f
   \,,
   \\
   \gamma_{K,q}^{(1)} &= C_F \ms C_A \biggl( \frac{134}{9}
   - \frac{2}{3} \pi^2 \biggr)
   - \frac{20}{9} C_F \ms n_f
   \,,
   \\[0.1em]
   \label{eq:tmd-g-cusp}
   \gamma_{K,g}^{(1)} &= \frac{C_A}{C_F} \, \gamma_{K,q}^{(1)}
   \,.
\end{align}

With the implementation of \msbar subtraction described in \sect{\ref{sec:matching-RGE}}, the coefficients in the $a_s$ expansion \eqref{eq:expand-Z-tmd} of the renormalisation factor are sums of poles,
\begin{align}
   \label{eq:Z-tmd-poles}
   \widehat{Z}_a^{(n)}(\zeta/\mu^2, \epsilon)
   &= \sum_{i=1}^{\infty} \frac{1}{\epsilon^i} \,
   \widehat{Z}_a^{(n,i)}(\zeta/\mu^2)
   \,.
\end{align}
The analysis of these pole terms proceeds in the same way as was described for the DPD renormalisation factor $\pr{R\Rp}{Z}_{a b}$ in \sect{\ref{sec:Z-factor}}, with the simplification that Mellin convolutions are replaced by ordinary products.  As a result, we find that the highest pole at order $a_s^{N}$ is $\epsilon^{-2 N}$ and that the anomalous dimension can be obtained from the single pole as usual:
\begin{align}
   \label{eq:gamma-tmd-from-Z}
   \gamma_{F,a}^{(N-1)}(\zeta/\mu^2)
   &= - 2 N \ms \widehat{Z}_a^{(N,1)}(\zeta/\mu^2)
   \,.
\end{align}
The LO and NLO coefficients in \eqref{eq:Z-tmd-poles} read
\begin{align}
   \widehat{Z}_a^{(1,1)}(\zeta/\mu^2)
   &= - \frac{1}{2}\ms \gamma_{F,a}^{(0)}(\zeta/\mu^2)
   \label{eqn:Z_hat_(1,1)}
   \,,
   \\[0.2em]
   \widehat{Z}_a^{(1,2)}
   &= - \frac{1}{4}\ms \gamma_{K,a}^{(0)}
   \label{eqn:Z_hat_(1,2)}
   \intertext{and}
   \widehat{Z}_a^{(2,1)}(\zeta/\mu^2)
   &= - \frac{1}{4}\ms \gamma_{F,a}^{(1)}(\zeta/\mu^2)
   \label{eqn:Z_hat_(2,1)}
   \,,
   \\[0.2em]
   \widehat{Z}_a^{(2,2)}(\zeta/\mu^2)
   &= \frac{1}{8}\, \biggl\{ \Bigl(\gamma_{F,a}^{(0)}(\zeta/\mu^2)\Bigr)^2
   + \beta_0 \ms \gamma_{F,a}^{(0)}(\zeta/\mu^2)
   - \frac{1}{2}\ms \gamma_{K,a}^{(1)}\biggr\} \notag\\
   &= \frac{1}{8}\, \biggl\{ \frac{1}{4} \Bigl( \gamma_{K,a}^{(0)}
   \ln \frac{\zeta}{\mu^2} \Bigr)^2 - \Bigl( \frac{1}{2} \ms \beta_0
   + \gamma_a^{(0)} \Bigr) \ms \gamma_{K,a}^{(0)} \ln\frac{\zeta}{\mu^2}
   \notag\\[0.2em]
   &\qquad\ \
   + \Bigl(\gamma_a^{(0)}\Bigr)^2
   + \beta_0 \ms \gamma_a^{(0)} - \frac{1}{2}\ms \gamma_{K,a}^{(1)} \biggr\}
   \,,
   \\[0.2em]
   \widehat{Z}_a^{(2,3)}(\zeta/\mu^2)
   &= \frac{1}{8}\ms \gamma_{K,a}^{(0)} \biggl\{ \gamma_{F,a}^{(0)}(\zeta/\mu^2)
   + \frac{3}{4} \ms \beta_0 \biggr\}
   \,,
   \label{eqn:Z_hat_(2,3)}
   \\[0.2em]
   \widehat{Z}_a^{(2,4)}
   &= \frac{1}{32} \ms \Bigl( \gamma_{K,a}^{(0)} \Bigr)^2
   \,.
\end{align}

%%%%%%u%%%%%%%%%%%%%%%%%%%%%%%%%%%%%%%%%%%%%%%%%%%%%%%%%%%%%%%%%%%%%%%%%%%%%%%%%

% the following lines create an entry in the table of contents
\phantomsection
\addcontentsline{toc}{section}{References}

\bibliographystyle{JHEP}
\bibliography{dglap-nlo.bib}

\end{document}